\documentclass[preprint,3p, times, sort&compress]{elsarticle}

\usepackage{amsmath}      
\DeclareMathOperator*{\argmin}{arg\,min}

\usepackage{mathtools, cuted}
\usepackage{amsfonts}         
\usepackage{tikz} 
\usepackage{caption}
\usepackage[labelformat=simple]{subcaption}    
\usepackage[Procedure]{algorithm}
\usepackage[noend]{algpseudocode}
\usepackage{xfrac}
\usepackage{siunitx}

\makeatletter
\renewcommand\p@subfigure{\thefigure\,}

\makeatother

\usepackage{amssymb}
\usepackage{mathrsfs}
\DeclareMathAlphabet\mathbfcal{OMS}{cmsy}{b}{n}

\makeatletter
\def\ps@pprintTitle{%
  \let\@oddhead\@empty
  \let\@evenhead\@empty
  \let\@oddfoot\@empty
  \let\@evenfoot\@oddfoot
}
\makeatother

\begin{document}

\begin{frontmatter}

\title{An isogeometric analysis framework for ventricular cardiac mechanics}

\author[1]{Robin Willems\corref{cor1}}
\cortext[cor1]{Corresponding author}
\ead{R.Willems@tue.nl}
\author[1]{Koen L.P.M. Janssens}
\author[1]{Peter H.M. Bovendeerd}
\author[2]{Clemens V. Verhoosel}
\author[3,4]{Olaf van der Sluis}
\address[1]{Faculty of Biomedical Engineering, Cardiovascular Biomechanics, Eindhoven University of Technology, The Netherlands}
\address[2]{Faculty of Mechanical Engineering, Energy Technology and Fluid Dynamics, Eindhoven University of Technology, The Netherlands}
\address[3]{Faculty of Mechanical Engineering, Mechanics of Materials, Eindhoven University of Technology, The Netherlands}
\address[4]{Philips Research, AI, Data Science and Digital Twin Department, High Tech Campus 34 Eindhoven, The Netherlands}

\begin{abstract}
The finite element method (FEM) is commonly used in computational cardiac simulations. For this method, a mesh is constructed to represent the geometry and, subsequently, to approximate the solution. To accurately capture curved geometrical features many elements may be required, possibly leading to unnecessarily large computation costs. Without loss of accuracy, a reduction in computation cost can be achieved by integrating geometry representation and solution approximation into a single framework using the Isogeometric Analysis (IGA) paradigm. In this study, we propose an IGA framework suitable for echocardiogram data of cardiac mechanics, where we show the advantageous properties of smooth splines through the development of a multi-patch anatomical model. A nonlinear cardiac model is discretized following the IGA paradigm, meaning that the spline geometry parametrization is directly used for the discretization of the physical fields. The IGA model is benchmarked with a state-of-the-art biomechanics model based on traditional FEM. For this benchmark, the hemodynamic response predicted by the high-fidelity FEM model is accurately captured by an IGA model with only 320 elements and 4,700 degrees of freedom. The study is concluded by a brief anatomy-variation analysis, which illustrates the geometric flexibility of the framework. The IGA framework can be used as a first step toward an efficient workflow for an improved understanding of, and clinical decision support for, the treatment of cardiac diseases like heart rhythm disorders.
\end{abstract}


\end{frontmatter}

\section{Introduction}\label{sec:Intro}
The integration of patient-specific computational models into clinical practice, which aid in understanding biophysical phenomena and provide decision support for clinicians, has become more frequent over the past decades \cite{PatientspecificBiventricleHermite, Patientspecific3Dprintedheart, Antiga2008AnIM, Taylor2010ImageBasedMO}. These models should be robust, provide meaningful, accurate and timely feedback, and limit the costs associated with (a.o.) the time spent by the clinician in diagnosis and treatment selection and application.

The long-term goal of our ongoing research is the development of a cardiac mechanics model for patients who have suffered from a Myocardial Infarction (MI) and are at high risk of developing Ventricular Tachycardias (VTs), a cardiac disease in which the ventricles show an abnormal increase in rhythm. An MI is caused by a blockage in the arteries that supply oxygenated blood to the myocardium, the muscular layer of the heart. It is expected that an MI located in the left ventricle has the potential to cause VTs over a time span of several years which involves (a.o.) growth and remodeling~\cite{Sutton2003}. VTs lead to decreased pumping performance, becoming life-threatening when left untreated. Current treatments involve using an implantable cardioverter-defibrillator (ICD) and preventive ablation therapy \cite{SANTANGELI20161552}. While ICDs prevent the occurrence of VTs, they have a detrimental effect on the quality of life \cite{ICDQOL}. Ablation is therefore a viable alternative, but is only successful for 50-80\% of the patients \cite{Ablation} due to the complexity of the disease and various treatment approaches employed between clinicians. A patient-specific electromechanical cardiac model that makes use of the existing clinical workflow data can provide more insight into the disease, thereby providing decision support for clinicians in a systematic way. This is believed to lead to an increased success rate of ablation therapy~\cite{deLepper2022,Arevalo2016,Sung2021}.

Various patient-specific cardiac models have been proposed over the years, focusing on a specific disease or phenomenon \cite{PatientspecificBiventricleHermite, DrugsHeartKuhl, WholeheartAtria}. These models are becoming increasingly more complex as they incorporate a broad range of physiological multiscale aspects, \emph{e.g.}, a computational model of the whole human heart was recently introduced in Ref.~\cite{FEDELE2023115983}. Such high-fidelity models are often based on high-resolution scan data to construct the anatomical model \cite{EMmodelTrayaReview,overviewmodels, IntegratedHeartQuart,ZHANG20121130}, \emph{i.e.}, the geometrical dimensions, fiber-sheet structure, and scar tissue location or other anomalies. Typical high-resolution imaging methods in clinics today are magnetic resonance imaging (MRI) and computer tomography (CT), which provide the cardiac geometry, and diffusion tensor MRI (DT-MRI), used to visualize the fiber-sheet structure. Methods such as marching cubes/volumes \cite{Marchingcubes} and immersed (isogeometric) analysis \cite{Verhoosel2015} exist to construct a discretized (meshed) geometry based on such high-resolution data. The obtained mesh serves as the basis for computational analysis using a discretized physical model.

High-resolution anatomic scan-data as input for a patient-specific computational analysis is, however, not always available. In the case of VT patients, for example, MRI and/or CT scans are not standard in the clinical workflow. Patients that are susceptible to VTs are systematically monitored to determine if and when treatment should be applied. Part of this clinical workflow is to monitor the patient's cardiac function (\emph{e.g.}, pump efficiency, valve efficiency) using a (2D) echocardiogram. This ultrasound imaging technique is relatively low-cost and low-resolution. The echo data only shows the general contours of the ventricles (and sometimes the atria). Furthermore, depending on the echo view, geometrical features may be lost or obscured by imaging artifacts. The lack of resolution and loss of information in echo data hinders the patient-specific computational workflow as applied to high-resolution scans, requiring an alternative modeling strategy in which the geometry is interpolated in the region where data is scant. Provided that this interpolated geometry is meshable, simulations can then be conducted in a manner similar to what is done for high-resolution data.

\begin{figure*}[ht]
\centering
\begin{tikzpicture}
    \node[anchor=south west,inner sep=0] at (0,0) {\includegraphics[width=\textwidth]{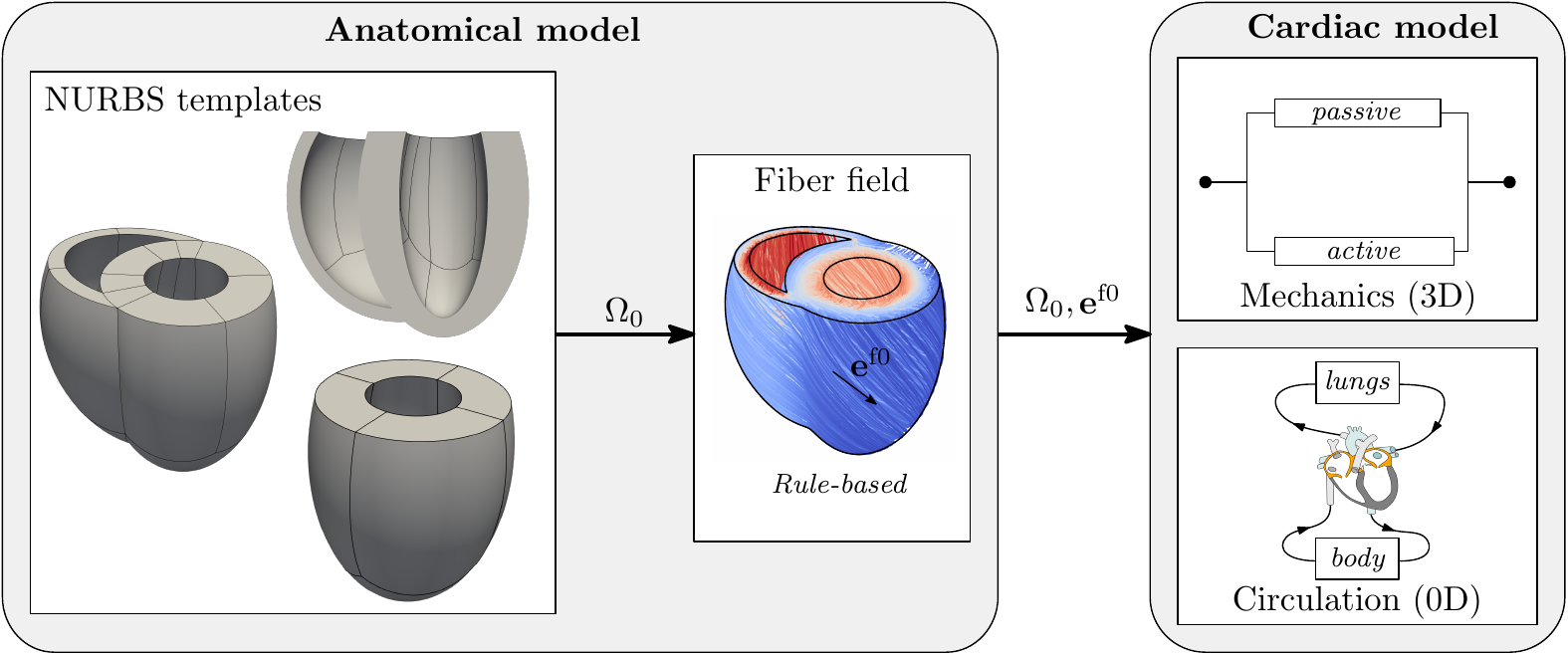}};
    \node at (5.8, -0.4) {\footnotesize (a)};
    \node at (14.5 , -0.4) {\footnotesize (b)};
\end{tikzpicture}
  \caption{\label{fig:workflow} The IGA model workflow consists of two main components: the anatomical model and the cardiac model. The anatomical component relies on the parametrized NURBS (Non-Uniform Rational B-Splines) template of the bi-ventricle and single left ventricle, which is assumed to be stress-free and denoted by $\Omega_0$. A rule-based fiber field is then constructed based on the cardiac shape of which the fiber direction denoted by $\mathbf{e}^{\rm{f0}}$ is of importance to the cardiac model.} The geometrical properties, ${\Omega_0}$ and $\mathbf{e}^{\rm{f0}}$, will be used as input for the cardiac model which solves the relevant physics and consists of a three-dimensional (3D) mechanical and zero-dimensional (0D) circulatory component.  
\end{figure*}

Splines are frequently used in industry for geometry representation in computer-aided design (CAD) tools, on account of their smoothness~\cite{NURBSbook}. In the setting of anatomical features, splines are not used for design but rather for interpolation of segmented anatomical data~\cite{ukwatta2015}. In particular, for low-resolution or sparse (\emph{e.g.}, only a limited number of two-dimensional cross-sections are observed, rather than the complete three-dimensional object) geometrical data, splines have the ability to extract a realistic geometry. When the spline-based geometry is meshed for analysis purposes using linear elements, a significant number of elements may be introduced to capture curved geometries (typical for biophysical pathologies). Depending on the goal of the simulation, these elements introduced to capture the geometry might not be essential to representing the solution. In this contribution, we propose a robust patient-specific computational workflow suitable for low-resolution scan data, which does not introduce such unnecessary additional elements. The proposed model is based on the Isogeometric Analysis (IGA) paradigm \cite{Hughes2005,Bazilevs2009}, in which splines are used to both parametrize the geometry (\emph{i.e.}, the anatomical model) and to discretize the physical model (\emph{i.e.}, the cardiac mechanics model), without the need for potentially laborious geometry clean-up and meshing operation~\cite{FEDELE2021Mesh}. By circumventing the introduction of a large number of elements to capture the geometry, the isogeometric approach can have a competitive edge over approaches based on (low-order, piecewise polynomial) meshes.

The proposed workflow in which we leverage the advantages of splines for geometry construction is visualized in Figure~\ref{fig:workflow}. In this work, we develop an idealized parametrized bi- and left-ventricle NURBS (Non-Uniform Rational B-Splines) template geometry, ${\Omega_0}$, suitable for nonlinear mechanics. The splines allow for capturing strong gradients in tissue properties that are typical for cardiac pathologies while limiting the degrees of freedom (DOFs). Since the global behavior of the cardiac model -- which comprises a coupled three-dimensional (3D) mechanics component and a zero-dimensional (0D) circulatory component  (Figure~\ref{fig:workflow}~(b)) -- can be approximated well using relatively few DOFs, the ability of IGA to solve the physical model directly on the spline geometry parameterization avoids the introduction of mesh refinements (and thereby additional DOFs) to capture the geometry with sufficient detail. Since in the considered data-scan scenario information on the fiber field, $\mathbf{e}^{\rm{f0}}$, is generally missing, a rule-based-method (calibrated on averaged population data) \cite{PIERSANTI2021113468} is used. Our work provides the basis for a patient-specific computational methodology based on a sparse point cloud (\emph{e.g.}, obtained from echocardiogram data), but the development of this methodology -- including a fitting algorithm that deforms the NURBS template -- is beyond the scope of this manuscript.

Over the past years, the suitability of IGA for biomedical applications has been demonstrated for a broad range of problems, including blood flow in arteries \cite{FSIarteryIGA, DIVI2022114648}, the fluid dynamics of an idealized left-ventricle \cite{IGAfluid2017Tagliabue}, the structural analysis of aortic valves \cite{MORGANTI2015508,KAMENSKY20151005}, and trabecular bone structural analysis \cite{Verhoosel2015}. However, the literature on IGA cardiac electro-mechanical models is limited, especially with respect to the mechanical analysis. Bucelli~\emph{et al.} \cite{IGAquarteroni} used a multi-patch NURBS approach to analyze the electrophysiology (neglecting the mechanical response) of the ventricles, based on high-resolution data. The benefit of higher-order basis functions with high continuity across mesh element boundaries has also been exploited in Refs.~\cite{PATELLI2017248, PEGOLOTTI201952}, where the electrophysiology of the atria is studied. An alternative approach to the spline-based parametrization that is employed in IGA is the use of Hermite elements, which, unlike IGA, has already been considered in electromechanical cardiac simulations~\cite{GeomEcho2011Kerckhoffs,GeomHermit2016Krishnamurthy,ZHANG20121130,PatientspecificBiventricleHermite}. Such Hermite elements allow for a smooth geometry and solution approximation, and also limit the number of elements required to capture curved features. In this sense, Hermite elements are similar to the spline-based elements used in IGA. However, in contrast to IGA, Hermite elements require the computation of nodal derivatives in addition to the nodal displacements. IGA provides a more generic framework for higher-order continuous discretizations, in the sense that it gives control over the order and regularity of the spline basis. Furthermore, due to the availability of spline operations in CAD software~\cite{NURBSbook}, IGA is flexible with respect to the parametrization of complex geometries. As part of these operations, various mesh refinement and order-elevation techniques have been developed.

The novelty of our research resides in the development of a parameterized NURBS bi-ventricle template geometry, suitable for nonlinear electro-mechanical analysis. A generalized template construction procedure based on multiple patches is proposed and can be used to produce a broad range of analysis-suitable geometries. The developed cardiac mechanics model uses a monolithic solver to couple the spline-based 3D nonlinear mechanics model to the 0D circulatory system model. The isogeometric analysis results are benchmarked using a state-of-the-art biomechanics finite element model~\cite{bovendeerdlvstrain}, which also illustrates some important differences between the two analysis paradigms. Subsequently, the benchmarked isogeometric simulation framework is used to investigate the influence and sensitivity of geometrical variations on the mechanical response of the heart. 

The structure of this paper is as follows. In Section~\ref{sec:AnatomicalModel} we start by explaining the anatomical model of the heart, focusing on the parameterized NURBS geometry and the corresponding fiber field description. Next, in Section~\ref{sec:CardiacModel} the cardiac model is discussed, where we focus on the mechanics and simplify the electrophysiology using a constant activation time model. Section~\ref{sec:IGA} then discusses the IGA  discretization, along with some essential computational aspects. The IGA results are then compared to finite element analysis (FEA) results and other benchmark problems in Section~\ref{sec:Results}. The results section is concluded with a bi-ventricle anatomy-variation analysis, illustrating the geometric flexibility of the IGA workflow. Conclusions and recommendations are presented in Section~\ref{sec:ConcRec}.

\section{The anatomical model} \label{sec:AnatomicalModel}
\begin{figure*}[ht]
\begin{tikzpicture}
    \node[anchor=south west,inner sep=0] at (0,0) {\includegraphics[width=\textwidth]{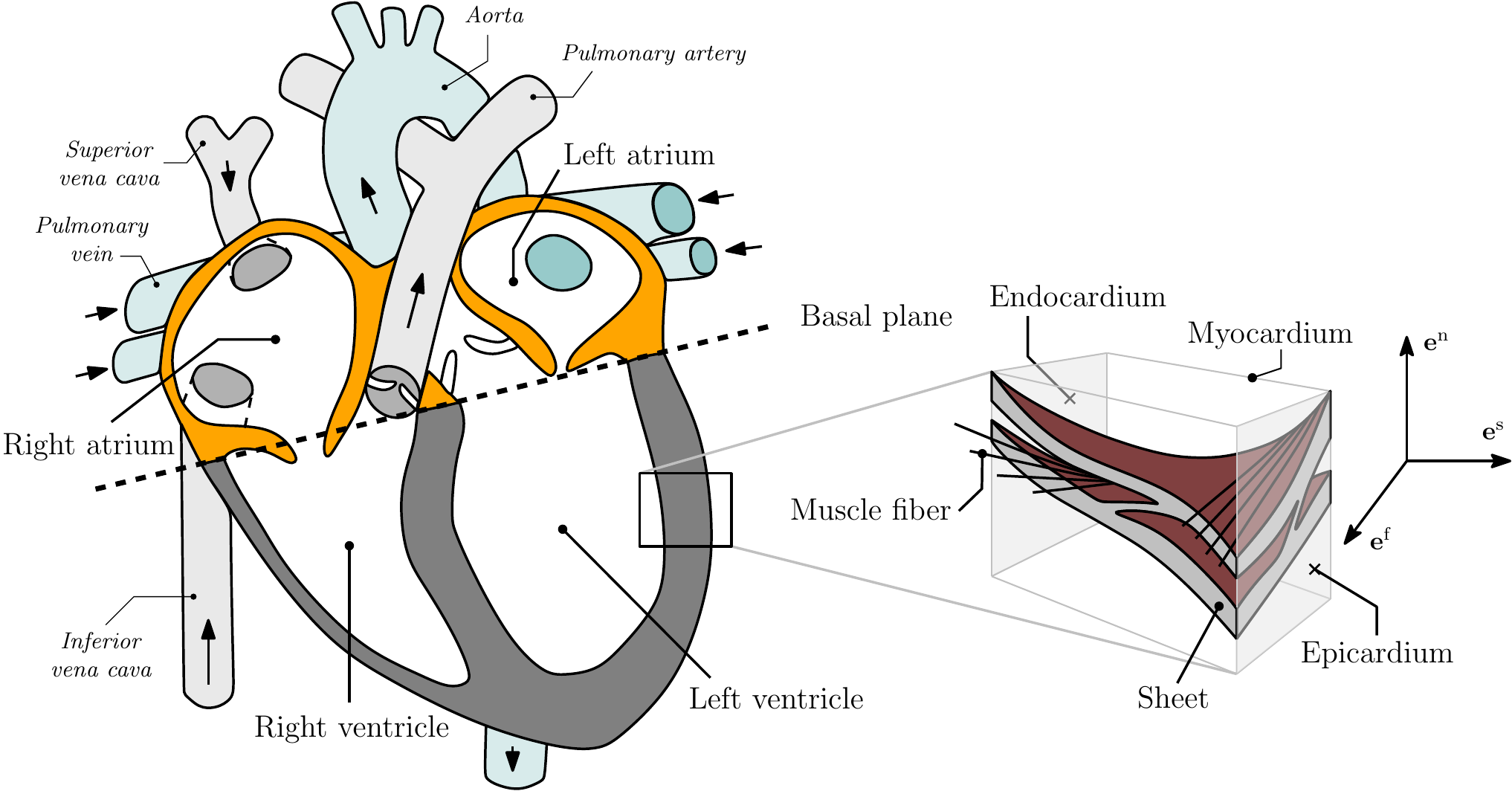}};
    \node at (5.5, -0.5) {\footnotesize  (a)};
    \node at (12.6 , -0.5) {\footnotesize  (b)};
\end{tikzpicture}
  \caption{\label{fig:heart_illustration} (a) Anatomical illustration of the human heart, indicating the two atria (in orange) and ventricles (in dark grey). The arrows show the blood flow direction, where oxygenated blood is flowing from the pulmonary veins into the left atrium and subsequently into the left ventricle, and deoxygenated blood is flowing from the inferior and superior vena cava into the right atrium to the right ventricle. (b) Zoom of a transmurally-cut myocardium block, which consists of sheets and muscle fibers that change in orientation across the thickness \cite{CardiacStructure}. The directions are defined by a local basis $\{\mathbf{e}^{\rm{f}},\mathbf{e}^{\rm{s}},\mathbf{e}^{\rm{n}}\}$ consisting of the fiber direction, $\mathbf{e}^{\rm{f}}$, the sheet direction, $\mathbf{e}^{\rm{s}}$, and the sheet-normal direction, $\mathbf{e}^{\rm{n}}$.}    
\end{figure*}

The workflow pursued in this paper (see Figure~\ref{fig:workflow}) relies on the construction of an (idealized) anatomical model of the human heart. This section discusses the procedure to construct this template geometry. To set terminology, in Section~\ref{sec:theheart} we commence with a brief description of the aspects of the anatomy of the human heart which are essential to our work. We then define the parameters of the template geometry in Section~\ref{sec:idealized}, after which a general NURBS multi-patch construction procedure is explained in Section~\ref{sec:multipatch}. Section~\ref{sec:RRBM} finally discusses the representation of the  fiber field.

\subsection{Essential anatomical aspects of the human heart}\label{sec:theheart}
The heart is a four-chambered muscular organ, visualized in Figure~\ref{fig:heart_illustration}, which is responsible for maintaining a constant supply of oxygenated blood and nutrients to the organs while relieving them from waste products. It is positioned inside the pericardium, \emph{i.e.}, a double-walled sac filled with pericardial fluid. The heart can be divided into a right and left side, each consisting of an atrium and ventricle chamber. The left side pumps blood to the systemic circulation via the aorta and the right side pumps blood to the pulmonary circulation via the pulmonary artery. The left atrium collects the blood from the pulmonary circulation via the pulmonary vein and consequently fills the left ventricle again. The right atrium collects the blood from the systemic circulation via the two large veins (vena cavae) after which it fills the right ventricle. When both ventricles are filled, a depolarization wave spreads across the myocardial wall, initiating muscle contraction. For in-depth information regarding the (multi-)physics and scales involving the cardiac cycle, \emph{i.e.}, electrophysiology, molecular mechanics, sub-cellular processes, etc., the reader is referred to Ref.~\cite{IntegratedHeartQuart}. Since the considered cardiac disease affects the ventricles and no geometrical information is known for the atria, it is a common approach in cardiac modeling to truncate the ventricles at the basal plane, Figure~\ref{fig:heart_illustration} (a) \cite{GeomEcho2011Kerckhoffs,bovendeerdlvstrain, MRI2010Niederer}.

Considering the ventricles' myocardium, it is composed of a complex macroscopic morphological structure that has been studied extensively \cite{Heartstructure1995, Heartstructure2005,Lombaert20121436}. The myocardium consists of myocardial cells or myocytes, which have a directional distribution. Aligned myocytes form muscle fibers or myofibers and are typically idealized as cylindrical objects, indicating the general direction of the myocardial cells. The myofibers contain numerous sarcomeres which are responsible for the contraction. The muscle fibers are located and connected in sheets. The sheets are interconnected by collagen layers and are, therefore, only loosely coupled to each other when compared to the coupling of muscle fibers within the sheet. The muscle fiber and sheet directions also show a transmural variation, \emph{i.e.}, variation through the thickness of the wall, visualized in Figure~\ref{fig:heart_illustration} (b) \cite{CardiacStructure,Anatomy1,Anatomy2}, which make the myocardium a locally orthotropic material.

\subsection{Idealized geometry}\label{sec:idealized}
\begin{figure*}[ht]
\centering
\includegraphics[width=1.\textwidth]{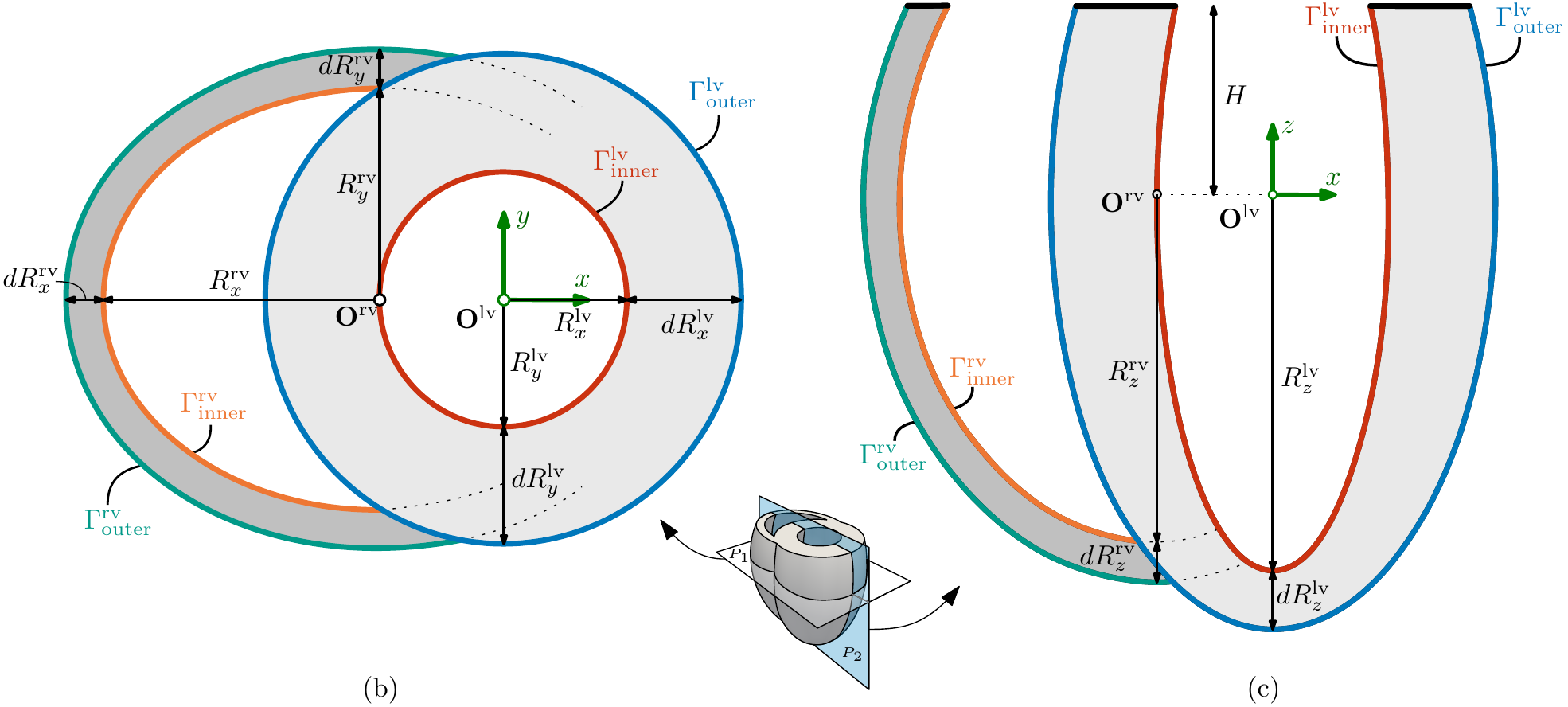}
  \caption{\label{fig:Dimensions} Idealized representation of the left and right ventricles. The 3D ventricles are defined by parameters on two orthogonal planes, $P_1: \ xy-\text{plane}$ and $P_2: xz-\text{plane}$, according to the global coordinates ${x,y,z}$. (a) Parameters defined on the $P_1-\text{plane}$, with $R$ the radius, $dR$ the wall thickness, and $O$ the origin of the considered ellipsoid. (b) Parameters defined on the $P_2-\text{plane}$, where $H$ is the truncation height of the basal plane.}    
\end{figure*}

We herein employ a common simplification of the ventricles (see Figure~\ref{fig:Dimensions}), using ellipsoids for both the left and right ventricles which are truncated at the basal plane~\cite{bovendeerdlvstrain,Pluijmert2017,GeomEcho2011Kerckhoffs}. Moreover, the considered idealized model assumes a smooth representation of the epi- and endocardium and neglects the papillary muscles and trabeculations inside the ventricles which are responsible for opening the mitral and tricuspid valves. This simplification has been widely used for understanding cardiac mechanics and also allows for analytically defined fiber fields that closely represent experimental data \cite{bovendeerdlvstrain,HolzapfelFiber}.

The idealized geometry is parametrized by 14 parameters if we constrain the origin of the global coordinate system. Parameter values are specified by the user but are subject to geometrical constraints, \emph{e.g.}, the left and right ventricles should intersect, or physiological constraints, \emph{e.g.}, the maximum heart size. It is therefore common to make certain parameters interdependent, \emph{e.g.}, $R^{\rm{lv}}_x=R^{\rm{lv}}_y$, or dependent on a \textit{secondary} parameter, \emph{e.g.}, a relation between the wall and cavity volumes~\cite{Pluijmert2017}. Ultimately, the parametrization visualized in Figure~\ref{fig:Dimensions} allows for a wide variety of geometrical variation, \emph{e.g.}, variable wall thickness is achieved by adjusting $dR^{\rm{lv}}_{i},dR^{\rm{rv}}_{i}$ for $i\in\{x,y,z\}$. These radii parameters could be derived from low-resolution image data such as echocardiograms as an initial guess of the patient's anatomy.
Note that the parametrization of the bi-ventricle, as shown in Figure~\ref{fig:Dimensions}, could be extended in future work by introducing an independent inner and outer radius for the septum wall, similar to Ref.~\cite{Pluijmert2017}. 
\subsection{Multi-patch NURBS geometry}\label{sec:multipatch}
To perform an isogeometric analysis, the curved and smooth idealization of the ventricles should be converted to an IGA-suitable geometry. We make use of Non-Uniform Rational B-splines (NURBS), which, in contrast to regular B-splines, enable the exact representation of circles and ellipses. The physical geometry is obtained by mapping the NURBS defined in a parametric domain to the physical domain, \emph{i.e.}, the ellipsoid, using a geometrical map. The topological representation of both domains is defined as a patch that is $C^0$-continuous at its boundaries. The complex topology of the bi-ventricles does not allow for a single patch representation without distorting the elements \cite{IGAquarteroni}. Such a distortion, which is related to the Jacobian of the geometrical map, would especially be detrimental to our numerical simulations that will involve large (mesh) deformations \cite{IntegratedHeartQuart}. To avoid such issues, we employ a multi-patch IGA approach, in which multiple patches are used to represent the geometry. As a consequence, we have to handle the $C^0$-continuity at patch boundaries with care.

To formalize the geometric setting of our problem, let $\Omega \subset \mathbb{R}^{d}$ be an open, bounded computational domain defined in a $d$-dimensional space, which consists of $n_{\rm{patch}}$ conforming subdomains, $\Omega^{\rm{P}}$, referred to as patches, where $\rm{P}=\{1,...,n_{\rm{patch}}\}$. The closed domain, $\overline{\Omega}$, is then defined as the union of the individual closed subdomains $\overline{\Omega}^{\rm{P}}$: 
\begin{equation}
\label{eq:MULTIPATCH}
    \overline{\Omega} =\bigcup^{n_{\rm{patch}}}_{\rm{P}=1} \overline{\Omega}^{\rm{P}}   \quad \Omega^i \cap \Omega^j = \varnothing \quad i \neq j.
\end{equation}
The individual patch domains, $\overline{\Omega}^{\rm{P}}$, are all mapped from a single parametric domain, $\widehat{\Omega} \subset \mathbb{R}^{\hat{d}}$ where $\hat{d} \leq d$. In the following, we will only consider patches in 3D (physical) spaces, $d=3$, and denote the closed domain that is mapped from a parametric domain of dimension $\hat{d}$, by a subscript, $\overline{\Omega}_{\hat{d}}$. 

To construct a trivariate multi-patch geometry, we first consider a univariate ($\hat{d}$=1) spline basis, $\{R_{k,p}\left( \xi \right)\}$ with $k=\{1,...,n_{\mathrm{cps}}\}$, of degree $p$, corresponding to patch $\rm{P}$, with parametric coordinate, $\xi$. The mapped patch-domain, $\mathbf{C}_{p}\left(\xi \right)$, is then defined as
\begin{equation}
\label{eq:MAPP}
    \overline{\Omega}^{\rm{P}}_{\hat{d}=1} := \mathbf{C}_{p}\left(\xi \right) = \sum^{n_{\rm{cps}}}_{k=1} R_{k,p}\left( \xi \right) \mathbf{P}_{k} ,
\end{equation}
where ${\{\mathbf{P}_{k}\} \subset \mathbb{R}^{d}}$ are the control point coordinates which are defined in the physical space and $n_{\rm{\mathrm{cps}}}$ are the number of control points (cps). The univariate basis functions defined on the parametric domain on a single patch are chosen to be rational (NURBS) of degree $p$ and are given as
\begin{equation}
\label{eq:NURBS}
    R_{k,p}\left( \xi \right) = \frac{N_{k,q} \left(\xi \right) w_k }{\sum^{n_{\rm{cps}}}_{j=1} N_{j,p}\left(\xi \right) w_{j}}\qquad k=\{1,...,n_{\mathrm{cps}}\},
\end{equation}
where $\{N_{k,p}\left(\xi \right)\}$ are univariate B-splines of degree $p$ and $\{w_k\}$ are the weights of control point $k$.

The use of univariate basis functions results in a curve in the physical domain, Equation~\eqref{eq:MAPP}. This can be easily extended to surfaces and solids if the basis functions are chosen to be bivariate and trivariate, respectively. The reader is referred to Ref.~\cite{NURBSbook} for details on how these surfaces and volumes are constructed.
From Equation~\eqref{eq:MAPP} it is apparent that we have to find the control point coordinates in the physical domain, $\{\mathbf{P}_{k}\}$, as well as the corresponding weights $\{w_k\}$, that fit the domain defined in Equation~\eqref{alg:NURBS}. In the remainder of this section, we propose a general construction procedure using multi-patch NURBS for complex shapes that produces non-distorted elements.  

\begin{figure*}[ht]
     \centering
     \begin{subfigure}[b]{0.3\textwidth}
         \centering
         \includegraphics[width=\textwidth]{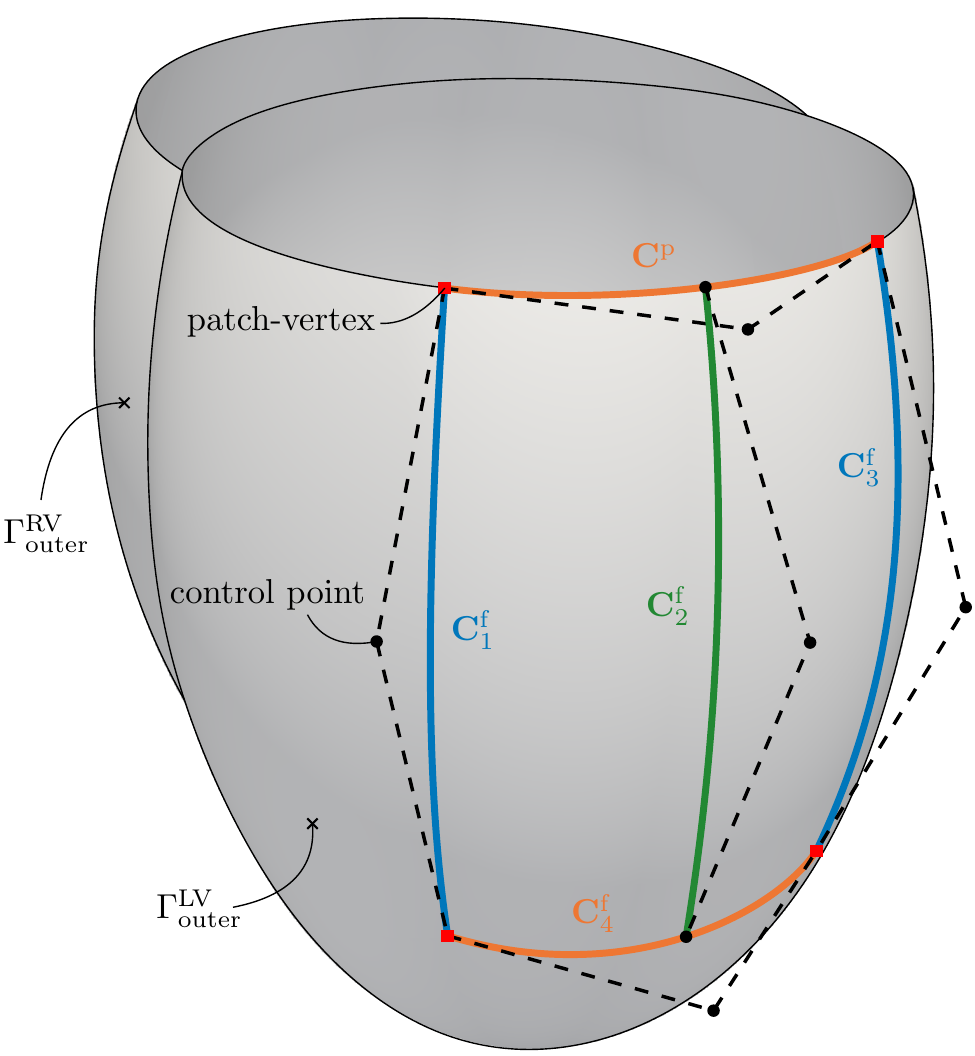}
         \caption{}
         \label{fig:step1}
     \end{subfigure}
     \hfill
     \begin{subfigure}[b]{0.28\textwidth}
         \centering
         \includegraphics[width=\textwidth]{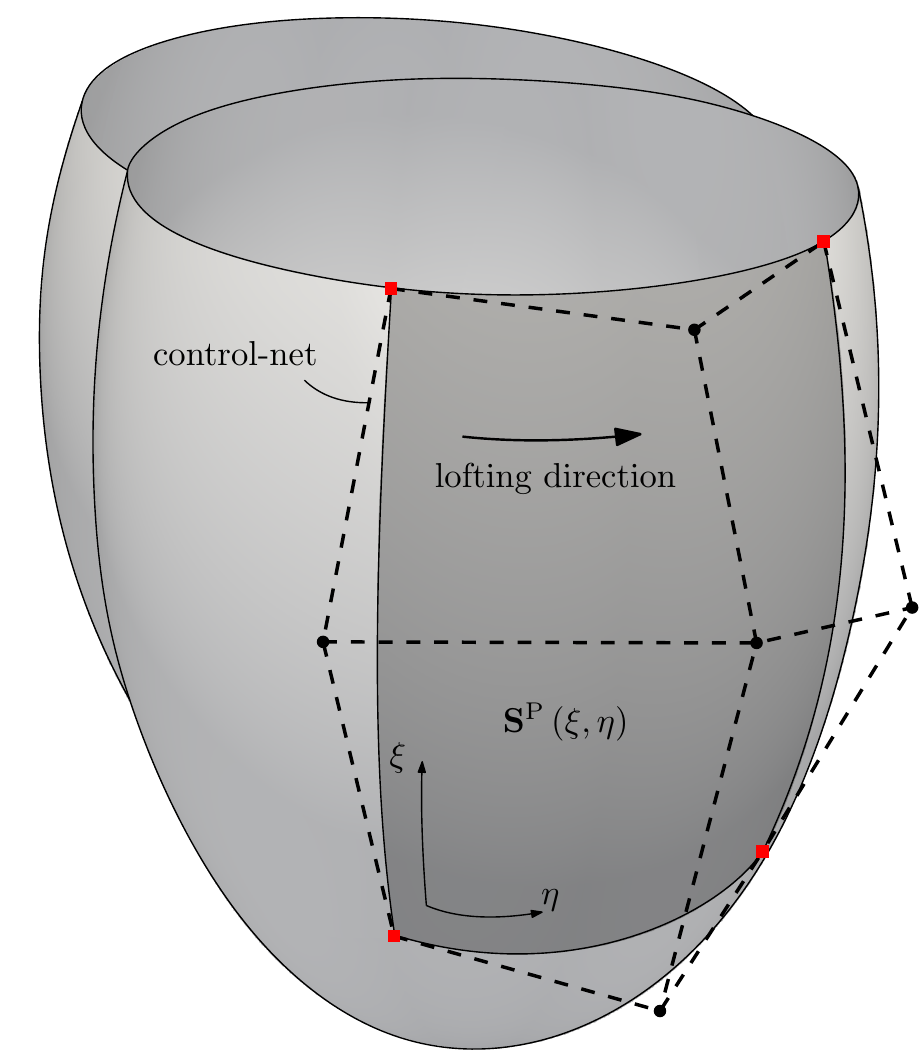}
         \caption{}
         \label{fig:step2}
     \end{subfigure}
     \hfill
     \begin{subfigure}[b]{0.268\textwidth}
         \centering
         \includegraphics[width=\textwidth]{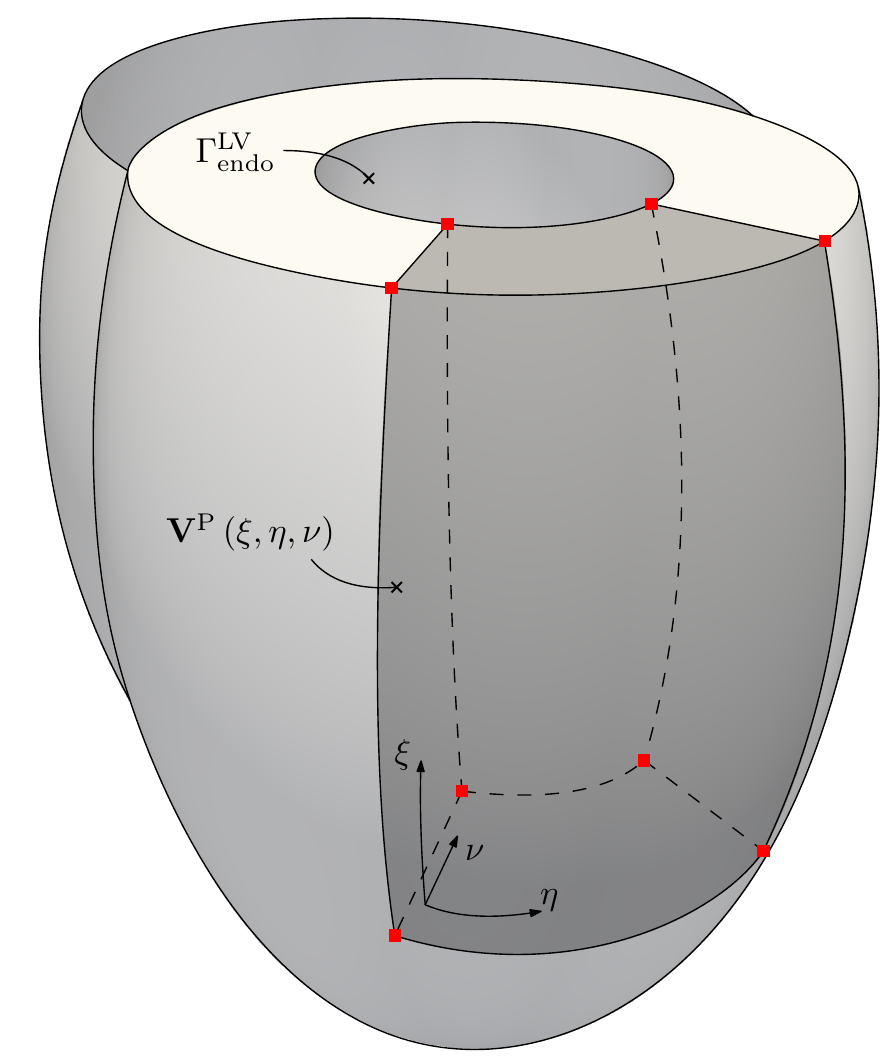}
         \caption{}
         \label{fig:step3}
     \end{subfigure}
        \caption{Construction procedure for a single patch on the left outer ventricle (ellipsoid) surface, $\Gamma^{\rm{LV}}_{\rm{outer}}$, as explained in Procedure~\ref{alg:NURBS}. (a) Patch vertices are positioned on the truncated ellipsoid in \emph{Step~(i)}, between which NURBS curves are constructed following \emph{Step~(ii)}. The constructed curves, colored in orange and blue, represent the patch boundaries. An additional curve is positioned within the patch, colored in green, of which boundary control points are positioned on the Greville points of the corresponding patch boundary curve. (b) The $\rm{P}-$th surface patch, $\overline{\Omega}^{\rm{P}}_{\hat{d}=2}:=S^{\rm{P}}_{i}\left( \xi, \eta \right)$, is constructed from the curves of \emph{Step~(ii)} by means of lofting (or skinning) \cite{NURBSbook}. The lofting direction is determined by the orientation chosen for the green curve inside the surface patch as seen in~(a). (c) The solid $\rm{P}-$th patch $\overline{\Omega}^{\rm{P}}_{\hat{d}=3}:=V^{\rm{P}}_{i}\left( \xi, \eta, \nu \right)$, is obtained by interpolating the inner and outer patch surfaces. Interpolation is such that the control points within the wall correspond to the Greville points. The visualized procedure is then repeated for every patch to form the final multi-patch geometry.}
        \label{fig:proceduresteps}
\end{figure*}

\subsubsection*{Construction procedure}
The procedure of constructing the multi-patch bi-ventricle geometry consists of 4 steps. The geometry can be constructed in a generic way by traversing through the dimensions by the subsequent construction of: \\

\emph{(i)~points} $\rightarrow$~\emph{(ii)~curves} $\rightarrow$~\emph{(iii)~surfaces} $\rightarrow$~\emph{(iv)~solids}. \\

\noindent
The construction routine is listed in Procedure~\ref{alg:NURBS} and visualized in Figure~\ref{fig:proceduresteps}.
\begin{algorithm}
\centering
\caption{The multi-patch NURBS construction procedure, consisting of 4 steps.}\label{alg:NURBS}
    \begin{algorithmic}
        \State $\tilde{\Gamma}_{\rm{b}} \gets \{ \Gamma^{\rm{lv}}_{\rm{inner}}, \Gamma^{\rm{lv}}_{\rm{outer}} ,\Gamma^{\rm{rv}}_{\rm{inner}}, \Gamma^{\rm{rv}}_{\rm{outer}} \}$
        \vspace{0.5em}
        \For{$\Gamma_{\rm{b}}$ \textbf{in} $\tilde{\Gamma}_{\rm{b}}$ }
            \vspace{1em}
            \State \parbox[t]{400pt}{\textit{Step~(i):} Specify vertices of patch $\overline{\Omega}_{\hat{d}= 2}$ on $\Gamma_{\rm{b}}$} 
            \vspace{1em}
                \For{\rm{P} \textbf{in} $[ 1,..., n_{\rm{patch}} ]$  }
                    \vspace{1em}
                    \State \parbox[t]{400pt}{\textit{Step~(ii):} Construct curves between vertices, $\mathbf{C}_i(\xi)$}
                    \vspace{1em}
                    \State \parbox[t]{400pt}{\textit{Step~(iiia):} Construct patch surface $\overline{\Omega}^{\rm{P}}_{\hat{d}=2}:= \mathbf{S}^{\mathrm{P}}(\xi,\mu)$ from curves} 
                    \vspace{1em}
                \EndFor
            \vspace{1em}
            \State \parbox[t]{400pt}{\textit{Step~(iiib):} Construct multi-patch surface $\overline{\Omega}_{\hat{d}= 2}$}
            \vspace{1em}
        \EndFor
        \State \parbox[t]{450pt}{\textit{Step~(iv):} Construct multi-patch volume $\overline{\Omega}_{\hat{d}= 3}:=\mathbf{V}^{\mathrm{P}}(\xi,\mu,\nu)$ from multi-patch surfaces $\overline{\Omega}_{\hat{d}= 2}$}
    \end{algorithmic}
\end{algorithm}

The proposed procedure depends on a spatial mathematical description of the shape that describes the surface boundary of the geometry in the physical space, \emph{i.e.}, an ellipsoid, sphere, plane \emph{etc.} In our case, the parametrized ventricles, illustrated in Figure~\ref{fig:Dimensions}, are described by ellipsoidal surface objects that are linearly interpolated through the wall to obtain a volumetric representation. As a result, adjustments to the parameters defined in Figure~\ref{fig:Dimensions} consequently alter the mathematical description of the ventricle on which the procedure is executed. The procedure consists of the following steps (listed in Procedure~\ref{alg:NURBS}): \emph{Step~(i)}, the procedure is initiated by specifying the patch-vertices on these boundary descriptions, denoted by $\tilde{\Gamma}_{\rm{b}} \subset \mathbb{R}^3 $ (Figure~\ref{fig:Dimensions}). The vertex locations are dependent on the parameter values and are specified \emph{a priori} such that the mapped patch surface patch, $\overline{\Omega}^{\rm{P}}_{\hat{d}= 2}$, yields satisfactory element shapes. The exact vertex locations were empirically determined. 
\emph{Step~(ii)}, by constraining these patch-vertices and evaluating each surface patch individually, we first construct a quadratic single element NURBS curve at the four boundaries that define the considered surface patch (Figure~\ref{fig:step1}). Additional curves are constructed inside the surface patch as well, which is required when constructing the NURBS surface. The curve construction is achieved by a minimization problem, which is explained in the next section. 
\emph{Step~(iiia)}, the constructed curves of \emph{Step~(ii)} are then used to construct the patch surface, $\overline{\Omega}^{\rm{P}}_{\hat{d}= 2}$. This is achieved using lofting (or skinning) \cite{NURBSbook} (Figure~\ref{fig:step2}), which is dependent on the number of inner curves specified inside the patch (minimum of 1). 
\emph{Step~(iiib)}, the individual patch surfaces are combined to form a $C^0$-continuous multi-patch surface, $\overline{\Omega}_{\hat{d}= 2}$, for each boundary listed in $\tilde{\Gamma}_{\rm{b}}$. 
\emph{Step~(iv)}, the NURBS solid, $\overline{\Omega}_{\hat{d}=3}$, is then obtained by linear interpolation of the multi-patch surfaces through the wall (Figure~\ref{fig:step3}).

\begin{figure*}[ht]
     \centering
     \begin{subfigure}[b]{0.24\textwidth}
         \centering
         \includegraphics[width=\textwidth]{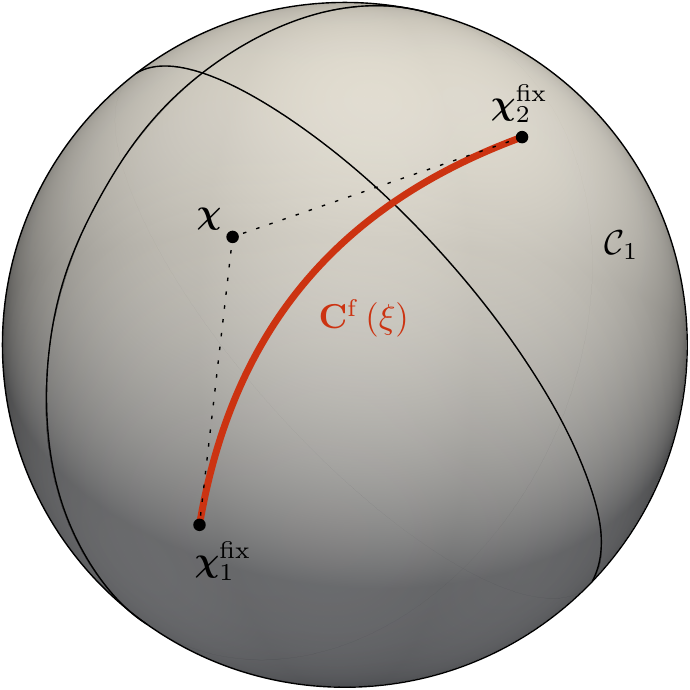}
         \caption{}
         \label{fig:curve}
     \end{subfigure}
     \hfill
     \begin{subfigure}[b]{0.3\textwidth}
         \centering
         \includegraphics[width=\textwidth]{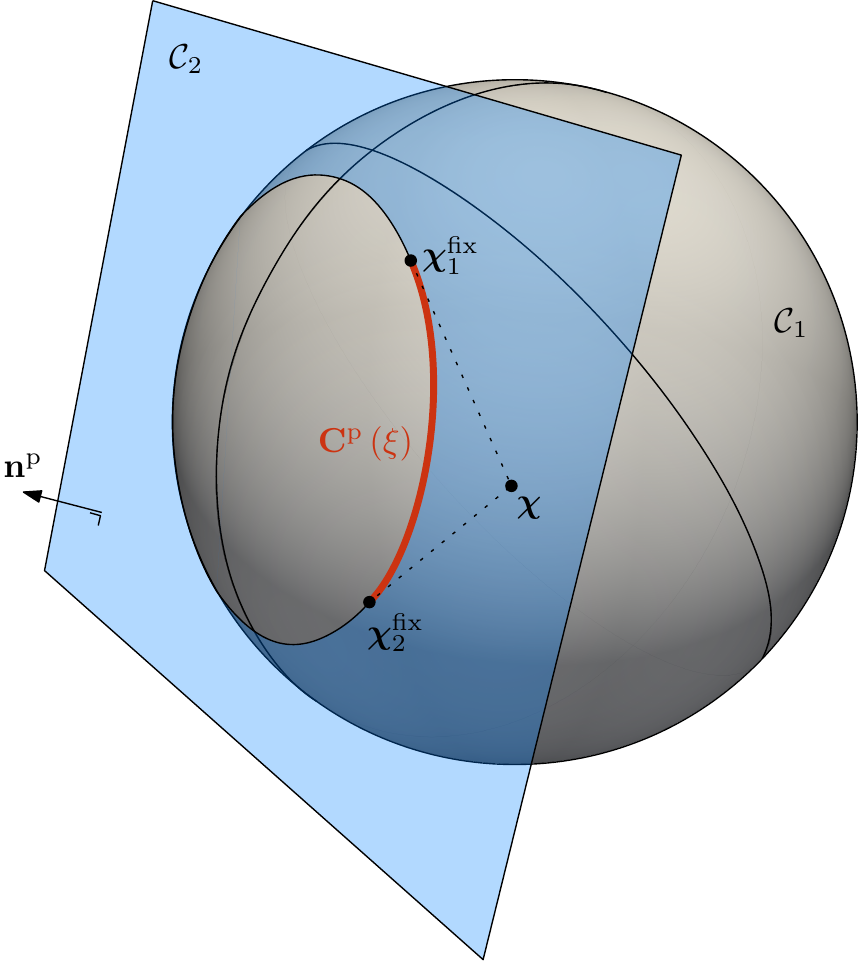}
         \caption{}
         \label{fig:surface}
     \end{subfigure}
     \hfill
     \begin{subfigure}[b]{0.3\textwidth}
         \centering
         \includegraphics[width=\textwidth]{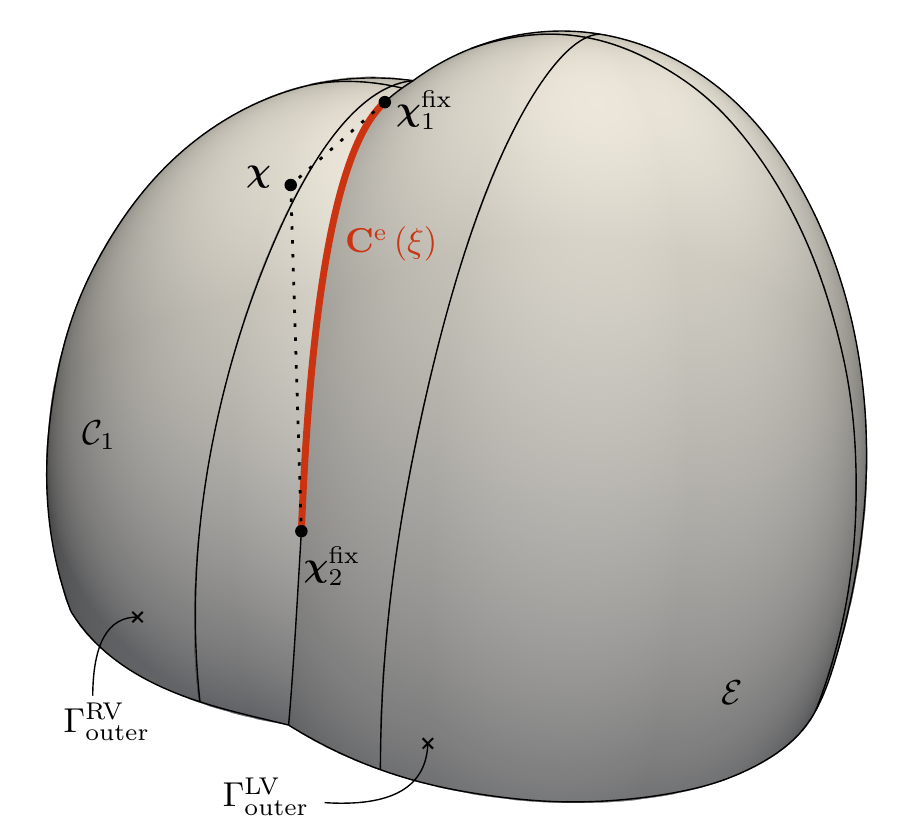}
         \caption{}
         \label{fig:ellips}
     \end{subfigure}
        \caption{Illustration of the three different types of curves, $C^{\rm{f}}_i,C^{\rm{p}}_i,C^{\rm{e}}_i$, specified in Table~\ref{tab:functions}. All three types of curves are used to construct the entire bi-ventricle multi-patch geometry. (a) $C^{\rm{f}}_i$ NURBS curve located on the unit sphere, \emph{i.e.}, an ellipsoid with unit radii. (b) $C^{\rm{p}}_i$ NURBS curve located on the intersection curve of a unit sphere, $\mathcal{C}_1$, and a plane, $\mathcal{C}_2$. The orientation of the plane is defined according to the normal, $n^{\rm{p}}_i$. (c) $C^{\rm{e}}_i$ NURBS curve located on the intersection curve between two ellipsoids, $\mathcal{E}$ and $\mathcal{C}_1$, representing the left and right ventricles respectively.}
        \label{fig:curvesfit}
\end{figure*}

\subsubsection*{NURBS curve construction, Step~(ii)}
The idealized geometry, illustrated in Figure~\ref{fig:Dimensions}, consists of two ellipsoids on which, as explained in Procedure~\ref{alg:NURBS}, NURBS curves are to be constructed between two constrained patch-vertices. Provided that the patch vertices are positioned correctly, we can construct a single-element quadratic NURBS curve that is positioned exactly on the ellipsoid or sphere (see Figure~\ref{fig:curve}). This is because NURBS provide an exact representation of circles and ellipses (ellipses are linear transformations of circles). This can be visualized by an intersection curve between an ellipsoid and a plane (a single element quadratic NURBS curve is always located in a plane in 3D space), which is an ellipse (see Figure~\ref{fig:surface}). The position and weight of the remaining control point and weight in 3D space must still be determined, satisfying the constraints on the boundary cps and weights (patch-vertices).

To determine this final control point position and weight, consider a single element quadratic NURBS curve as in Equation~\eqref{eq:MAPP} and \eqref{eq:NURBS}, with degree $p=2$ and $n_{\rm{cps}}=3$ and $\xi\in \left[ 0,1\right]$, as a function of the unknown control point, $\boldsymbol\chi$, and weight, $\omega$, at index $k=2$ or $\mathbf{P}_2 = \boldsymbol\chi$ and $w_2 = \omega$:
\begin{equation}
\label{eq:NURBS_adj}
    \mathbf{C}_i\left(\xi; \boldsymbol\chi, \omega\right) = \sum^3_{k=1} R_{k}\left(\xi, \omega \right) \mathbf{P}_{k}\left( \boldsymbol\chi \right),
\end{equation}
where
\begin{equation}
    R_{k}\left(\xi; \omega \right) = \frac{N_k \left(\xi \right) w_k\left(\omega \right)}{\sum^3_{j=1} N_{j}\left(\xi \right) w_{j}\left(\omega \right)},
\end{equation}
subject to the following constraints
\begin{subequations}
\begin{align}
    \mathbf{P}_{1} &= \boldsymbol\chi^{\rm{fix}}_1, \quad  w_1 = 1 \quad \text{at} \quad \xi =0, \\
    \mathbf{P}_{3} &= \boldsymbol\chi^{\rm{fix}}_2, \quad w_3 = 1 \quad \text{at} \quad \xi =1, 
\end{align}
\end{subequations}
in which $\{N_k \left(\xi \right)\}$ are the univariate B-spline basis functions of degree two with a single-element knot vector. The constrained control points of the curve boundaries that correspond to the patch-vertices of \emph{Step~(i)} are denoted by $\boldsymbol\chi^{\rm{fix}}_1$ and $\boldsymbol\chi^{\rm{fix}}_2$.

For the current bi-ventricular application, three curve types are defined: $\mathbf{C}^{\rm{f}}$, $\mathbf{C}^{\rm{p}}$, $\mathbf{C}^{\rm{e}}$ (see Figure~\ref{fig:curvesfit}), where each curve should satisfy a set of geometrical constraints. These curves are obtained by solving for the unknown control point coordinate, $\boldsymbol\chi$, and the corresponding weight, $\omega$. Let $\mathbfcal{X}$ be the space that contains the solution for $\boldsymbol\chi_i$ and $\omega$.

The curves are then defined by the minimization problem
\begin{equation}
\label{eq:argmin}
    \mathbf{C}^{\rm{c}} = \mathbf{C}\left(\xi; \boldsymbol\chi,\omega\right), \quad \boldsymbol\chi, \omega = \argmin_{\left( \tilde{\boldsymbol\chi}, \tilde{\omega}, \tilde{\lambda}_j \right) \in \mathbfcal{X}\times\mathcal{W}\times \mathcal{V}}\left( \mathcal{L}\left(\tilde{\boldsymbol\chi}, \tilde{\omega} , \tilde{\lambda}_j \right)  \right),
\end{equation}
where the general functional, $\mathcal{L}\left( \cdot \right)$, is defined as
\begin{equation}\label{eq:functional}
        \mathcal{L}\left(\boldsymbol\chi, \omega , \lambda_j\right) = \int \frac{1}{2}\mathcal{E}^2\left(\xi; \boldsymbol\chi, \omega \right) \text{d} \xi  + \sum^{n_{\rm{cons}}}_{j=1} \int \lambda\left(\xi\right) \cdot \mathcal{C}_j\left(\xi; \boldsymbol\chi, \omega \right) \text{d} \xi,
\end{equation}
and where, $\mathcal{E}\left( \cdot \right)$, is the target function which is subject to the constraint functions $\mathcal{C}_j\left( \cdot \right)$ that are enforced by the Lagrange multiplier fields $\lambda_j\left( \xi \right)$. These Lagrange multiplier fields share the same (B-spline) basis functions, but have different coefficients, and are given by 
\begin{equation}
    \lambda_j \left( \xi \right)= \sum^3_{i=1} N_i \left( \xi \right)\hat{\lambda}_{ij} \quad \text{with} \quad j = \{1,..,n_{cons}\},
\end{equation}
where $\{\hat{\lambda}_{ij}\}$ is the set of Lagrange multiplier coefficients at the $i$-th control point and $j$-th constraint function\footnote{The introduction of Lagrange multipliers results in a saddle point problem that might be subject to numerical instabilities. However, the computations for the considered cases all yield stable solutions.}. The energy term and constraint functions in Equation~(\ref{eq:functional}) are different for each curve type, as illustrated in Figure~\ref{fig:curvesfit} and listed in Table~\ref{tab:functions}. In general, we solve Equation~(\ref{eq:argmin}) directly on the ellipsoid, but one could choose to solve it on the unit sphere and map the resulting curve to the ellipsoid as a final step. However, we noticed that solving on the ellipsoid yield less distorted surface patches which are constructed using the result Equation~(\ref{eq:argmin}). This effect is most noticeable in highly curved areas of the ellipsoid since the resulting curve is a geodesic due to the local strain-based target function.

The nonlinear minimization problem of Equation~(\ref{eq:argmin}) is solved monolithic and yields exact results for canonical test-cases, \emph{i.e.}, NURBS curve located on an arbitrary circle. 
\begin{table*}[ht]
\centering
\caption{Target, $\mathcal{E}\left( \cdot \right)$, and constraint functions, $\mathcal{C}_i\left( \cdot \right)$, of the minimization problem specified in Equation~(\ref{eq:functional}). The radii of the left and right ventricles are respectively listed in a diagonal matrix, such that ${\mathbfcal{R}^{k}=\text{diag}\left[R^{\rm{k}}_{x}, R^{\rm{k}}_{y}, R^{\rm{k}}_{z} \right]^{-1}}$ with ${k = \rm{lv}, \rm{rv}}$. See Figure~\ref{fig:Dimensions} for the parameter definitions and Figure~\ref{fig:curvesfit} for a graphical illustration of the different curve types.}\label{tab:functions}
\begin{tabular}{lccl}
\hline
          & $\mathcal{E}\left( \cdot \right)$                                                   & $\mathcal{C}_1\left( \cdot \right)$                                                                      & \multicolumn{1}{c}{$\mathcal{C}_2\left( \cdot \right)$} \\ \hline
 $ \mathbf{C}^{\rm{f}}$ & $\left\| \frac{\partial \mathbf{C}\left( \cdot \right)}{\partial \xi } \right\|$                                                              & $ \left\| \mathbfcal{R}^{\rm{lv}}  \mathbf{C}\left( \cdot \right)\right\|^2 - 1$                                                                                                       &                                                                       \\
$\mathbf{C}^{\rm{p}}$ &  $ \left\| \frac{\partial \mathbf{C}\left( \cdot \right)}{\partial \xi }\right\|$                                                             & $\left\| \mathbfcal{R}^{\rm{lv}}  \mathbf{C}\left( \cdot \right) \right\|^2 - 1$                                                                                                      & \multicolumn{1}{c}{$\mathbf{n}^{\rm{P}} \mathbf{C}\left( \cdot \right) - d$}                            \\
$\mathbf{C}^{\rm{e}}$ & $\left\| \mathbfcal{R}^{\rm{lv}} \mathbf{C}\left( \cdot \right) \right\|^2 - 1$ & $\left\| \mathbfcal{R}^{\rm{rv}} \left( \mathbf{C}\left( \cdot \right) - \mathbf{O}^{\rm{rv}} \right) \right\|^2 - 1$ & \\    \hline                                                                 
\end{tabular}
\end{table*}

\subsubsection*{From curves to surface, Step~(iii)}
The NURBS curves determined in \emph{Step~(i)} are used to construct individual surface patches, which are consequently combined to form a multi-patch surface. The process of blending NURBS curves together to form a surface is referred to as \textit{lofting} (or \textit{skinning}) \cite{NURBSbook}. 
Lofting considers a set of NURBS curves, $\{\mathbf{C}^{\rm{c}}\left( \xi \right) \}$, referred to as section curves, with superscript $\rm{c}$ the curve constraint type as defined in Table~\ref{tab:functions}. A \textit{lofted} surface patch, $\overline{\Omega}^{\rm{P}}_{\hat{d}=2} := \{ \mathbf{S}^{\rm{P}}\left( \xi, \eta \right) \}$, with new parametric direction, $\eta$ (referred to as the longitudinal direction) can then be constructed by interpolating through the section curves such that they become isoparametric curves on the resulting lofted surface, $\mathbf{C}^{\rm{c}}\left( \xi \right) \subset \overline{\Omega}^{\rm{P}}_{\hat{d}=2}$. Although lofting is a powerful and widely used tool, the resulting surface shape depends on the position, shape, and number of section curves, as well as on the interpolation used in the {$\eta$-direction}. Additional challenges emerge when dealing with rational curves since the surface construction then takes place in the homogeneous space, \emph{i.e.}, $\mathbb{R}^{d+1}$, due to the weights associated with the control points. Because of this, we limit ourselves to single-element quadratic NURBS, as discussed in \emph{Step~(i)}, in which the shapes are mostly similar, \emph{i.e.}, circular. The position of the section curves is strongly related to the longitudinal interpolation choice. In this application, we position the section curves such that they intersect with the Greville abscissae \cite{Bazilevs2009} (averages of the knots) of the boundary curves, illustrated in orange in Figure~\ref{fig:curve}. The longitudinal interpolation, \emph{i.e.}, the parametric distance between the section curves in $\eta$-direction, is therefore chosen to be identical to the Greville points. A brief illustrative comparison of the number of section curves is given in \ref{app:testcase}.

This surface-construction procedure is repeated for every surface patch, $\overline{\Omega}^{\rm{P}}_{\hat{d}=2}$ with P=$\{1,...,n_{\rm{patch}}\}$, after which these patches are connected to form the multi-patch surface $\overline{\Omega}_{\hat{d}=2}$, in accordance with Equation~\eqref{eq:MULTIPATCH}. The patch interfaces, $\overline{\Omega}^{i}_{\hat{d}=2}\cap \overline{\Omega}^{j}_{\hat{d}=2} $ with $i \neq j$, are conforming, both in terms of the knot vector, the spline degree, and in terms of the geometry (identical control points and weights). This is inherent to \emph{Step~(i)}, in which the patch-boundary curves are first constructed, after which the patch surfaces are constructed in \emph{Step~(ii)}. The final \emph{Step~(iv)} involves interpolating the inner and outer multi-patch surfaces defined on $\Gamma^{i}_{\rm{outer}}$ and $\Gamma^{i}_{\rm{inner}}$ for $i=\rm{lv},\rm{rv}$, such that a volumetric multi-patch domain is obtained, $\overline{\Omega}_{\hat{d}=3}$. 

In the remainder, we will omit the parametric dimension subscript of the geometrical domain and always assume $\hat{d}=3$, unless stated otherwise. 
\begin{figure*}[ht]
     \centering
     \begin{subfigure}[b]{0.45\textwidth}
         \centering
         \includegraphics[width=\textwidth]{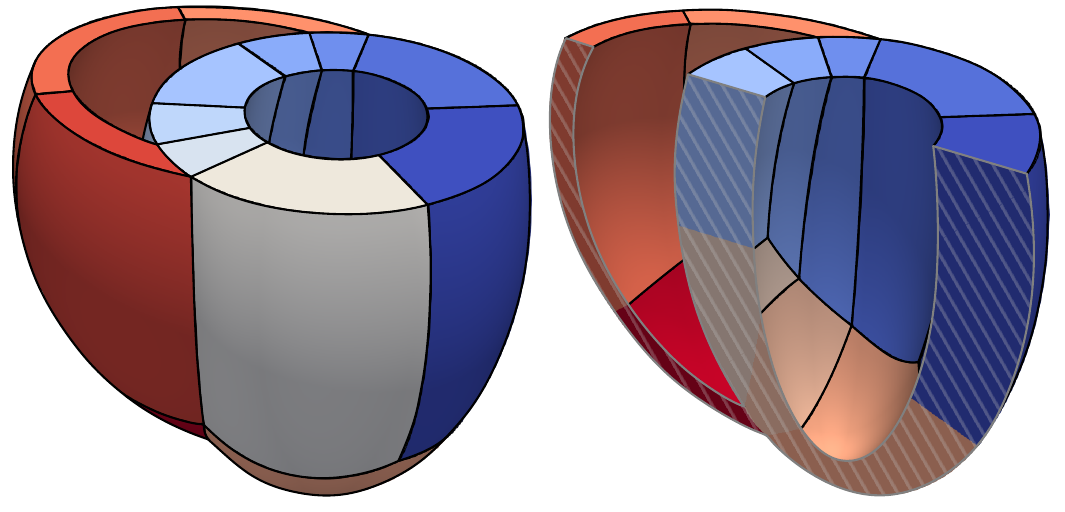}
                  \vspace{2em}
                  \caption{}
                  \label{fig:BVTemplate}
     \end{subfigure}
     \hfill
     \begin{subfigure}[b]{0.45\textwidth}
         \centering
         \includegraphics[width=\textwidth]{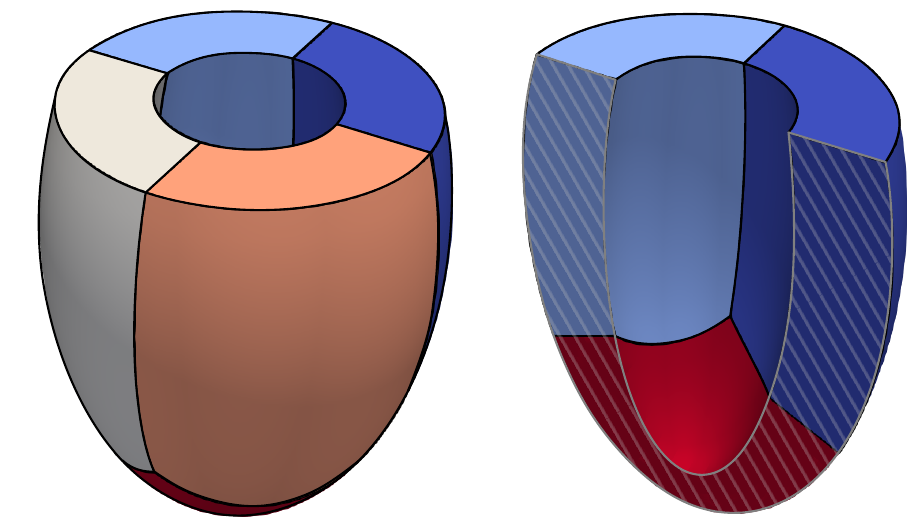}
                 \vspace{0.2em}
         \caption{}
         \label{fig:LVTemplate}
     \end{subfigure}
        \caption{Template representation of the bi-ventricles (a) and the left ventricle (b) which are constructed using the NURBS multi-patch procedure (different colors represent different patches). The bi-ventricle template (a) requires a minimum of $15$ single-element patches to represent the geometry, while the left ventricle template (b) only requires $5$ single-element patches. Additional elements can be used to construct the template, by specifying additional section curves during the lofting procedure, \emph{Step~(iii)}.}
        \label{fig:Templates}
\end{figure*}

\begin{figure*}[!tp]
\centering
\includegraphics[width=1.\textwidth]{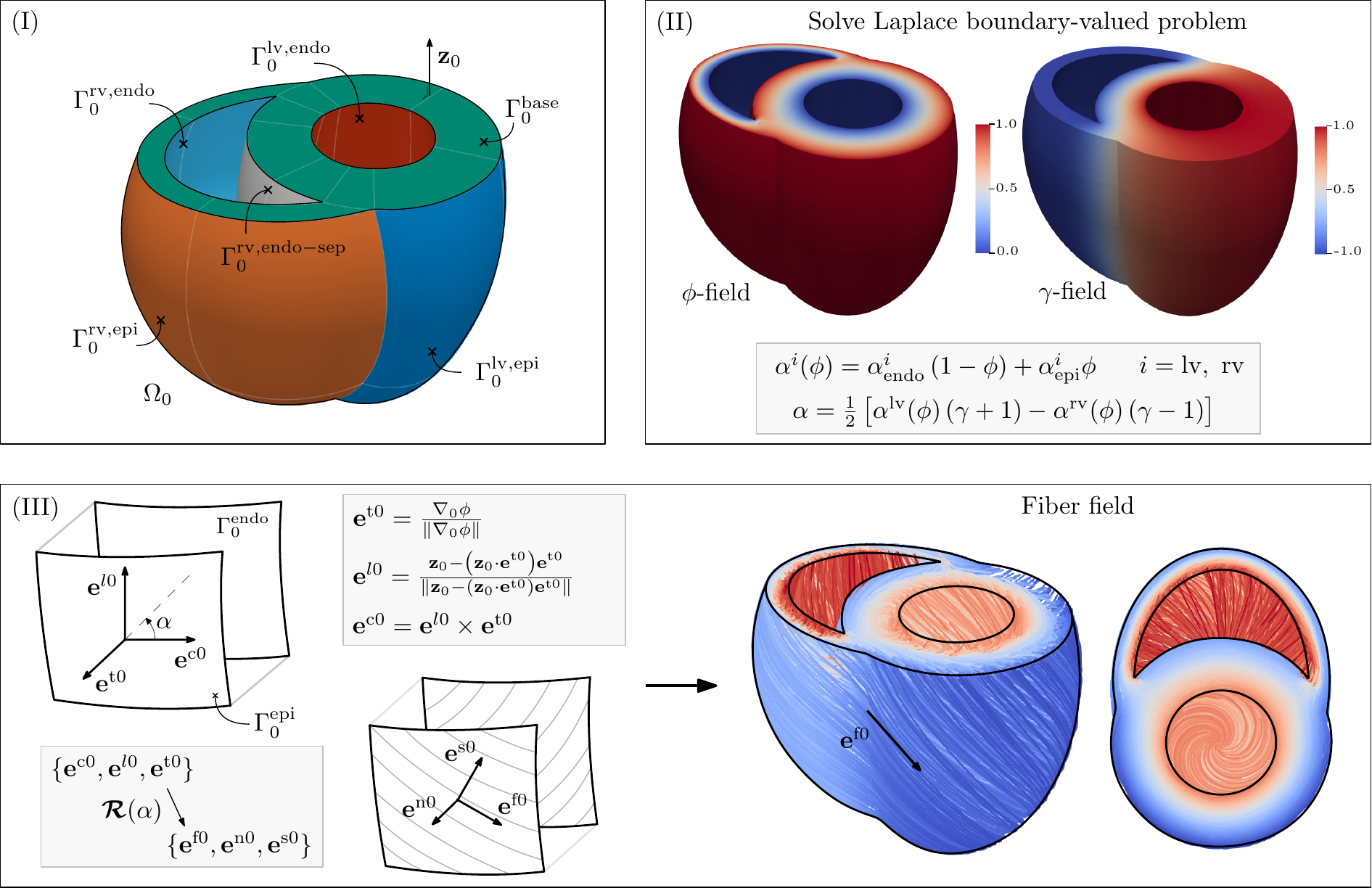}
  \caption{\label{fig:RRBM} Schematic overview of the fiber field method based on Rossi \emph{et al.}~\cite{ThermoQuarteroni}. The method consists of three steps: (I) Boundary tagging, (II) solving the Laplace boundary-valued problem, and (III) constructing and rotating the local coordinate system to obtain the fiber direction vector, $\mathbf{e}^{\rm{f0}}$. Rotation of the local coordinate system is performed using Rodriguez formula, $\boldsymbol{\mathcal{R}}(\alpha)$, and depends on the helix angle $\alpha$. The helix angle is obtained from experiments and is specified at the boundaries of the left and right ventricle, $\phi^i_{\rm{epi}}$ and $\phi^i_{\rm{endo}}$ with $i=\rm{lv},\rm{rv}$, and consequently interpolated using the results of the Laplace boundary-valued problem, $\phi$ and $\gamma$.} 
\end{figure*}
\subsection{Fiber orientation}\label{sec:RRBM}
When a patient is monitored for VT risk, information on the orthotropic fiber-sheet structure is irrelevant to the monitoring process and, therefore, generally missing. High-resolution DTI-MRI scans could provide this information, but these are expensive and not part of the current clinical workflow. Because of this, a rule-based method (RBM) is used, which enables the computation of a smooth continuous fiber field applicable to arbitrarily shaped atria geometries, single or bi-ventricles, and full heart geometries \cite{PIERSANTI2021113468,Bayer2012}. The considered type of RBM is based on the solution of Laplace boundary-value problems, of which we employ the method proposed by Rossi \emph{et al.}~\cite{ThermoQuarteroni,IntegratedHeartQuart}. This method is considered to be the simplest solution for constructing a fiber field on arbitrary bi-ventricle geometries.

The method proposed by Rossi \emph{et al.} consists of three distinct steps, visualized in Figure~\ref{fig:RRBM}. Step~(I) The domain boundaries are tagged which is required for step~(II) in which two Laplace boundary-value problems (LBV), $\phi$ and $\gamma$, are solved. The employed boundary conditions and weak formulations for solving the $\phi$- and $\gamma$-fields are elaborated in \ref{app:fiberfield}. The numerical solution of the LBV is then used in two ways: First, a local basis, $\{\mathbf{e}^{\rm{c0}},\mathbf{e}^{l0},\mathbf{e}^{\rm{t0}}\}$, is constructed based on the gradient, $\nabla_{0} \phi$ and the basal plane normal vector, denoted by $\mathbf{z}_{0}$. Next, both solutions, $\phi$ and $\gamma$, are used to obtain a homogeneous interpolation of fiber angles specified at the boundaries of the left and right ventricles, $\phi^i_{\rm{endo}}$ with $i=\rm{lv},\rm{rv}$. These fiber angles are referred to as helix angles and are based on histological (experimental) observations in the literature \cite{ExperimentFiberangles,CardiacStructure,Lombaert20121436} (summarized in Table~\ref{tab:fiberangles}). We limit ourselves to helical angles and neglect any transmural component of the fibers (transverse angle)~\cite{bovendeerdlvstrain}, \emph{i.e.}, fiber directions not parallel to the epi- and endocardium inside the myocardium. We also neglect differences in material properties in sheet $\mathbf{e}^{\rm{s0}}$ and sheet-normal $\mathbf{e}^{\rm{n0}}$ direction, \emph{i.e.}, we assume transversely isotropic material behaviour, elaborated on in Section~\ref{sec:CardiacModel}. The interpolation of the helix angle is given in Step~(II) of Figure~\ref{fig:RRBM}, after which the local circumferential vector, $\mathbf{e}^{\rm{c0}}$, is rotated about the transmural vector, $\mathbf{e}^{\rm{t0}}$, given the helix angle $\alpha$ using Rodriguez formula~\cite{IntegratedHeartQuart} (\ref{app:fiberfield}). 

It should be noted that alternative approaches such as the deformable mapping of DTI-MRI~\cite{PatientspecificBiventricleHermite,ZHANG20121130} could be a viable option when considering patient-specific simulations, provided that accurate data on the fiber and sheet orientations and the associated elastic properties are available. These conditions on the accuracy of the data are not met in our study, making this alternative approach unfeasible.

\begin{table}[ht]
\centering
\caption{Helix fiber angle $\alpha$ specified at the epicardium (epi) and endocardium (endo) boundaries of the left ventricle (lv) and the right ventricle (rv), visualized in Figure~\ref{fig:RRBM} and based on \cite{PIERSANTI2021113468}.}\label{tab:fiberangles}
\begin{tabular}{lcc}
\hline
& \multicolumn{1}{l}{epicardium (epi)} & endocardium (endo) \\ \hline
Left ventricle (lv)  & -60.0 [deg]                                & +60.0 [deg]             \\
Right ventricle (rv) & -25.0 [deg]                               & +90.0 [deg]  \\    \hline
\end{tabular}
\end{table}

\section{The cardiac model}\label{sec:CardiacModel}
\begin{figure*}[ht]
\includegraphics[width=\textwidth]{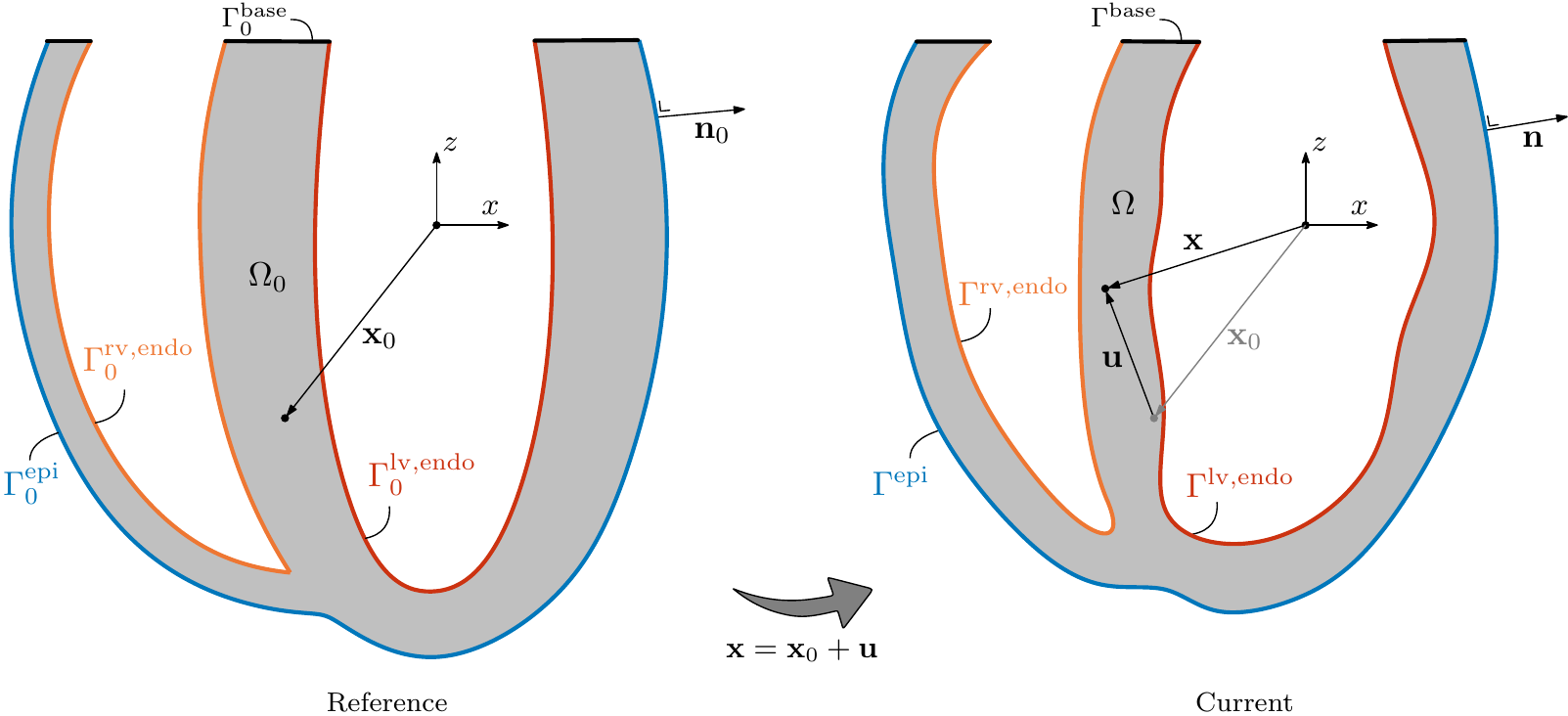}
  \caption{Two-dimensional illustration of the computational domain of the bi-ventricles, representing the myocardium. A distinction is made between the hypothetical stress-free \textit{reference} configuration and the deformed \textit{current} configuration. The reference configuration is mapped to the current configuration, based on the displacement vector, $\mathbf{u}$. The reference domain is bounded by a piecewise smooth boundary, ${{\Gamma}_0 = {\Gamma}_{\mathrm{D}} \cup \ {\Gamma}_{\mathrm{N}}\cup \ {\Gamma}_{\mathrm{R}}}$, which consists of a Dirichlet, ${\Gamma}_{\mathrm{D}}$, a Neumann, ${\Gamma}_{\mathrm{N}}$, and Robin boundary, ${\Gamma}_{\mathrm{R}}$, such that ${\Gamma}_{\mathrm{D}} \cap \ {\Gamma}_{\mathrm{N}} \cap \ {\Gamma}_{\mathrm{R}}  =\varnothing$. The domain boundaries consist of the epicardium (epi), the endocardium (endo) of the left (lv) and right (rv) ventricle, and the basal (base) plane. The reference boundaries are part of, ${\Gamma}^{\mathrm{base}}_0 \subset {\Gamma}_{\mathrm{D}} $, $\{ {\Gamma}^{\mathrm{rv,endo}}_0, \ {\Gamma}^{\mathrm{lv,endo}}_0 \}  \subset {\Gamma}_{\mathrm{N}}$, and ${\Gamma}^{\mathrm{epi}}_0 \subset {\Gamma}_{\mathrm{R}}$.} \label{fig:modeldomain}
\end{figure*}
For the cardiac model, we consider the open stress-free reference domain, $\Omega_0$, which is bounded by a piecewise smooth boundary consisting of a Dirichlet, Neumann, and Robin boundary, ${\Gamma_0 = \Gamma_{\rm{D}} \cup \ {\Gamma}_{\rm{N}} \cup {\Gamma}_{\rm{R}}}$, illustrated in Figure~\ref{fig:modeldomain}. The reference domain is mapped to the current or deformed domain, $\Omega$, as $x_i = x_{0_i} + u_i$, where ${x}_{0_i}$ and $x_{i}$ are the spatial coordinates of respectively the reference and current configuration, and $u_i$ is the displacement field. Note that for the problem formulation the index notation is employed according to the Cartesian coordinate system such that $i \in \{x,y,z\}$. Following this map, the deformation gradient is defined as
\begin{equation}\label{eq:defTensor}
    F^{\phantom{}}_{ij} = \frac{\partial x_{i}}{\partial {x}_{0_j}} = \delta_{ij} + \frac{\partial u_{i}}{\partial {x}_{0_j}},
\end{equation}
where $\delta_{ij}$ corresponds to the identity tensor. The determinant of the deformation gradient,
\begin{equation}
    J = \text{det}\left( \mathbf{F} \right),
\end{equation}
is a measure of volume change and an important quantity for (nearly) incompressible materials.

The employed cardiac model is based on Ref.~\cite{bovendeerdlvstrain} and solves for the unknown displacement vector, $u_i$, the scalar contractile length, $l^{\rm{c}}$, which is related to the dynamical response of muscle contraction, and a set of 0D pressure terms, $\tilde{p}=\{p^{\rm{0D}}\}$, representing the circulatory system. The model, defined on the current domain, $\Omega$, then states:
\begin{equation}
    \begin{cases}\text { Find } \left( u_i, l^{\rm{c}},\tilde{p} \right) \text { for } t\in \left( 0,T\right], & \text{such that: } \\
    \frac{\partial}{\partial x_i} \sigma^{\phantom{}}_{ij}( l^{\rm{c}},u_k, t )  = 0_j  & \text { in } \Omega \times(0, T], \\ 
    \mathcal{A}( l^{\rm{c}},u_i)  = 0 & \text { in } \Omega\times(0, T], \\ 
    \mathcal{M}(\tilde{p}^{\mathrm{0D}}, u_i)  = \left\{0\right\} & \text { in } (0, T],\\
    &\\
\text{Subject to:} & \\
    n_i \sigma^{\phantom{}}_{ij} = -p^{\mathrm{lv}}( u_k) \ n_j & \text { on } \Gamma^{\mathrm{lv,endo}} \times(0, T], \\
    n_i \sigma^{\phantom{}}_{ij} = -p^{\mathrm{rv}}( u_k ) \ n_j & \text { on } \Gamma^{\mathrm{rv,endo}} \times(0, T],\\
    n_i \sigma^{\phantom{}}_{ij} = -k^{\mathrm{peri}} u_j  & \text { on } \Gamma^{\mathrm{epi}} ,\\
    n_i u_i  = 0 & \text { on } \Gamma^{\mathrm{base}},\\
    \tilde{p} = \tilde{p}_{\mathrm{0}}, l^{\rm{c}}=l^{\rm{s0}}, \ u_i = 0_i & \text { at } t=0.
\end{cases}\label{eq:setofeqs}
\end{equation}
The complete cardiac model consists of three sub-models: the equilibrium equation regarding the myocardium mechanics ${\sigma_{ij} \left( \cdot \right)}$, the model related to muscle contraction dynamics, ${\mathcal{A}\left( \cdot \right)}$, and the 0D circulatory model, ${\mathcal{M}\left( \cdot \right)}$. The circulatory model approximates the dynamics of the human circulatory system by means of 0D pressure points which represent the pressure at representative locations in the arterial and venous part of the systemic and pulmonary circulation, and is coupled to the cardiac domain at the boundaries, $\Gamma^{\mathrm{lv,endo}}$ and $\Gamma^{\mathrm{rv,endo}}$, by pressure (Neumann) boundary conditions. The cardiac domain is constrained in the normal direction of the basal plane, $\Gamma^{\mathrm{base}}$, and at the epicardium, $\Gamma^{\mathrm{epi}}$, by a stiffness term that mimics the behavior of the pericardium (heart sac) in a simplified manner~\cite{pericard2019}. The cardiac model is initiated at time $t=0$, using initial conditions for each unknown quantity, and is terminated at the time $T$. 

In the next sections, we discuss each individual sub-model, \emph{i.e.}, the mechanical model, the contraction dynamics model, and the circulatory model, in more detail. We conclude with an overview of the cardiac cycle simulation settings.

\subsection{Constitutive model of the mechanical behavior of the myocardium}
The equilibrium equation is solved to compute the displacement field, $u_i$. From this displacement field the Cauchy stress tensor, $\sigma_{ij}$, can be derived, which characterizes the spatial and temporal mechanical response of the myocardium. Experimental studies have shown that the passive (absence of muscle contraction) myocardial response exhibits viscoelastic behavior \cite{SOMMER2015172,SHEN20113008}. However, we limit ourselves to the quasi-static behavior, which is commonly used in the literature~\cite{IntegratedHeartQuart,guccione,kerckhoffs2003homogeneity,bovendeerdlvstrain}, and allows us to describe the tissue as a hyperelastic material. It is also common to assume the tissue to be nearly incompressible~\cite{incomprHolzapfel,overviewmodels}.

The mechanical response of the myocardium involves nonlinear large deformations and is often numerically described using the total Lagrange formulation of the finite strain theory. In this formulation, we first relate the Cauchy stress, defined on the current domain, $\Omega$, to the reference domain, $\Omega_0$, using the second Piola-Kirchhoff tensor, $S_{ij}$, such that
\begin{equation}
\label{eq:secondPstress}
    \sigma^{\phantom{}}_{ij} = \frac{1}{J} F^{\phantom{}}_{ik} S^{\phantom{}}_{kn} F^{\phantom{}}_{jn}, \quad \text{with} \quad S^{\phantom{}}_{ij} = S^{\mathrm{pas}}_{ij} + S^{\mathrm{act}}_{ij},
\end{equation}
which is additively decomposed according to a passive, $S^{\mathrm{pas}}_{ij}$, and an active contribution, $S^{\mathrm{act}}_{ij}$, both defined on the reference domain, $\Omega_0$. The passive part describes the hyperelastic behavior of the myocardium and the active part describes the force generated by the myocytes. Both the passive and active components depend on the Green-Lagrange strain tensor, defined as

\begin{equation}\label{eq:GreenLagr}
    E^{\phantom{}}_{ij} = \frac{1}{2} \left( F^{\phantom{}}_{ki} F^{\phantom{}}_{kj} - \delta^{\phantom{}}_{ij} \right).
\end{equation}

\subsubsection*{Passive component}
The nearly incompressible passive myocardial response is described by defining the second Piola-Kirchhoff stress tensor as
\begin{equation}\label{eq:passivePK}
    S^{\mathrm{pas}}_{ij} = \frac{\mathrm{d} \psi\left(E^{\phantom{}}_{ij} \right)}{\mathrm{d} E^{\phantom{}}_{ij}},  \quad \text{with} \quad \psi = \psi^{\rm{S}} + \psi^{\rm{V}}, 
\end{equation}
where $\psi$ is the elastic strain energy density function, composed of a tissue deformation part, $\psi^{\rm{S}}$, and a volumetric part, $\psi^{\rm{V}}$, which imposes the nearly incompressible constraint. The shape change of myocardial or other biological tissues, which are assumed to be hyperelastic, is generally described by a Fung-type strain energy function~\cite{fung1979pseudoelasticity}, although exceptions exist~\cite{Nash2000}. Several modifications of the Fung-type model have been introduced~\cite{overviewmodels}, which are typically based on phenomenological observations or the morphology of the heart tissue \cite{guccione, HolzapfelConstitutive}. In our analyses, we limit ourselves to the transversely isotropic model of Bovendeerd \emph{et al.}~\cite{bovendeerdlvstrain}. This model requires a minimum of 5 parameters and is defined as
\begin{subequations}
\label{eq:EnergyDensity}
\begin{align}
      \psi^{\rm{S}}\left( E^{\phantom{}}_{ij} \right) &= C [ \mathrm{exp}\left( Q \right) - 1    ], \\
       \psi^{\rm{V}}\left( E^{\phantom{}}_{ij} \right) & = \frac{1}{2} \kappa [ J^2 - 1    ]^2,
\end{align}
\end{subequations}
with
\begin{equation}
\label{eq:Q}
    Q =  a_1 I_1^2 - a_2 I_2 + a_3  I_4^2 ,
\end{equation}
where $C$ relates to the stiffness of the myocardium, $\kappa$ is the bulk modulus, and $a_i$ are material constants (listed in Table~\ref{tab:params}). The invariants of the strain tensor are defined as
\begin{subequations}
\label{eq:I}
\begin{align}
       I_1 &= E^{\phantom{}}_{ii}, \label{I1} \\
       I_2 & = \frac{1}{2} \left( E^{\phantom{}}_{ii} E^{\phantom{}}_{jj} -  E^{\phantom{}}_{ji} E^{\phantom{}}_{ij}\right),    \label{eq:I2}\\
       I_4 &= {\mathrm{e}}^{\mathrm{f0}}_i E^{\phantom{}}_{ij}  {\mathrm{e}}^{\mathrm{f0}}_j \ , \label{I4}
\end{align}
\end{subequations}
where, $I_1, I_2$, are the first two principal invariants of the Green-Lagrange strain tensor\footnote{The third principal invariant, $I_3$, equals the Jacobian of the gradient of deformation tensor, $J$.} and $I_4$ is a quasi-invariant representing the Green-Langrange fiber strain. The quasi-invariant is commonly defined for orthotropic and transversely isotropic materials~\cite{HolzapfelConstitutive}. Equation~\eqref{eq:Q} is based on a transversely isotropic material, which assumes that the remaining two cross-fiber directions have similar material properties. The addition of a more complex orthotropic material is also possible but is not considered in the current study.

\subsubsection*{Active component}
Shortly before cardiac contraction, a depolarization wave is initiated throughout the myocardium, which initiates the contraction of the myocytes. In the current model, spatial variation during the moment of depolarization, related to the finite velocity of wave propagation, is neglected and contraction of the myocytes is assumed to be initiated simultaneously at time, $T^{\mathrm{act}}$, which circumvents the use of an additional electrophysiological model. The force generated by myocytes depends on the time elapsed, $t^{\mathrm{a}}=t - T^{\mathrm{act}}$, since the most recent moment of activation $T^{\mathrm{act}}$. The produced stress in the current fiber direction is then added as a separate component to the passive stress tensor, Equation~\eqref{eq:secondPstress}. The active stress component is defined in terms of the second Piola-Kirchhoff stress
\begin{equation}\label{eq:Sact}
    S^{\mathrm{act}}_{ij}= S^{\mathrm{a}}( t^{\mathrm{a}}, l^{\rm{c}}) \ {\mathrm{e}}^{\mathrm{f0}}_i {\mathrm{e}}^{\mathrm{f0}}_j,
\end{equation}
where $S^{\mathrm{a}}$ is the stress magnitude generated by the sarcomeres, $t^{\mathrm{a}}$ the time elapsed since activation, ${\mathrm{e}}^{\mathrm{f0}}_i$ the fiber-direction in the reference configuration. The sarcomere can be regarded as a serial connection of a contractile and elastic element, with lengths $l^{\mathrm{c}}$ and $\left( l^{\mathrm{s}} - l^{\mathrm{c}} \right)$, respectively, where $l^{\mathrm{s}}$ represents the total sarcomere length. According to physiological observations, the generated stress is dependent on these length scales, the shortening velocity of the contractile element, and the time elapsed since activation~\cite{shorteningVel1984}. We adopt the model of Bovendeerd \emph{et al.}~\cite{bovendeerdlvstrain,kerckhoffs2003homogeneity}, where these dependencies are multiplicatively decomposed as
\begin{equation}
    S^a( t^{\mathrm{a}}, l^{\rm{c}}) =  \frac{l^{\mathrm{s0}}}{l^{\mathrm{s}}}f^{\mathrm{iso}}\left( l^{\mathrm{c}} \right) \ f^{\mathrm{twitch}}\left(t^{\mathrm{a}}, l^{\mathrm{s}}  \right) \ E^{\mathrm{a}} \left( l^{\mathrm{s}} - l^{\mathrm{c}} \right),
\end{equation}
where $l^{\mathrm{s0}}$ is the sarcomere length in the stress-free state and $E^{\mathrm{a}}$ is the stiffness of the series elastic element. The isometric contraction function, $f^{\mathrm{iso}}$, and twitch function, $f^{\mathrm{twitch}}$, are defined as
\begin{equation}
    f^{\mathrm{iso}} \left( l^{\mathrm{c}} \right ) = \left\{\begin{matrix}
T^0 \text{tanh}^2 \left[ a^{l} \left( l^{\mathrm{c}} - l^{\mathrm{c0}}\right) \right] & l^{\mathrm{c}} \geq l^{\mathrm{c0}}, \\ 
 0 & l^{\mathrm{c}} < l^{\mathrm{c0}} ,
\end{matrix}\right.
\label{eq:fiso}
\end{equation}
and
\begin{equation}
    f^{\mathrm{twitch}} \left( t^{\mathrm{a}}, l^{\mathrm{s}} \right ) = \left\{\begin{matrix}
    0  &  t^{\mathrm{a}} < 0, \\
\mathrm{tanh}^2 \left( \frac{t^{\mathrm{a}}}{\tau^{\mathrm{r}}}  \right ) \mathrm{tanh}^2 \left( \frac{t^{\mathrm{max}}-t^{\mathrm{a}}}{\tau^{\mathrm{d}}}\right) & 0 \leq t^{\mathrm{a}} \leq t^{\mathrm{max}} ,\\
0  & t^{\mathrm{a}} > t^{\mathrm{max}} ,
\end{matrix}\right.
\end{equation}
with
\begin{equation}\label{eq:tmax}
    t^{\mathrm{max}} = b \left(  l^{\mathrm{s}} - l^{\mathrm{d}} \right).
\end{equation}
In \eqref{eq:fiso}, $T^{\mathrm{0}}$ represents the reference active stress level, $a^{l}$ a parameter that governs the steepness of the stress-length curve, and $l^{\mathrm{c0}}$ the contractile element length below which no active stress is generated. The rise and decay of a myocyte contraction, \emph{i.e.}, the twitch, are governed by $\tau^{\mathrm{r}}$ and $\tau^{\mathrm{d}}$, respectively. The twitch duration $t^{\mathrm{max}}$ is described according to a linear relation with the total sarcomere length, provided a slope $b$ and a constant (extrapolated) sarcomere length where the twitch duration is zero, $l^{\mathrm{d}}$. The sarcomere length of Equation~\eqref{eq:tmax} is related to the quasi-invariant $I_4$ of Equation~\eqref{eq:I}, such that
\begin{equation}\label{eq:sarclength}
    l^{\mathrm{s}} = l^{\mathrm{s0}} \sqrt{2 I_4 + 1}.
\end{equation}

\subsection{Contractile dynamics}
The contractile element length, $l^{\rm{c}}$, is related to the shortening velocity, as observed in experiments~\cite{shorteningVel1984}. It is modeled by an ordinary differential equation (ODE) \cite{kerckhoffs2003homogeneity,bovendeerdlvstrain} for all material points, which describes the length evolution over time as
\begin{equation}
\label{eq:lengths}
\mathcal{A}\left( l^{\rm{c}}, u_i \right)=\left\{\begin{aligned}
       &  \frac{d l^{\mathrm{c}}}{dt} - \left[ E^{\mathrm{a}} \left( l^{\mathrm{s}} - l^{\mathrm{c}} \right) - 1\right] v^{\mathrm{0}} = 0 &  \text { in } \Omega_0 \times (T^{\mathrm{act}}, T], \\
       & l^{\mathrm{c}} = l^s &  \text { in }  \Omega_0 \times [ 0, T^{\mathrm{act}}],
\end{aligned}\right.
\end{equation}
where $v^{\mathrm{0}}$ represents the unloaded shortening velocity. The contractile length is kept equal to the total sarcomere length, Equation~\eqref{eq:sarclength}, before muscle activation, $t<T^{\rm{act}}$,  while the ODE in Equation~\eqref{eq:lengths}, is only solved after activation has occurred, $t > T^{\mathrm{act}}$.

\begin{table}[]
\caption{\label{tab:params} Model parameter values for the passive and active components.}
\centering
\begin{tabular}{llSl}
\hline
                  & \multicolumn{1}{c}{Parameters} & \multicolumn{1}{c}{Value} & \multicolumn{1}{c}{Unit}                    \\ \hline
Passive component & $C$                          & 0.4                       & kPa                     \\
                  & $\kappa$                          & 110.0                       & kPa                     \\
                  & $a_1$                          & 3.0                         & -                       \\
                  & $a_2$                          & 6.0                         & -                       \\
                  & $a_3$                          & 3.0                        & -                       \\ \hline
Active component  & $T^{\mathrm{0}}$                   & 185.0                       & kPa                     \\
                  & $E^{\mathrm{a}}$                   & 20.0                        & $\mu\text{m}^{-1}$      \\
                  & $a^{l}$                   & 2.0                         & $\mu\text{m}^{-1}$      \\
                  & $l^{\mathrm{c0}}$                  & 1.5                       & $\mu\text{m}$           \\
                  & $l^{\mathrm{s0}}$                  & 1.9                       & $\mu\text{m}$           \\
                  & $l^{\mathrm{d}}$                   & -1.0                      & $\mu\text{m}$           \\
                  & $v^{\mathrm{0}}$                & 7.5                        & $\mu\text{m} \ \text{s}^{-1}$                    \\
                  & $b$                            & 160.0                       & $\text{ms} \ \mu\text{m}^{-1}$ \\
                  & $T^{\mathrm{act}}$                & 300.0                        & ms                    \\
                  & $\tau^{\mathrm{r}}$                & 75.0                        & ms                      \\
                  & $\tau^{\mathrm{d}}$                & 150.0                       & ms                      \\ \hline
Boundary condition & $k^{\rm{peri}}$                   & 10.0                       & kPa/m                     \\
\hline
\end{tabular}
\end{table}

\subsection{Circulatory system} 
\begin{figure*}[!tp]
\centering
\includegraphics[width=1.\textwidth]{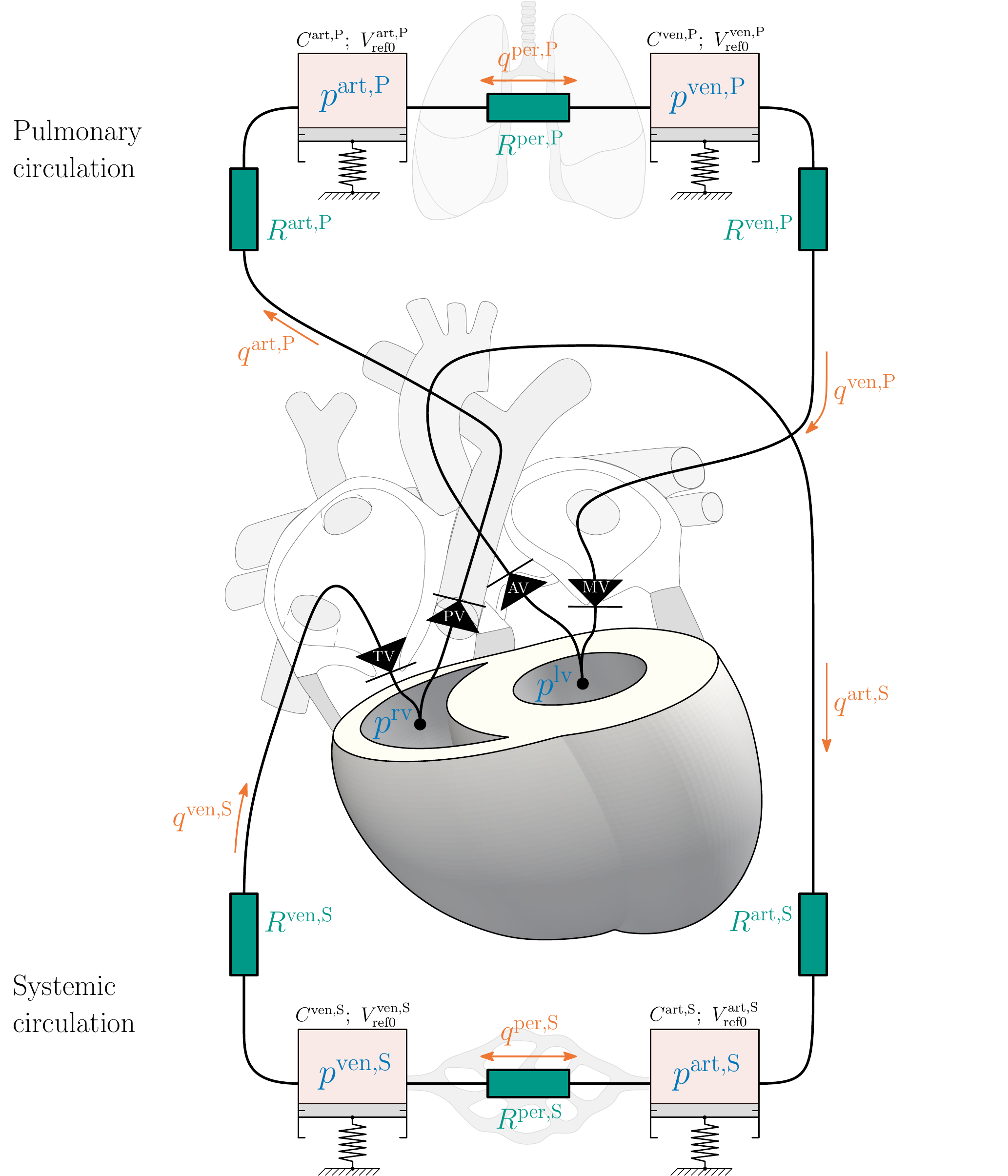}
  \caption{\label{fig:lumpedscheme} Schematic overview of the closed-loop lumped parameter model, representing the circulatory system of the human body. The circulatory system consists of a pulmonary (P) and systemic (S) circulation, with arterial (art), venous (ven), and peripheral (per) flows. The pulmonary and systemic circulation are modeled as two Windkessel compartments in series, characterized by a compliance, $C^i$, and a volume at zero reference pressure, $V^i_{\rm{ref0}}$. The compartments and ventricles are linked via resistances, $R^i$ (green), through which volume flow occurs, $q^i$ (orange), caused by the compartment pressure, $p^i$ (blue). The four heart valves: pulmonary valve (PV), tricuspid valve (TV), mitral valve (MV), and aortic valve (AV), are idealized such that they open and close instantaneously and prevent back-flow. The compartment and ventricle pressures, illustrated in blue, are the quantities that are solved for in the circulatory model.} 
\end{figure*}

The work performed by the heart is used to circulate oxygenated blood from the lungs to different parts of the human body and back, as illustrated in Figures \ref{fig:heart_illustration} and \ref{fig:lumpedscheme}. The fluid-structure interaction of the circulating blood through the human body is often simplified by a Windkessel or lumped-parameter model~\cite{IntegratedHeartQuart,bovendeerdlvstrain,PIERSANTI2022}. These models represent the pressure at representative locations in the arterial and venous part of the systemic and pulmonary circulation, as 0D pressure terms which are coupled by conservation of mass.

A schematic overview of the closed-loop circulatory model is visualized in Figure~\ref{fig:lumpedscheme}. In this study, we model the systemic and pulmonary circulations using two lumped Windkessel compartments in series (4 compartments in total). These compartments are considered compliant, which means that they allow for a volume change given an internal pressure. Because the compartments are connected in series, the difference in compartment pressure initiates a volume flow, which resembles the flow inside the arteries and veins. 

Given the closed-loop circulatory model $\mathcal{M} \left( \tilde{p}, u_i \right)$ in Equation~(\ref{eq:setofeqs}), the set of 0D pressure points, \\
$\tilde{p} = \{p^{\mathrm{lv}}, p^{\mathrm{rv}}, p^{\mathrm{art,P}} , p^{\mathrm{art,S}}, p^{\mathrm{ven,S}}\} \in \left(0, T\right]$, follows from solving
\begin{equation}\label{eq:circulat}
    \mathcal{M} \left( \tilde{p}, u_i \right) =
    \left\{\begin{aligned} 
          \frac{d V^{\mathrm{lv}}\left( u_i \right)}{dt} - q^{\mathrm{ven,P}} + q^{\mathrm{art,S}} & = 0, \\
          \frac{d V^{\mathrm{rv}}\left( u_i \right)}{dt} - q^{\mathrm{ven,S}} + q^{\mathrm{art,P}} & = 0, \\
          C^{\mathrm{art,P}}\frac{d p^{\mathrm{art,P}}}{dt} - q^{\mathrm{art,P}} + q^{\mathrm{per,P}} & = 0, \\
          C^{\mathrm{art,S}}\frac{d p^{\mathrm{art,S}}}{dt} - q^{\mathrm{art,S}} + q^{\mathrm{per,S}} & = 0, \\
          C^{\mathrm{ven,S}}\frac{d p^{\mathrm{ven,S}}}{dt} - q^{\mathrm{per,S}} + q^{\mathrm{art,S}} & = 0, \\
          C^{\mathrm{ven,P}} p^{\mathrm{ven,P}}  - V^{\mathrm{ven,P}} +  V^{\mathrm{ven,P}}_{\rm{ref0}}   & = 0,
         \end{aligned}\right.
\end{equation}
with
\begin{equation}
\label{eq:initp}
    \tilde{p} = \tilde{p}_{0} = \left\{ p^{\mathrm{lv}}_{0}, p^{\mathrm{rv}}_{0}, p^{\mathrm{art,P}}_{0} , p^{\mathrm{art,S}}_{0}, p^{\mathrm{ven,S}}_{0} \right\} \quad \text{at} \quad t = 0,
\end{equation}
where $C^i$ is the compartment compliance, $V^i$ the compartment volume at time $t$, and $q^i$ the volume flow between the compartments. The fixed compartment volume at zero reference pressure is denoted by $V^i_{\rm{ref0}}$, while the initial pressure values at time $t=0$, are given by $p^i_{0}$. Note that the left and right ventricle pressure, $p^{\mathrm{lv}}$ and $p^{\mathrm{rv}}$, are coupled to the cavity boundaries according to Equation~\eqref{eq:setofeqs}. The venous pressure of the pulmonary circulation, $p^{\rm{ven,P}}$, is not required to be solved for because it follows directly from the final relation in Equation~\eqref{eq:circulat} in combination with the conservation of mass when assuming blood to be incompressible~\cite{KWAK2005283}:
\begin{equation}\label{eq:conservationmass}
    V^{\mathrm{ven,P}} = V^{\mathrm{tot}} - V^{\mathrm{lv}} - V^{\mathrm{rv}} - V^{\mathrm{art,P}} - V^{\mathrm{art,S}}- V^{\mathrm{ven,S}},
\end{equation}
where $V^{\mathrm{tot}}$ is the total volume of blood inside the entire circulation. Inserting this expression in Equation~\eqref{eq:circulat}, yields a relation between the venous pulmonary pressure, $p^{\rm{ven,P}}$, and the total blood volume. The remaining pressure terms are linearly dependent on the individual compartment volumes, given as
\begin{equation}
    V^{i} = C^i p^i + V^i_{\rm{ref0}} \quad \text{for} \quad i \in \{ \mathrm{art,P}; \ \mathrm{art,S}; \ \mathrm{ven,S} \}.  
\end{equation}
The model is concluded by defining the volume flows between compartments as 
\begin{equation}\label{eq:volumeflow}
    \left\{\begin{aligned} 
          q^{\mathrm{art,P}} & = \langle \frac{p^{\mathrm{rv}} - p^{\mathrm{art,P}}}{R^{\mathrm{art,P}}} \rangle , \\
          q^{\mathrm{ven,P}} & = \langle \frac{p^{\mathrm{ven,P}} - p^{\mathrm{lv}}}{R^{\mathrm{ven,P}}} \rangle, \\
          q^{\mathrm{art,S}} & = \langle \frac{p^{\mathrm{lv}} - p^{\mathrm{art,S}}}{R^{\mathrm{art,S}}} \rangle , \\
          q^{\mathrm{ven,S}} & = \langle \frac{p^{\mathrm{ven,S}} - p^{\mathrm{rv}}}{R^{\mathrm{ven,S}}} \rangle , \\
          q^{\mathrm{per,P}} & = \frac{p^{\mathrm{art,P}} - p^{\mathrm{ven,P}}}{R^{\mathrm{per,P}}} , \\
          q^{\mathrm{per,S}} & =  \frac{p^{\mathrm{art,S}} - p^{\mathrm{ven,S}}}{R^{\mathrm{per,S}}} ,
         \end{aligned}\right.
\end{equation}
where $R^i$ is the flow resistance and Macauley brackets $\langle \cdot \rangle$ are used to model the cardiac valves, \emph{i.e.}, the mitral, tricuspid, pulmonary, and aortic valves illustrated in Figure \ref{fig:lumpedscheme}. These valves are idealized such that they open or close instantaneously and prohibit backflow.

The ventricle cavity volumes, $V^{\rm{lv}}\left( u_i\right)$ and $V^{\rm{rv}}\left( u_i\right)$, in Equation~\eqref{eq:circulat} are calculated based on the displacement field of the cavity domain boundaries, $\Gamma^{\rm{lv,endo}}$ and $\Gamma^{\rm{rv,endo}}$, defined in the current configuration, such that
\begin{equation}
    V^{i}\left( u_i \right) = \int \frac{1}{3} \left( x_{i} - H \right) n_i \ \text{d}\Gamma^{i,\rm{endo}} , \quad \text{with} \quad i = \rm{lv}, \rm{rv} ,
\end{equation}
where $x_i$ is the spatial location of the computational domain, Equation~\eqref{eq:defTensor}, and $H$ is the truncation height of the ventricles, Figure~\ref{fig:Dimensions}. The pressure-volume relation of the ventricles is dictated by the constitutive relation, Equation~\eqref{eq:EnergyDensity}, which is nonlinear. As a result, the pressure terms, $p^{\rm{lv}}$ and $p^{\rm{rv}}$ in Equation~\eqref{eq:circulat}, act as Lagrange multipliers such that the resulting cavity volume satisfies the conservation of mass of the circulatory model.

\begin{table*}[]
\renewcommand{\arraystretch}{1.2}
\caption{\label{tab:paramslumped} Circulatory system model parameters.}
\centering
\begin{tabular}{lllll}
\hline
                           & Parameter                     &   Systemic (i=S)     & Pulmonary (i=P)      & Unit            \\ \hline
Arterial resistance        & $R^{\mathrm{art,i}}$          & $8.58\times 10^{-3}$ & $8.57\times 10^{-4}$ & kPa s ml$^{-1}$ \\
Peripheral resistance      & $R^{\mathrm{per,i}}$          & $1.39\times 10^{-1}$ & $2.77\times 10^{-2}$ & kPa s ml$^{-1}$ \\
Venous resistance          & $R^{\mathrm{ven,i}}$          & $8.57\times 10^{-4}$ & $8.57\times 10^{-4}$ & kPa s ml$^{-1}$ \\
Arterial compliance         & $C^{\mathrm{art,i}}$          & $1.76\times 10^{1}$  & $5.06\times 10^{1}$  & ml kPa$^{-1}$   \\
Venous compliance          & $C^{\mathrm{ven,i}}$          & $3.53\times 10^{1}$  & $1.26\times 10^{2}$  & ml kPa$^{-1}$   \\
Arterial stressfree volume & $V_{\mathrm{ref0}}^{\mathrm{art,i}}$ & $6.30\times 10^{2}$  & $7.00\times 10^{1}$  & ml              \\
Venous stressfree volume   & $V_{\mathrm{ref0}}^{\mathrm{ven,i}}$ & $2.52\times 10^{3}$  & $2.80\times 10^{2}$  & ml              \\ \cline{3-4}
Total blood volume         & $V^{\mathrm{tot}}$            & \multicolumn{2}{c}{$5.00\times 10^{3}$}     & ml  \\
\hline
\end{tabular}
\end{table*}

\subsection{Cardiac cycle}
The cardiac model is initiated by estimating the initial pressures defined in Equation~\eqref{eq:initp} and the remaining quantities in Equation~\eqref{eq:setofeqs}. By specifying $p^{\mathrm{ven,P}}_{0}>p^{\mathrm{lv}}_{0}=0$ and $p^{\mathrm{ven,S}}_{0}>p^{\mathrm{rv}}_{0}=0$ at $t=0$, the model starts by filling the ventricles without muscle activation (passive filling). Once the ventricles are filled, \emph{i.e.}, the circulatory model approaches steady-state, muscle contraction is activated at $t=T^{\mathrm{act}}=300~[\text{ms}]$, which initiates the active component of the constitutive relation and the sarcomere dynamics model, Equation~\eqref{eq:secondPstress} and~\eqref{eq:sarclength}, respectively. The parameters used for the active tension model, listed in Table~\ref{tab:params}, determine the cardiac cycle duration, which corresponds to $T=800~[\text{ms}]$. Once the cardiac cycle is finished, $t>T$, a new cycle is initiated with the final results of the previous cycle specified as the initial values. It is required to run the model for several cardiac cycles so that the closed-loop circulatory model can reach dynamic equilibrium. The required number of cardiac cycles is typically $6$-$8$, depending on the estimated initial conditions, the ventricle cavity volumes, and the specified circulatory parameters in Table~\ref{tab:paramslumped}. The number of cardiac cycles will therefore be specified when considering a specific problem.

\section{Isogeometric analysis discretization}\label{sec:IGA}
The strong formulation of the coupled 3D-0D cardiac model in Section~\ref{sec:CardiacModel} is discretized according to the IGA paradigm, which is explained in this section. We first discuss the weak formulation which is derived according to the total Lagrange formulation, after which the temporal and spatial discretization is elaborated on. The obtained monolithic system is provided in matrix notation, after which the solving routine is discussed. 

\subsection{Weak formulation and temporal discretization}
The strong formulation of the cardiac model consists of a coupled 3D-0D model, Equation~\eqref{eq:setofeqs}, which solves for the displacement field, the contractile length field, and the windkessel pressures, \emph{i.e.}, $\{u_i,l^{\mathrm{c}},\tilde{p}\}$. The weak formulation for each model component is derived according to Galerkin's method and based on the total Lagrange formulation (\ref{app:discretization}), where appropriate trial spaces are defined as ${\{u_i,l^{\mathrm{c}}\}\in} { \boldsymbol{\mathcal{U}}\times\mathcal{L}} $ with ${\boldsymbol{\mathcal{U}}=\{ u_i \ | \ u_i \in \boldsymbol{H}^1\left( \Omega \right) , \ u_i n_i = 0 \ \text{on} \ \Gamma^{\rm{base}} \}}$ and ${ \mathcal{L} = \{ l^{\rm{c}} \ | \ l^{\rm{c}} \in {L}^2\left( \Omega \right)\} }$, where $L^2\left(\Omega\right)$ is the space of square integrable functions and $H^1\left(\Omega\right)$ is the Sobolev space of order one. The infinite-dimensional test spaces are chosen identical, modulo inhomogeneous boundary conditions, such that ${\{ w_i,q\}\in} {\boldsymbol{\mathcal{W}}\times\mathcal{Q}}$ where $ {\boldsymbol{\mathcal{W}} = \{ w_i \ | \ w_i \in \boldsymbol{H}^1\left( \Omega \right) , \ w_i n_i = 0 \ \text{on} \ \Gamma^{\rm{base}} } \}$ and $ {\mathcal{Q}  = \{ q_i \ | \ q_i \in L^2\left( \Omega \right)\}}$. This yields the following residual for the momentum balance at the current time increment, $n+1$, such that $t^{n+1}=(n+1)\Delta t$,
\begin{equation}\label{eq:resU}
\begin{aligned}
    \mathscr{R}^{u}_{n+1}\left( w_j, u_l, l^{\mathrm{c}}, p^{\mathrm{lv}}, p^{\mathrm{rv}}, t^{n+1} \right) & = \int  \frac{\partial w_j}{\partial x_{0_k}} \ S^{\mathrm{pas}}_{ki} \left( u_l \right) F^{\phantom{}}_{ki} \ \text{d}{\Omega_0} \\
    &+ \int  \frac{\partial w_j}{\partial x_{0_k}} \ S^{\mathrm{act}}_{ki} \left( u_l, l^{\mathrm{c}}, t^{n+1} \right) F^{\phantom{}}_{ki} \ \text{d}{\Omega_0} \\
     &- \int w_j p^{\mathrm{lv}} n_{0_k} F^{-1}_{kj} J \ \text{d}{\Gamma^{\mathrm{lv,endo}}_0} \\ & - \int w_j p^{\mathrm{rv}} n_{0_k} F^{-1}_{kj} J \ \text{d}{\Gamma^{\mathrm{rv,endo}}_0}  \\ &- \int w_j k^{\mathrm{peri}} u_j J \ \text{d}{\Gamma^{\mathrm{epi}}_0}  \qquad \forall w_j \in \boldsymbol{\mathcal{W}}.
\end{aligned}  
\end{equation}
We employ the $\theta$-method~\cite{HughesFEAbook} for the temporal discretization of the model, where the following definition is used,
\begin{equation}\label{eq:thetamethod}
\frac{\left(\cdot\right)_{n+1} - \left(\cdot\right)_{n} }{\Delta t} = \underbrace{\theta \left( \cdot \right)_{n+1} + \left( 1-  \theta \right) \left( \cdot \right)_{n}}_{\left( \cdot\right)_{n+\theta}}, \quad  \theta \in [0,1],
\end{equation}
where $\theta$ is the weight that controls the integration method, \emph{i.e.}, $\theta$=0 for explicit or forward Euler, $\theta$=1 for implicit or backward Euler, and $\theta$=0.5 for the Crank-Nicolson scheme. Applying this definition to the evolution equations of the sarcomere dynamics, $\mathcal{A}\left( \cdot \right)$, and circulatory model, $\mathcal{M}\left( \cdot \right)$, of Equation~\eqref{eq:setofeqs}, results in the following residuals
\begin{equation}\label{eq:resLc}
\begin{aligned}
    \mathscr{R}^{l^{\mathrm{c}}}\left(q, u_i, l^{\mathrm{c}} \right) &=\int q \left( l^{\mathrm{c}}_{n+1} - l^{\mathrm{c}}_n \right) \ \text{d}{\Omega_0} \\
    & - \Delta t \int q \  \left[ f   \left( u_{i}, l^{\mathrm{c}} \right) \right]_{n+\theta^{l^{\mathrm{c}}}} \ \text{d}{\Omega_0} \qquad \forall q \in {\mathcal{Q}},
\end{aligned}
\end{equation}
with
\begin{equation}
    f\left( u_{i}, l^{\mathrm{c}} \right) = \left[ E^{\mathrm{a}} \left( l^{\mathrm{s}}(u_i) - l^{\mathrm{c}} \right) - 1\right] v^{\mathrm{0}},
\end{equation}
according to Equation~\eqref{eq:sarclength} and \eqref{eq:lengths}. The circulatory model is discretized in time as follows
\begin{equation}\label{eq:resP}
    \begin{aligned}
    \mathscr{R}^{\tilde{p}}\left( u_i, \tilde{p} \right) =
    \left\{\begin{aligned} 
          V^{\mathrm{lv}}_{n+1}\left( u_i \right) - V^{\mathrm{lv}}_{n}\left( u_i \right) -  \Delta t \left( q^{\mathrm{ven,P}} - q^{\mathrm{art,S}}\right)_{n+\theta^{\tilde{p}}}, & \\
          V^{\mathrm{rv}}_{n+1}\left( u_i \right) - V^{\mathrm{rv}}_{n}\left( u_i \right) -  \Delta t \left( q^{\mathrm{ven,S}} - q^{\mathrm{art,P}}\right)_{n+\theta^{\tilde{p}}}, & \\
          C^{\mathrm{art,P}} \left( p^{\mathrm{art,P}}_{n+1}- p^{\mathrm{art,P}}_{n} \right) -  \Delta t \left( q^{\mathrm{art,P}} - q^{\mathrm{per,P}}\right)_{n+\theta^{\tilde{p}}}, & \\
          C^{\mathrm{art,S}} \left( p^{\mathrm{art,S}}_{n+1}- p^{\mathrm{art,S}}_{n} \right) -  \Delta t \left( q^{\mathrm{art,S}} - q^{\mathrm{per,S}}\right)_{n+\theta^{\tilde{p}}}, & \\
          C^{\mathrm{ven,S}} \left( p^{\mathrm{ven,S}}_{n+1}- p^{\mathrm{ven,S}}_{n} \right) - \Delta t \left( q^{\mathrm{per,S}} - q^{\mathrm{art,S}}\right)_{n+\theta^{\tilde{p}}}, & 
         \end{aligned}\right.
\end{aligned}
\end{equation}
where the volume flow definitions, $q^i$, are listed in Equation~\eqref{eq:volumeflow}. 
Note that the pressure terms of the circulatory model, $\tilde{p}=\{p^{\mathrm{lv}}, p^{\mathrm{rv}}, p^{\mathrm{art,P}} , p^{\mathrm{art,S}}, p^{\mathrm{ven,S}}\}$, are not defined on the 3D computational domain, $\Omega_0 \subset \mathbb{R}^3$, since these are point values, $\tilde{p} \subset \mathbb{R}^0$, associated with the Windkessel model and need not be discretized in space. 

For convenience, we decompose the temporal residuals defined in Equation~\eqref{eq:resLc} and \eqref{eq:resP} based on the previous, $\mathscr{P}$, and current, $\mathscr{C}$, time-increment, $n$ and $n+1$ respectively, such that
\begin{equation}\label{eq:resdecomp}
    \mathscr{R}^{i}= \mathscr{P}^{i}_{n} + \mathscr{C}^{i}_{n+1}, \quad \text{with} \quad i=l^{\mathrm{c}}, \tilde{p}.
\end{equation}
The definition of each component in the residual, \emph{i.e.}, $\mathscr{R}^{l^{\mathrm{c}}},\mathscr{R}^{\tilde{p}}$ , is elaborated on in \ref{app:discretization}.

\subsection{Spatial discretization}\label{sec:spatialdisc}
The formulations of Equation~\eqref{eq:resU} and \eqref{eq:resLc} are discretized in space using isogeometric analysis based on multi-patch NURBS. The infinite-dimensional test spaces are first approximated by discrete subspaces, $\{w^h_i,q^h\}\in  \boldsymbol{\mathcal{W}}^h\times\mathcal{Q}^h \subset   \boldsymbol{\mathcal{W}}\times\mathcal{Q}$, where the superscript $h$ is used to indicate that discrete quantities are concerned. Next, we chose the discrete trial spaces to be identical to the discrete test spaces, such that ${\{u^h_i,l^{\mathrm{c}^h}\}\in} {\boldsymbol{\mathcal{W}}^h\times\mathcal{Q}^h }$. The discrete subspaces, $\boldsymbol{\mathcal{W}}^h$ and $\mathcal{Q}^h$, are spanned by the vector-valued B-spline shape functions ${N}^{u}_{ij} \in \boldsymbol{\mathcal{W}}^h$ for the displacement and scalar-valued B-spline shape functions for the contractile sarcomere length, ${N}^{\tilde{p}}_{i} \in \mathcal{Q}^h$. The approximate displacement $u_i^h$ and contractile sarcomere length $l^{\mathrm{c}^h}$ are then given by
\begin{equation}\label{eq:discret}
    \begin{cases} 
    u^{h}_j\left( x_{0_k} \right) &= \sum_{i}{N}^{u}_{ij}\left( {x}_{0_k} \right) \tilde{u}_i,  \\
    l^{\mathrm{c}^h}\left( x_{0_k} \right) &= \sum_{i}{N}^{l^{\mathrm{c}}}_{i}\left( {x}_{0_k} \right) \tilde{l}^{\mathrm{c}}_i,
    \end{cases} \qquad \forall x_{0_k} \in \Omega_0,
\end{equation}
where $\tilde{u}_i$ and $\tilde{l}^{\mathrm{c}}$ denote the displacement and contractile length degrees-of-freedom. We leverage the advantageous continuity properties of B-splines for the displacement and contractile length, by employing cubic $(p=3)$ and quadratic $(p=2)$ B-splines respectively. This ensures $C^{p-1}$-continuity across element boundaries within a patch. However, patch interfaces remain $C^{0}$-continuous. The B-spline degree difference between the displacement and contractile length is motivated by the definition of Equation~\eqref{eq:sarclength}, which shows that the contractile sarcomere length is related to the gradient of the deformation.
Substituting the discrete space in the weak residual formulation of Equation~\eqref{eq:resU} and \eqref{eq:resLc} results in the following (abstract) Galerkin discretization at time-increment $t^{n+1}=(n+1)\Delta t$:
\begin{equation}\label{eq:galerkinDiscr}
    \begin{cases}\text{Find } ({u}^h_{i}, l^{\mathrm{c}^h}) \in \boldsymbol{\mathcal{W}}^h \times \mathcal{Q}^h \text{ and } \tilde{p} \text { such that: } & \\ 
    \mathscr{R}^{u^h}_{n+1}( w^h_j, u_l, l^{\mathrm{c}^h}, p^{\mathrm{lv}}, p^{\mathrm{rv}}, t^{n+1} ) = 0& \hspace{-2em} \forall w^h_j \in \boldsymbol{\mathcal{W}}^h,\\ 
    \mathscr{P}^{l^{\mathrm{c}^h}}_{n}(q^h, u^h_i, l^{\mathrm{c}^h}) + \mathscr{C}^{l^{\mathrm{c}^h}}_{n+1}(q^h, u^h_i, l^{\mathrm{c}^h})=0 & \hspace{-2em} \forall q^h_{\phantom{}} \in \mathcal{Q}^h,\\
    \mathscr{P}^{\tilde{p}^h}_{n}( u^h_i, \tilde{p})+ \mathscr{C}^{\tilde{p}^h}_{n+1}( u^h_i, \tilde{p})=0. &
    \end{cases}
\end{equation}
For notation, we only indicate the time-increment, $n$ or $n+1$, as a subscript of the residual and not of the solution variables on which the residual depends. However, it still holds that the solution variables of a specific residual share the same time-increment as indicated by the residual.

\subsection{Solving the 3D-0D coupled problem}
We solve the coupled 3D-0D nonlinear Galerkin model of Equation~\eqref{eq:galerkinDiscr} with the Newton-Raphson method at each time-increment $n+1$. The coupled system of equations is solved in a monolithic fashion, for increments in displacement, $\Delta \tilde{u}$, the contractile sarcomere length, $\Delta \tilde{l}^{\mathrm{c}}$, and the circulatory pressures, $\Delta \tilde{p}$. The resulting set of equations can be represented in matrix form,   
\begin{equation}
    \underbrace{\left[\begin{array}{ccc}
\frac{\partial \mathscr{R}^{u}}{\partial \tilde{u}} & \frac{\partial \mathscr{R}^{u}}{\partial \tilde{l}^{\mathrm{c}}} & \frac{\partial \mathscr{R}^{u}}{\partial \tilde{p}}  \\ 
\frac{\partial \mathscr{C}^{l^\mathrm{c}}}{\partial \tilde{u}} & \frac{\partial \mathscr{C}^{l^\mathrm{c}}}{\partial \tilde{l}^{\mathrm{c}}} & \tilde{0}  \\
\frac{\partial \mathscr{C}^{\tilde{p}}}{\partial \tilde{u}}  &
\tilde{0}  & \frac{\partial \mathscr{C}^{\tilde{p}}}{\partial \tilde{p}} \\
\end{array}\right]_{n+1}^{i}}_{K^{i}_{n+1}} 
 \
\left[\begin{array}{c}
\Delta \tilde{u} \\
\Delta \tilde{l}^{\mathrm{c}} \\
\Delta \tilde{p}
\end{array}\right]_{n+1}^{i+1}=
\underbrace{-\left[\begin{array}{c}
\mathscr{R}^{u} \\
\mathscr{C}^{l^\mathrm{c}} \\
\mathscr{C}^{\tilde{p}}
\end{array}\right]_{n+1}^{i}
-\left[\begin{array}{c}
\tilde{0} \\
\mathscr{P}^{l^\mathrm{c}} \\
\mathscr{P}^{\tilde{p}}
\end{array}\right]_{n}^{i}}_{\tilde{r}^{i}},
\end{equation}
where the tangential stiffness matrix, $K^{i}_{n+1}$, is constructed at each Newton iteration $i$ for time-increment $n+1$. The system is considered to be converged when the norm of the residual column is below a specific tolerance, {$\| \tilde{r}^i \| < \mathrm{tol}$}. Once the system has converged, we proceed to the next time-increment while using the converged solution of the previous time-step as an initial guess for the Newton procedure. The parameters used for time-integration are user-defined and will be specified for the problem considered.

\section{Results}\label{sec:Results}
In this section, we benchmark the cardiac model by comparing the results of a left ventricle simulation with an established finite element analysis (FEA) solver implemented in FEniCS~\cite{FEniCSBook}. The benchmarked model is then used to showcase the proposed IGA workflow, visualized in Figure~\ref{fig:workflow}, where anatomical variations are applied to the bi-ventricle geometry on which numerical analyses are conducted directly. All simulations considered in this section assume an initial stress-free geometry at zero internal pressure, following the approach in \emph{e.g.}, Refs.~\cite{bovendeerdlvstrain,kerckhoffs2003homogeneity,Pluijmert2017}. Depending on the considered case, considering the impact of pre-stresses can be relevant. However, this is beyond the scope of the current manuscript.

\begin{figure*}[!t]
\includegraphics[width=\textwidth]{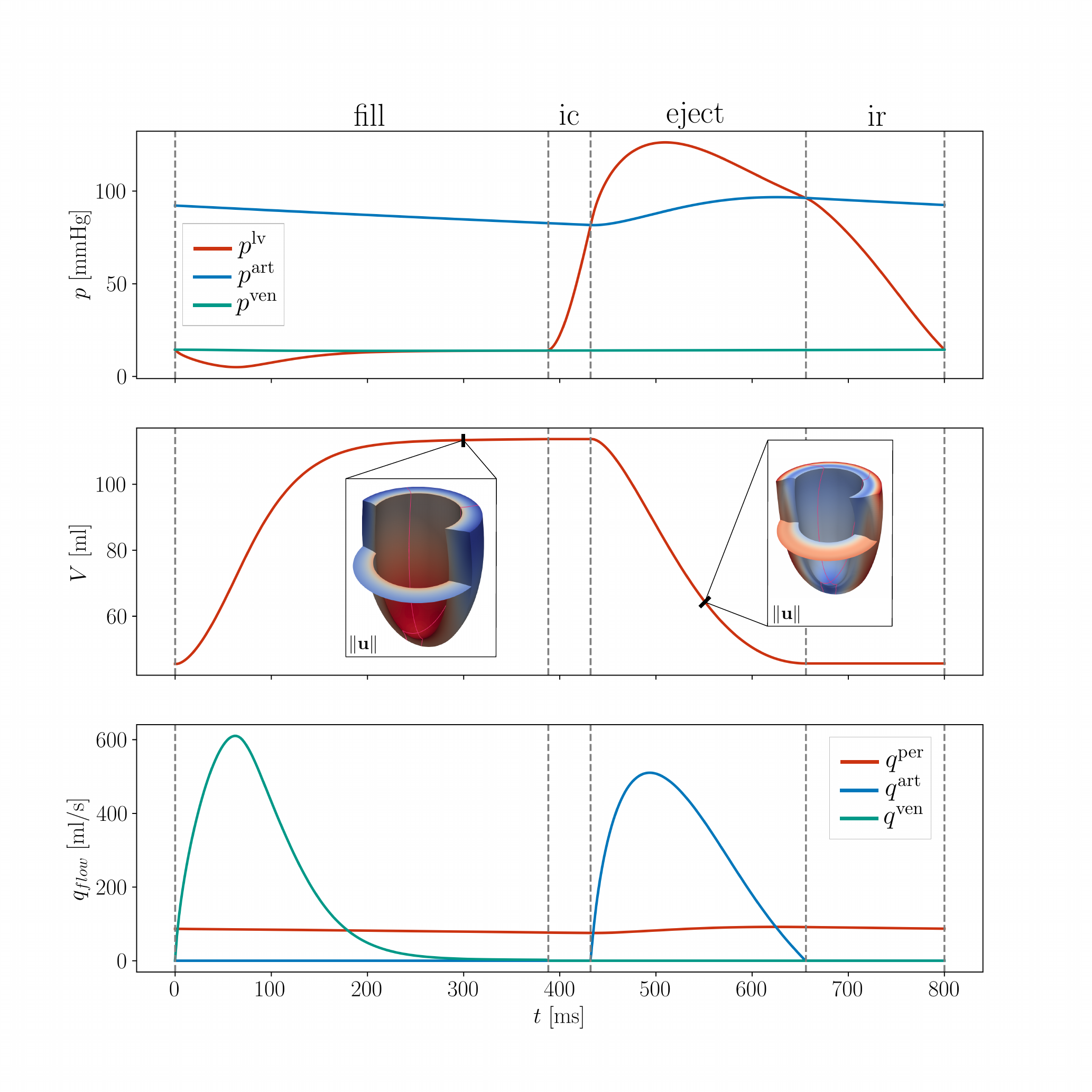}
  \caption{Hemodynamics result of a left ventricle single cardiac cycle. The cardiac cycle is divided into a diastolic phase, consisting of isovolumetric relaxation (ir) and filling (fill), and a systolic phase, consisting of isovolumetric contraction (ic) and ejection (eject). Top row: Pressure evolution of the left ventricle cavity, $p^{\mathrm{lv}}$, the arterial compartment, $p^{\mathrm{art}}$, and the venous compartment, $p^{\mathrm{ven}}$. Middle row: Evolution of the left ventricle cavity volume, with contour plots showing the displacement magnitude during filling and ejection projected on the current (deformed) configuration. The magenta-colored lines are straight in the reference configuration and show the torsional behavior in the current configuration. Bottom row: Arterial $q^{\mathrm{art}}$, venous $q^{\mathrm{ven}}$, and peripheral $q^{\mathrm{per}}$ volume flows, indicating the opening and closure of the left ventricle valves.} \label{fig:typicalcycle}
\end{figure*}

\subsection{Left ventricle model}\label{subsec:leftventricle}
Before the IGA cardiac model detailed in Section~\ref{sec:CardiacModel} is used to study anatomical variations of the bi-ventricle, it is first benchmarked against an established FEA solver~\cite{bovendeerdlvstrain} to identify potential limitations. Our benchmark is limited to the left ventricle geometry. In the extension to the bi-ventricles scenario at the end of this section, it is assumed that this alternative anatomical model does not substantially compromise the validity of the cardiac model (\emph{i.e.}, the physical model benchmarked for the single ventricle is expected to remain valid). Details regarding the left ventricle boundary conditions are given in \ref{app:lvmodel}.

A typical cardiac cycle of the left ventricle is illustrated in Figure~\ref{fig:typicalcycle}, in which the results of the IGA model (computed on a sufficiently refined mesh, as discussed in detail below) are presented. The cardiac cycle is subdivided into a diastolic, \emph{i.e.}, relaxing, and systolic, \emph{i.e.}, contracting, phase. During diastole, the heart has finished ejection and briefly enters the isovolumetric relaxation phase (ir) before it starts filling (fill) with oxygenated blood via the mitral valve, $q^{\mathrm{ven}}$. The systolic phase is entered when muscle contraction starts after filling. The pressure inside the left ventricle increases during isovolumetric contraction (ic), after which the left ventricle pressure exceeds the arterial pressure and the blood is ejected (eject), $q^{\mathrm{art}}$. Figure~\ref{fig:typicalcycle} shows the different pressures, cavity volume, and volume flows as a function of time. The duration of the cardiac cycle is an input of the numerical model and set to be $800$~[ms] or 75 beats-per-minute (bpm) and is controlled by the parameters of the active Cauchy stress component in Section~\ref{sec:CardiacModel}.

In the remainder of this section, we first investigate the convergence behavior of the IGA left ventricle model, before comparing it with the FEA results. The presented comparison pertains to the hemodynamics as well as to the mechanical behavior, \emph{i.e.}, strains and stresses.

\subsubsection*{Convergence}
The coupled 0D-3D cardiac model does not allow for an analytic solution to be derived. However, there exist analytic solutions for the decoupled hyperelastic behavior of the cardiac model on a simplified geometry, \emph{i.e.}, the inflation of an isotropic thick-walled sphere. The numerical results of the thick-walled sphere problem are compared to the derived analytic solution in \ref{app:testcase}, where the considered spherical multi-patch geometry is also discussed. In summary, the results of the IGA cardiac model for this analytical test case exhibit optimal $L^2$ and $H^1$ convergence rates for both quadratic and cubic B-splines given the mixed ($u_i-p$) formulation, and nearly optimal rates for the single field ($u_i$) formulation (Figure~\ref{fig:Sphereconvergence}). Even though the single IGA field formulation exhibits a slightly lower rate, the obtained accuracy is still within an acceptable range for our application. The IGA thick-walled sphere convergence rates are also compared to the FEA convergence rates. It is observed that IGA exhibits superior behavior over the FEA single field formulation in terms of accuracy \emph{vs.} element size, which is mainly attributed to the geometrical approximation inaccuracies introduced during the meshing phase of the FEA procedure. 

\begin{figure*}[!t]
     \centering
     \begin{subfigure}[b]{0.42\textwidth}
         \centering
             \includegraphics[width=\textwidth]{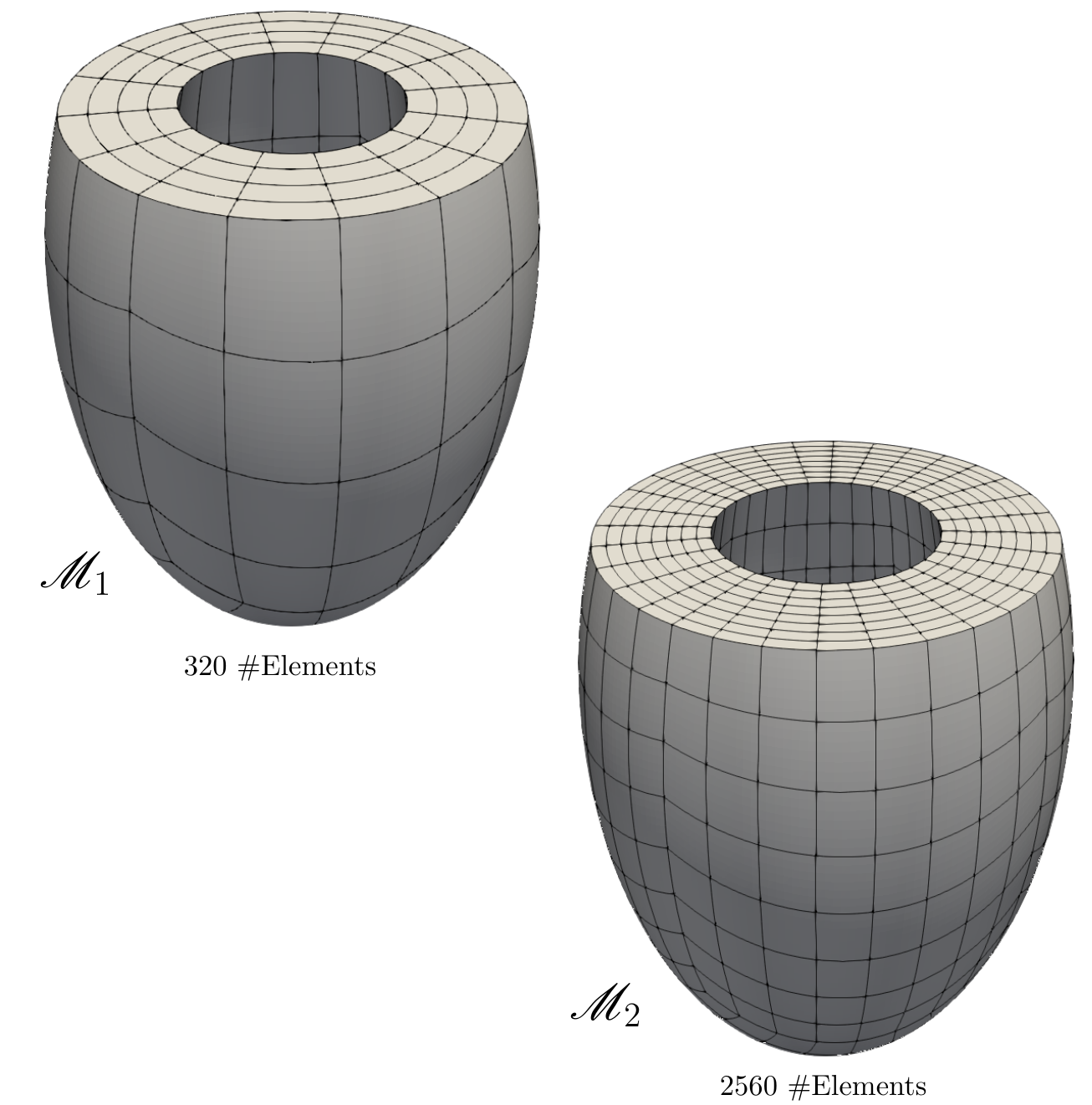}
             \vspace{2em}
         \caption{}
         \label{fig:meshes} 
     \end{subfigure}
     \hfill
     \begin{subfigure}[b]{0.55\textwidth}
         \centering
    \includegraphics[width=\textwidth]{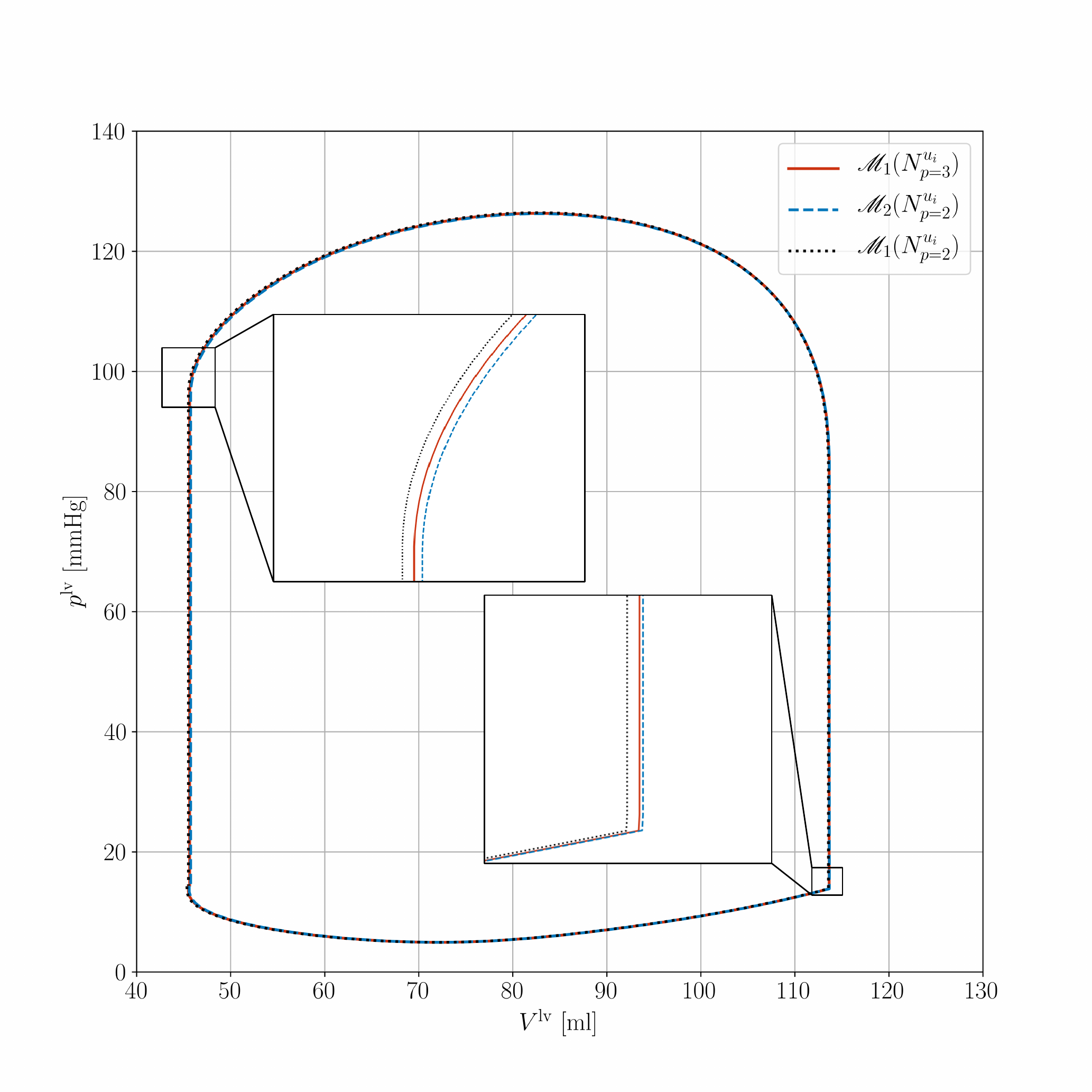}
                  \caption{}
                  \label{fig:PVconvergence}
     \end{subfigure}
        \caption{(a) Two left ventricle meshes, $\mathscr{M}_1$ and $\mathscr{M}_2$, used for the convergence analysis of the model. The meshes are obtained by uniform refinement of the coarsest mesh depicted in Figure~\ref{fig:Templates}, and consist of 320 and 2560 quadratic elements, respectively. (b) The pressure-volume (PV) curves for different spatial discretizations of the displacement field, $\mathscr{M}_1(N^{u_i}_{p=3})$, where subscript $p$ denotes the degree of the B-spline.}
        \label{fig:ComparisonLVHemo}
\end{figure*}

Given that the nearly incompressible hyperelastic response is rigorously verified in terms of mesh convergence rates, we now analyze the convergence of the coupled 0D-3D cardiac model in terms of hemodynamics and mechanical stress-strain behavior. Convergence of the spatial discretization is studied using $h$- and $p$-refinement, \emph{i.e.}, refinement in element size and spline degree, while the temporal discretization is kept constant. The numerical results are analyzed for two meshes, illustrated in Figure~\ref{fig:meshes}, and two spline degrees, quadratic and cubic. The two meshes, $\mathscr{M}_1$ and $\mathscr{M}_2$, are obtained by uniform refinement (knot insertion) of the left ventricle template, which consists of 5 elements and is visualized in Figure~\ref{fig:LVTemplate}. The idealized template allows for an analytic fiber field description explained in \ref{app:lvmodel}, in contrast to the rule-based fiber field method discussed in Section~\ref{sec:RRBM}. This analytic field is insensitive to mesh refinement and is therefore beneficial for convergence analyses. Results are visualized for a fixed time step of $\Delta t=2$~[ms] at the $5th$ cardiac cycle, which is where the model exhibits a cyclic steady-state behavior. A complete overview of the simulation settings is provided in \ref{app:lvmodel}.

The hemodynamical results are presented in Figure~\ref{fig:PVconvergence}, in which the left ventricle cavity pressure-volume (PV) loop is illustrated. Minor differences are observed between the different spatial discretizations. The 0D circulatory model is dependent on the cavity volume, which remains unchanged under mesh refinement given the IGA paradigm. The observed differences are therefore related to the spatial approximation of the cardiac model and not the parametrization of the geometry. It is also observed that the results of the finer mesh with quadratic basis functions for the displacement discretization, $\mathscr{M}_2(N^{u_i}_{p=2})$, and the coarser mesh with cubic B-splines, $\mathscr{M}_1(N^{u_i}_{p=3})$, are closely positioned. The other hemodynamic quantities, \emph{i.e.}, volume changes and flows, are not presented since there was no distinct difference noticeable. From these results, it is concluded that the hemodynamical response of the FEA benchmark is closely captured by all considered IGA discretizations.

\begin{figure*}[!t]
     \centering
     \begin{subfigure}[b]{0.42\textwidth}
         \centering
             \includegraphics[width=\textwidth]{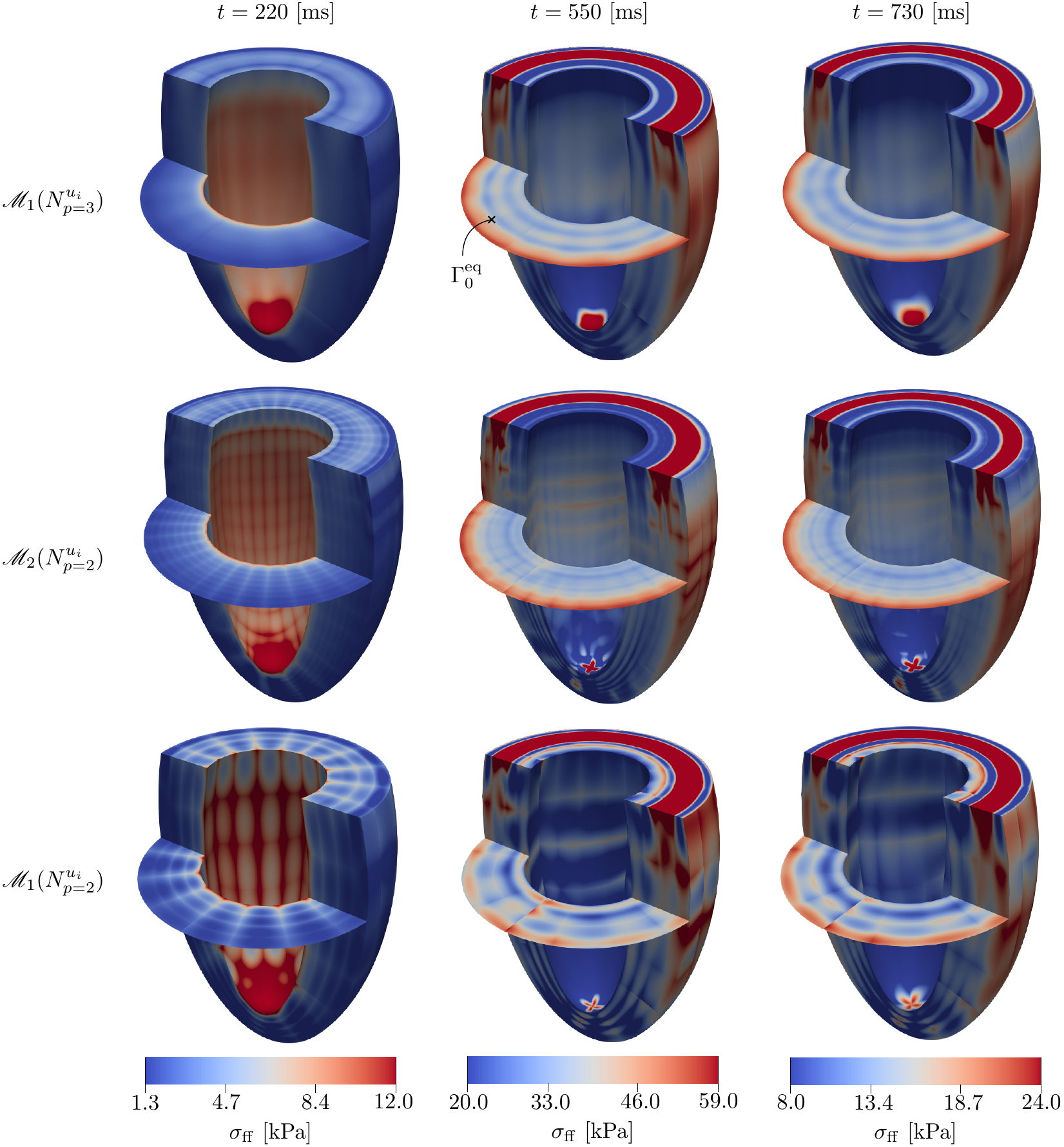}
         \caption{}
         \label{fig:ContourStressLV} 
     \end{subfigure}
     \hfill
     \begin{subfigure}[b]{0.55\textwidth}
         \centering
    \includegraphics[width=\textwidth]{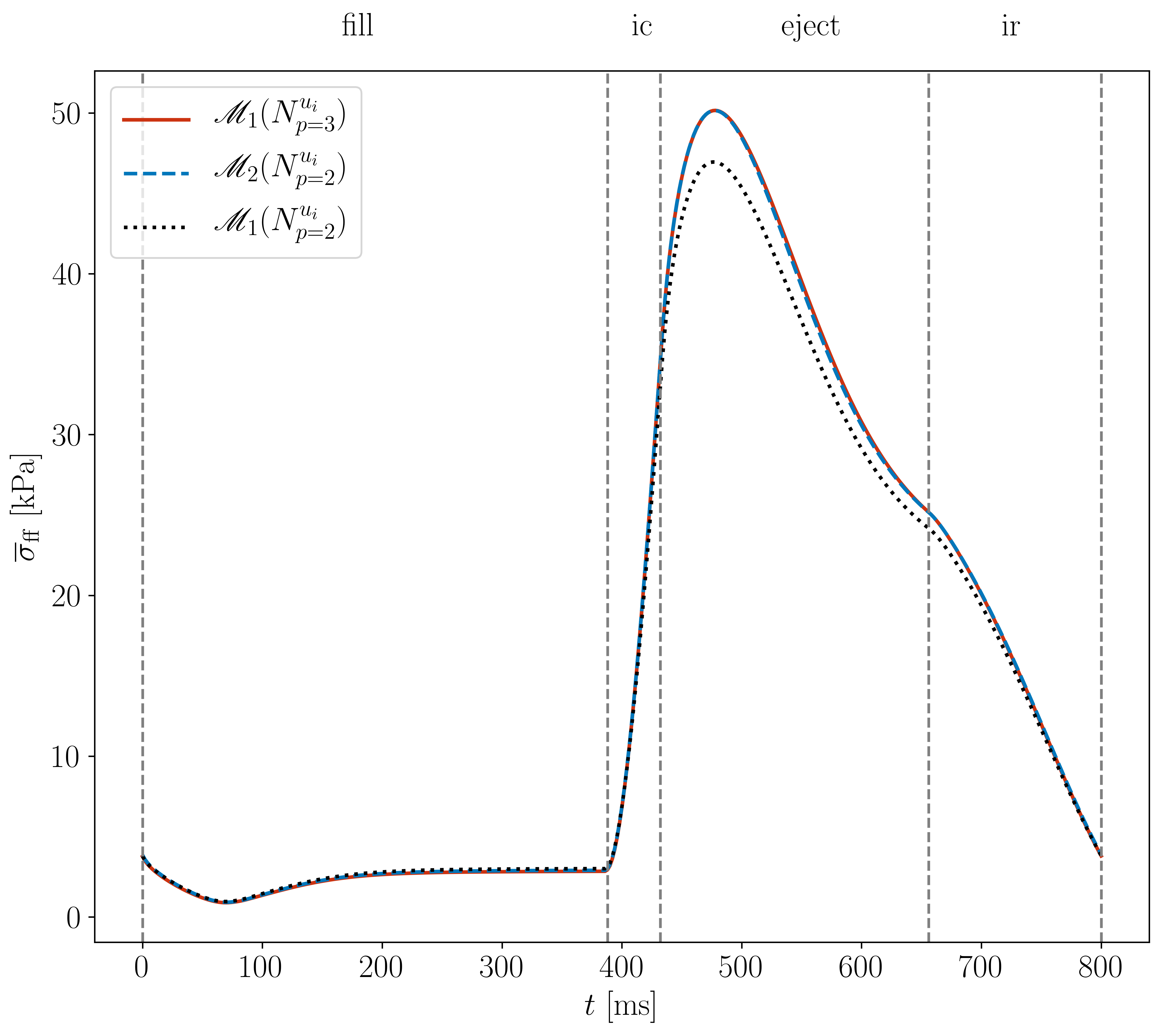}
                  \caption{}
                  \label{fig:StressComparison}
     \end{subfigure}
        \caption{(a) Contour plots of the total Cauchy fiber stress component $\sigma_{\mathrm{ff}}$ visualized on the left ventricle half in the reference configuration, $\Omega_0$. The equator plane is denoted by $\Gamma^{\mathrm{eq}}_0$ and located at $z=0$. The rows indicate the spatial discretization for the displacement field and the columns the time-instances. (b) The total Cauchy fiber stress component is spatially averaged over the equatorial plane $\overline{\sigma}_{\mathrm{ff}}$ and monitored in time. The vertical dashed lines indicate the different cardiac phases.}
        \label{fig:StressConvergence}
\end{figure*}

\begin{figure*}[!t]
     \centering
     \begin{subfigure}[b]{0.42\textwidth}
         \centering
             \includegraphics[width=\textwidth]{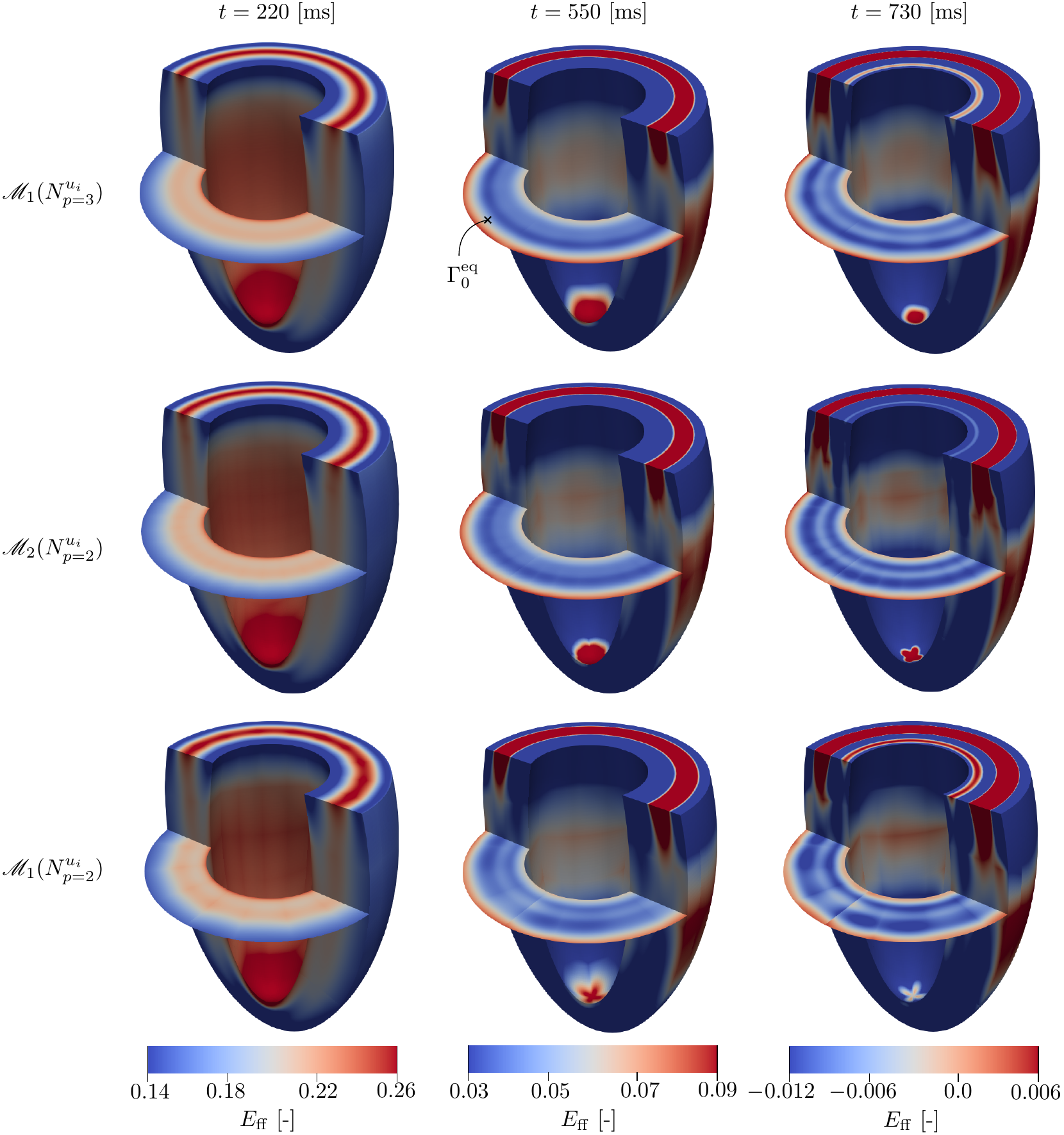}
         \caption{}
         \label{fig:ContourStrainLV} 
     \end{subfigure}
     \hfill
     \begin{subfigure}[b]{0.55\textwidth}
         \centering
    \includegraphics[width=\textwidth]{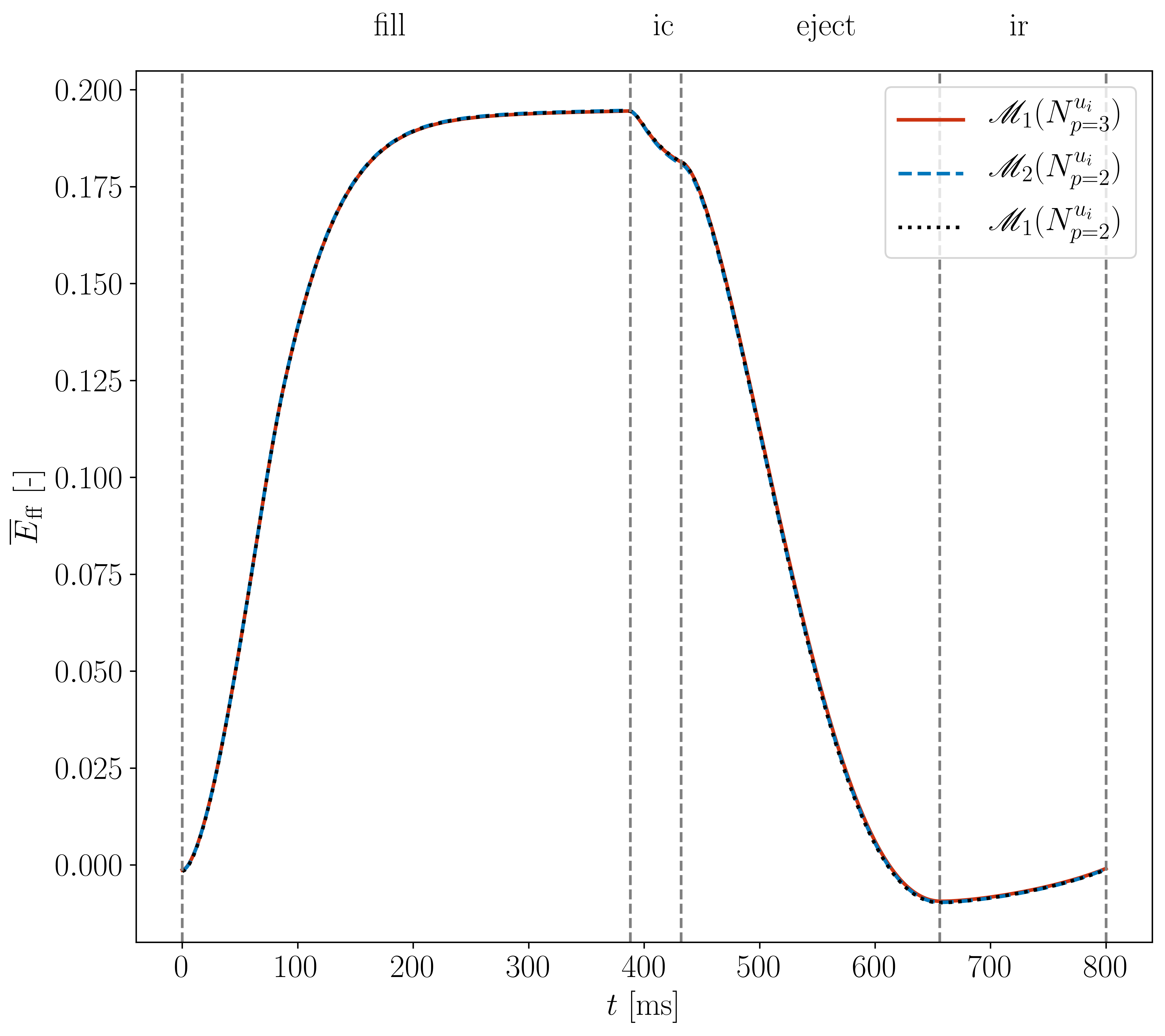}
                  \caption{}
                  \label{fig:StrainComparison}
     \end{subfigure}
        \caption{(a) Contour plots of the Green-Lagrange strain fiber component $E_{\mathrm{ff}}$ visualized on the left ventricle half in the reference configuration, $\Omega_0$. The equator plane is denoted by $\Gamma^{\mathrm{eq}}_0$ and located at $z=0$. The Green-Lagrange strain tensor is defined with respect to the reference configuration. The rows indicate the spatial discretization for the displacement field and the columns the time-instances. (b) The Green-Lagrange strain fiber component is spatially averaged over the equatorial plane $\overline{E}_{\mathrm{ff}}$ and monitored in time. The vertical dashed lines indicate the different cardiac phases.}
        \label{fig:StrainConvergence}
\end{figure*}

The mechanical results are compared for the fiber component of the total Cauchy fiber stress and Green-Lagrange strain tensor, and the circumferential-longitudinal shear component of the Green-Lagrange strain tensor, defined as
\begin{subequations}
\begin{align}
     \sigma_{\mathrm{ff}} & = \mathrm{e}^{\mathrm{f0}}_i     \sigma_{ij}     \mathrm{e}^{\mathrm{f0}}_j, \\
       E_{\mathrm{ff}}    & = \mathrm{e}^{\mathrm{f0}}_i E^{\phantom{}}_{ij} \mathrm{e}^{\mathrm{f0}}_j, \\
       E_{\mathrm{cl}}    & = \mathrm{e}^{\mathrm{c0}}_i E^{\phantom{}}_{ij} \mathrm{e}^{l\mathrm{0}}_j,
\end{align}
\end{subequations}
where $\sigma_{ij}$ is given in Equation~\eqref{eq:secondPstress} and $E^{\phantom{}}_{ij}$ is the Green-Lagrange strain tensor with respect to the (undeformed) reference configuration, Equation~\eqref{eq:GreenLagr}. The quantities are integrated over the equatorial plane, $\Gamma^{\mathrm{eq}}_0$, defined in the reference configuration $\Omega_0$ where $z=0$, at each time-instance $t$ according to 
\begin{equation}
    \overline{q}(t) = \frac{1}{\int \text{d} \Gamma^{\mathrm{eq}}}_0 \int q(x_i, t) \ \text{d} \Gamma^{\mathrm{eq}}_0 \quad \text{for} \quad q = \sigma_{\mathrm{ff}}, E_{\mathrm{ff}},  E_{\mathrm{cl}}.
\end{equation}
The results are compared in Figures~\ref{fig:StressConvergence} and~\ref{fig:StrainConvergence}.

\begin{figure*}[!t]
     \centering
    \begin{subfigure}[b]{0.42\textwidth}
         \centering
    \includegraphics[width=\textwidth]{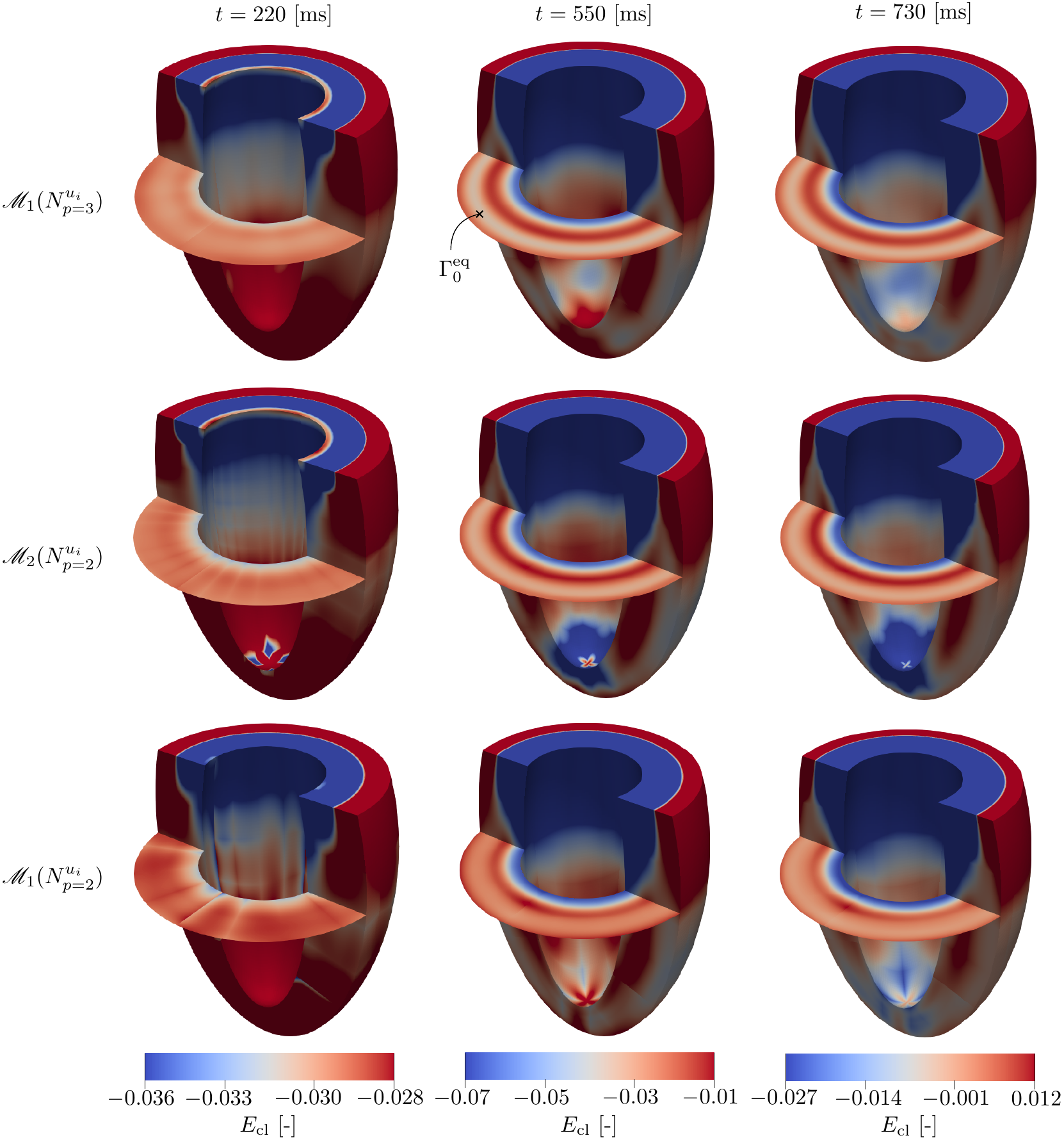}
                  \caption{}
                  \label{fig:ContourShearLV}
     \end{subfigure}
     \hfill
    \begin{subfigure}[b]{0.55\textwidth}
         \centering
    \includegraphics[width=\textwidth]{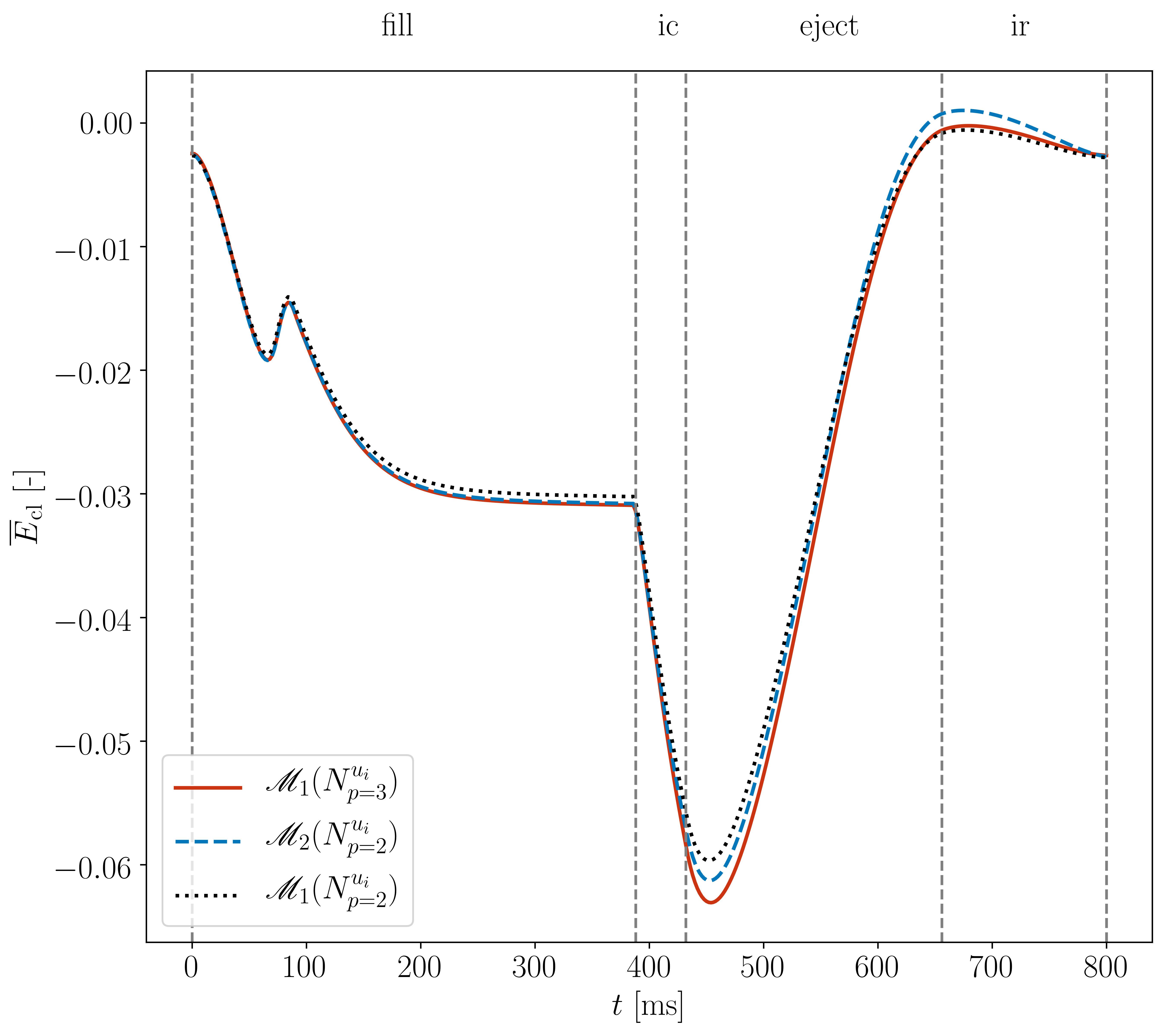}
                  \caption{}
                  \label{fig:ShearComparisonLV}
     \end{subfigure}\\
          \begin{subfigure}[b]{0.5\textwidth}
         \centering
             \includegraphics[width=\textwidth]{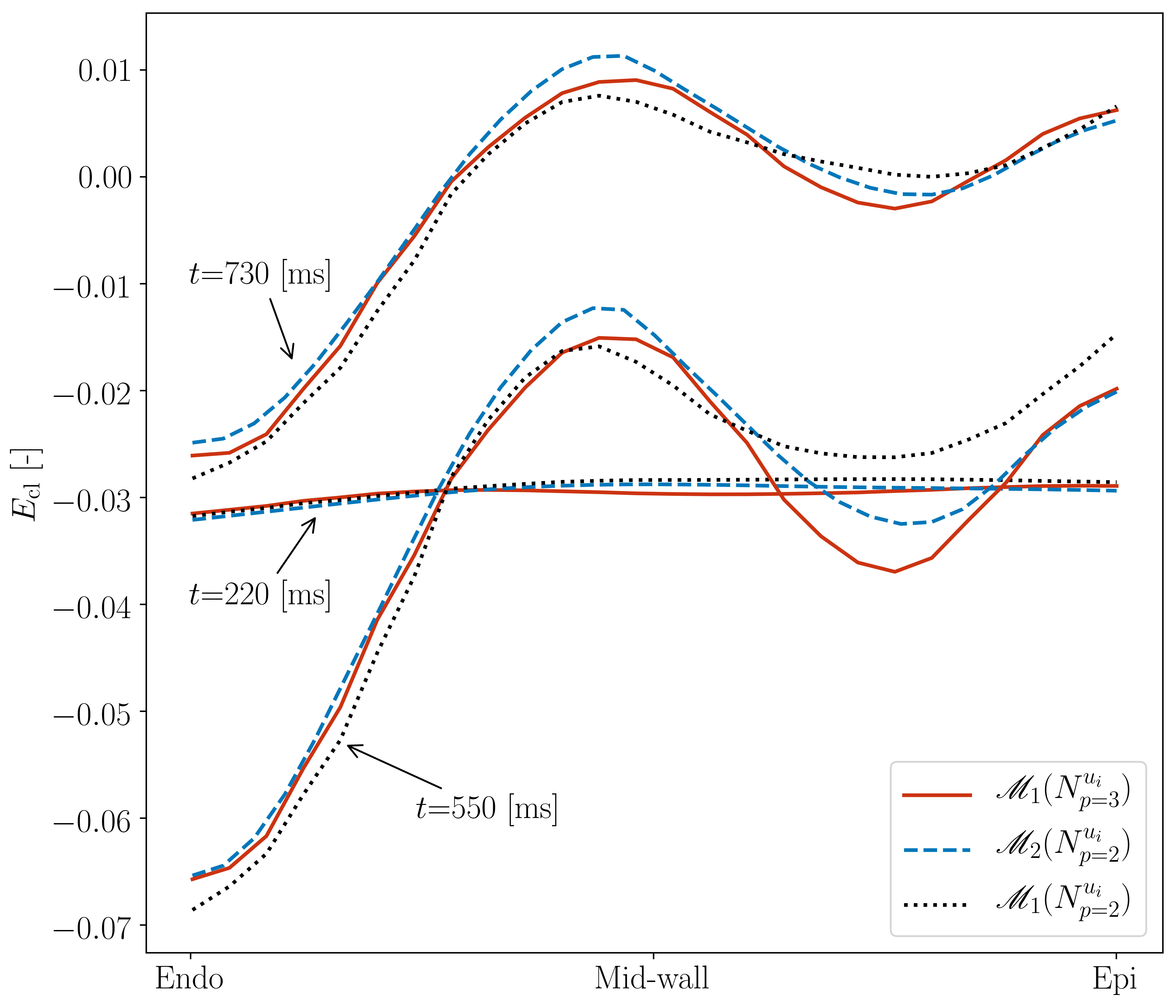}
         \caption{}
         \label{fig:ShearComparisonRadialLV} 
     \end{subfigure}
        \caption{(a) Contour plots of the Green-Lagrange strain longitudinal-circumferential shear component $E_{\mathrm{cl}}$. The shear component is visualized on the left ventricle half in the reference configuration, $\Omega_0$, where the equator plane is denoted by $\Gamma^{\mathrm{eq}}_0$ and located at $z=0$. The Green-Lagrange strain shear component is defined with respect to the reference configuration. The rows indicate the spatial discretization for the displacement field and the columns the time-instances. (b) The Green-Lagrange strain shear component is spatially averaged over the equatorial plane $\overline{E}_{\mathrm{cl}}$ and monitored in time. The vertical dashed lines indicate the different cardiac phases. (c) Radial distribution of the shear component at three time-instances. The shear components were averaged in the circumferential direction to obtain the radial distribution.}
        \label{fig:ShearConvergence}
\end{figure*}

The considered mechanical quantities show a clear converging behavior for the stress and strain fiber component, Figure~\ref{fig:StressConvergence} and \ref{fig:StrainConvergence}. This observation is in line with the thick-walled sphere, in which the principal components in stress and strain converge rapidly for coarse meshes with high order (cubic) basis functions, see \ref{app:testcase}. However, the averaged shear component of the strain, Figure~\ref{fig:ShearComparisonLV}, exhibits slightly different behavior, in which no clear convergence is observed between the $\mathscr{M}_1(N^{u_i}_{p=3})$ and $\mathscr{M}_2(N^{u_i}_{p=2})$ discretizations. This is most noticeable when comparing these two results to the coarsest discretization, $\mathscr{M}_1(N^{u_i}_{p=2})$, with the difference clearly visible at the global maxima and minima of the curve, Figure~\ref{fig:ShearComparisonLV}. By investigating the radial distribution at three time-instances, Figure~\ref{fig:ShearComparisonRadialLV}, it is, however, observed that the distribution does converge, albeit rather slowly. The radial distribution also exhibits an oscillating behavior, which can explain the observed discrepancy in the convergence of the integrated average quantity in Figure~\ref{fig:ShearComparisonLV}. 

The contour plots of the fiber stress and strain quantities, {Figures~\ref{fig:StressConvergence}-\ref{fig:ShearConvergence}}, show a similar trend for all quantities for varying discretizations. First, the approximate solution of the cardiac model exhibits undesired boundary effects near the basal plane, caused by the normal displacement constraint, and near the apex, caused by a singularity in the fiber field \ref{appS:fiberfield}. It is therefore common to only analyze the solution sufficiently far away from these two locations. Second, the approximate solution results with a quadratic B-spline discretization, $\mathscr{M}_2(N^{u_i}_{p=2})$ and $\mathscr{M}_1(N^{u_i}_{p=2})$, do not allow for $C^{1}$-continuity (or higher) across element interfaces for quantities which are one order of regularity lower than the displacement, \emph{i.e.}, stresses and strains. This is solved by employing cubic B-splines, which are $C^{2}$-continuous and yield smooth stresses and strains, which is expected to be closer to the physiological behavior in comparison to the $C^{1}$-continuity of the quadratic B-splines. However, inter-patch continuity remains unaffected by the spline order and is restricted to $C^{0}$-continuity

Based on the convergence study presented in this section, in the remainder of this contribution, we will employ cubic B-splines, $N^{u_i}_{p=3}$, for the displacement field in combination with the coarsest mesh, $\mathscr{M}_1$. While a finer mesh yields more accurate results, on the coarsest mesh the numerical errors are anticipated to be small in comparison to the parametric uncertainties (both anatomical and physiological).

\subsubsection*{Comparison to FEA}\label{sec:FEniCSvsNutils}
The developed IGA model is compared to an established FEA cardiac model~\cite{bovendeerdlvstrain} to build confidence in the model results and identify potential limitations. The IGA model is constructed using the Nutils Python toolbox~\cite{nutils7}, while the FEA model is constructed using the FEniCS Python toolbox~\cite{FEniCSBook}. Both models solve the same system of equations as specified in Section~\ref{sec:CardiacModel}, but with a fundamentally different implementation on account of the employed software frameworks. A detailed discussion of the differences between both frameworks is considered impractical and therefore omitted here. However, the main difference with regard to the solver is noteworthy: The IGA model solves the entire set of coupled equations in a monolithic approach, which allows for fully and semi-implicit time-integration. The FEniCS implementation relies on a staggered approach and for every time step employs an explicit one-step method \cite{BOVENDEERD19921129}. To limit potential result differences, we employ identical boundary conditions and temporal conditions, \emph{i.e.}, initial conditions, time step size, and time-integration scheme (\ref{app:lvmodel}), for both the IGA and FEA models. Additionally, the fiber field is derived analytically and the epicardium boundary condition in Equation~\eqref{eq:setofeqs} is replaced by a zero-traction condition in accordance with Ref.~\cite{bovendeerdlvstrain}. The remaining parameter values are discussed in \ref{app:lvmodel}.

\begin{figure}[]
\centering
\includegraphics[width=0.5\textwidth]{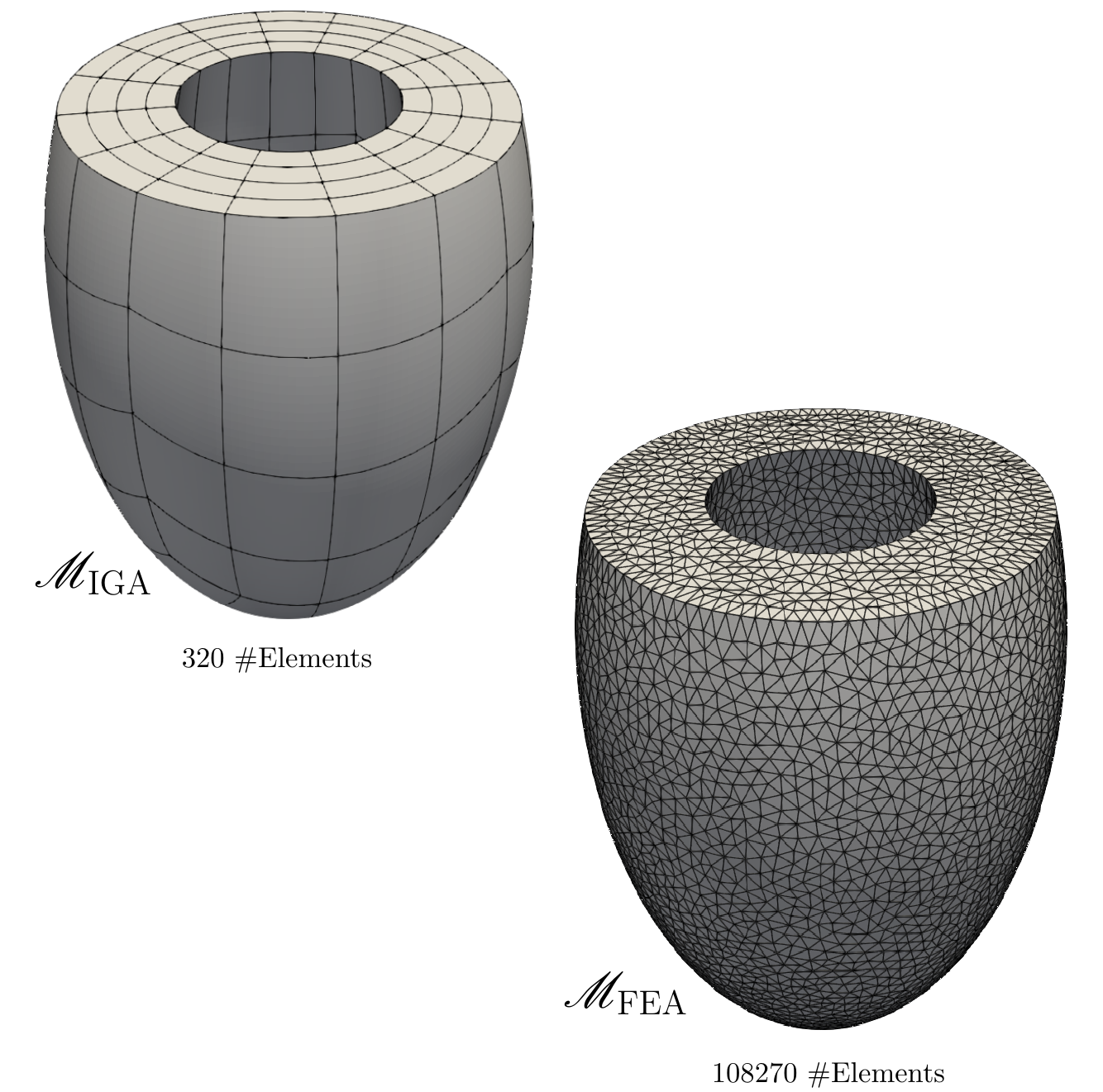}
  \caption{The meshes that are used in the IGA, $\mathscr{M}_{\mathrm{IGA}}$, and FEA model, $\mathscr{M}_{\mathrm{FEA}}$. The IGA mesh is obtained by uniform refinement of the coarse template geometry, Figure~\ref{fig:LVTemplate}, which consists of 320 quadratic NURBS elements resulting in a total of 4,700 DOFs. The FEA mesh approximates the analytic ellipsoid shape and consists of 108,270 linear tetrahedron elements resulting in 469,734 DOFs. The IGA model is constructed and solved with the Nutils Python toolbox\cite{nutils7}, while the FEA model employs the FEniCS Python toolbox~\cite{FEniCSBook}. } \label{fig:meshFEA}
\end{figure}

We use two different meshes for the comparison, denoted by $\mathscr{M}_{\mathrm{IGA}}$ and $\mathscr{M}_{\mathrm{FEA}}$, which are shown in Figure~\ref{fig:meshFEA}. The IGA mesh is obtained by uniform refinement of the quadratic NURBS template, visualized in Figure~\ref{fig:LVTemplate}, while the unstructured FEA mesh consists of linear tetrahedrons. The IGA model employs cubic B-splines for the displacement vector $u_i$ and quadratic B-splines for the contractile length $l^{\mathrm{c}}$, resulting in a nonlinear monolithic system with a total number of 4,700 degrees-of-freedom (DOFs). The FEA model discretizes the displacement vector using quadratic Lagrangian basis functions, resulting in a nonlinear staggered system with 469,734 DOFs. We do want to emphasize that the FEniCS approximate solution is not considered to be the ground truth, since it is an approximation itself and should ideally have been subject to a conclusive convergence assessment. Although the FEniCS approximate solution is observed to be converged for the principal strain components, it is not yet fully converged for the shear-strain components. However, the accuracy of these shear components is still considered to be within acceptable accuracy for clinical research and thus considered to be sufficient for the comparison considered here.

\begin{figure*}[!t]
\includegraphics[width=\textwidth]{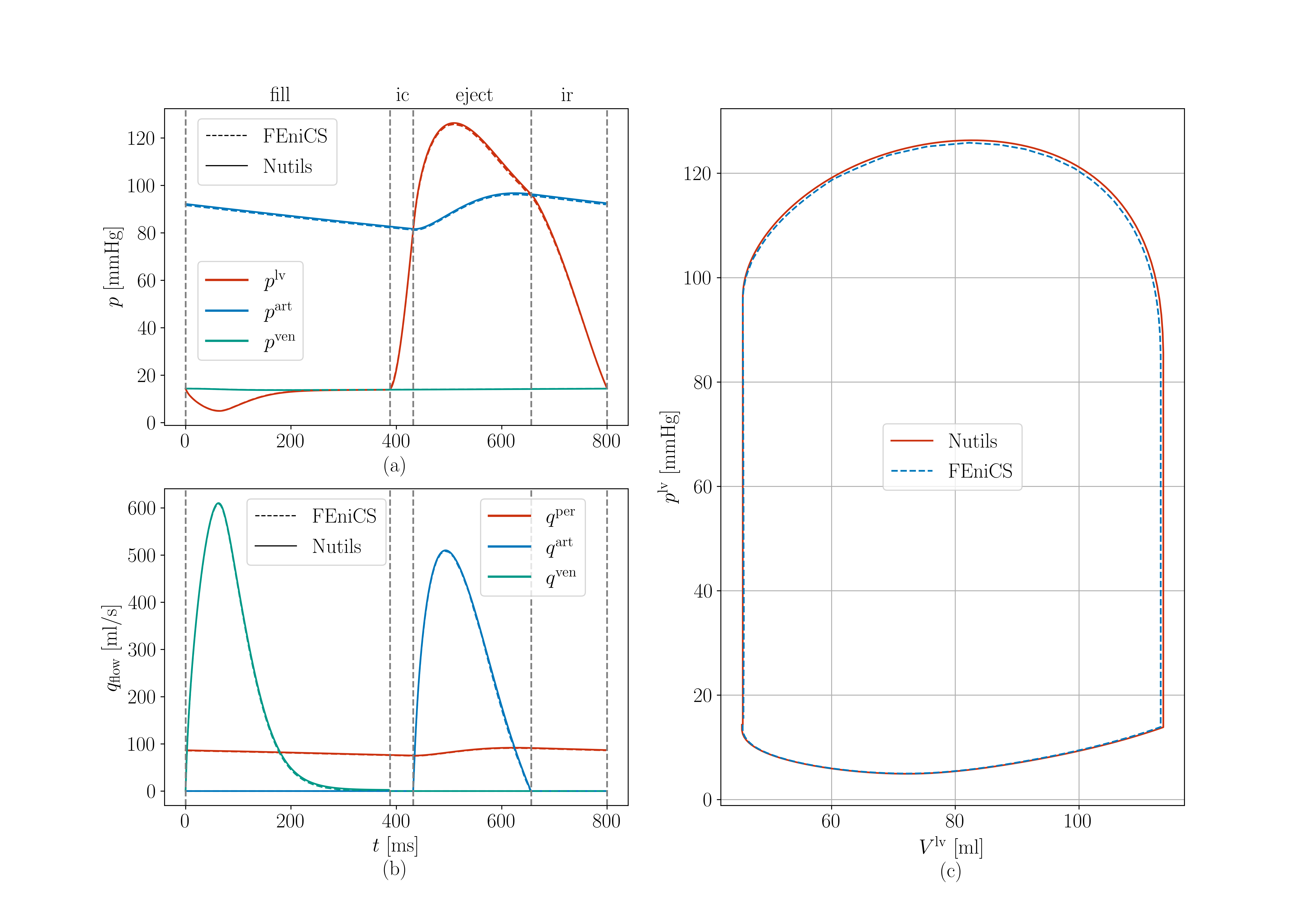}
  \caption{Comparison of the hemodynamical results for the IGA and FEA cardiac models given the $5^{\rm th}$ cardiac cycle. (a) Pressure profiles of the arterial, venous, and left ventricle cavities. (b) Volume flow profiles of the arterial, venous, and peripheral flow. (c) Pressure-volume loop of the left ventricle.} \label{fig:HemodynamicsValidation}
\end{figure*}

\begin{figure*}[!t]
     \centering
     \begin{subfigure}[b]{0.5\textwidth}
         \centering
             \includegraphics[width=\textwidth]{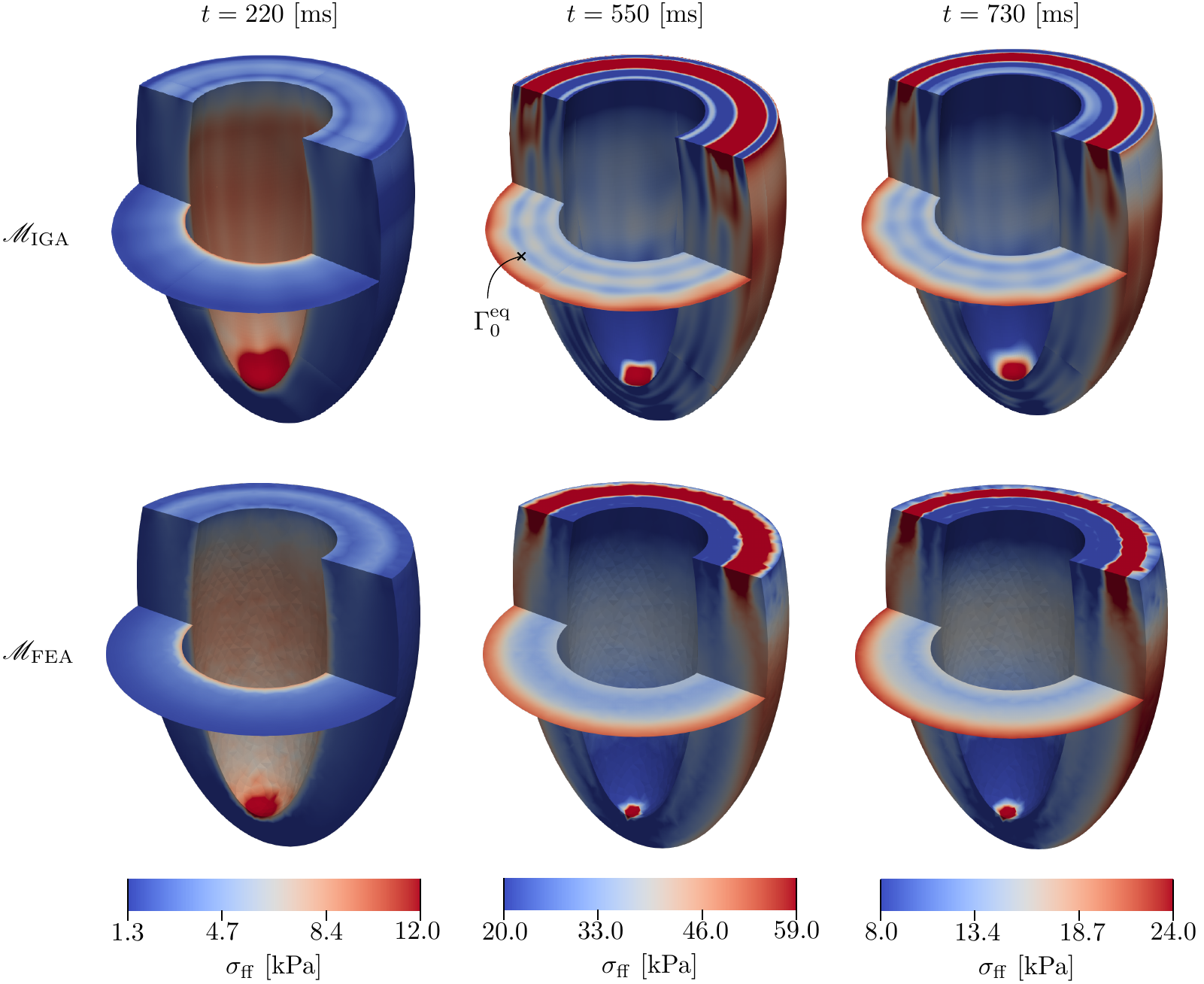}
         \caption{}
         \label{fig:ContourStressValidationLV} 
     \end{subfigure}
     \hfill
     \begin{subfigure}[b]{0.45\textwidth}
         \centering
    \includegraphics[width=\textwidth]{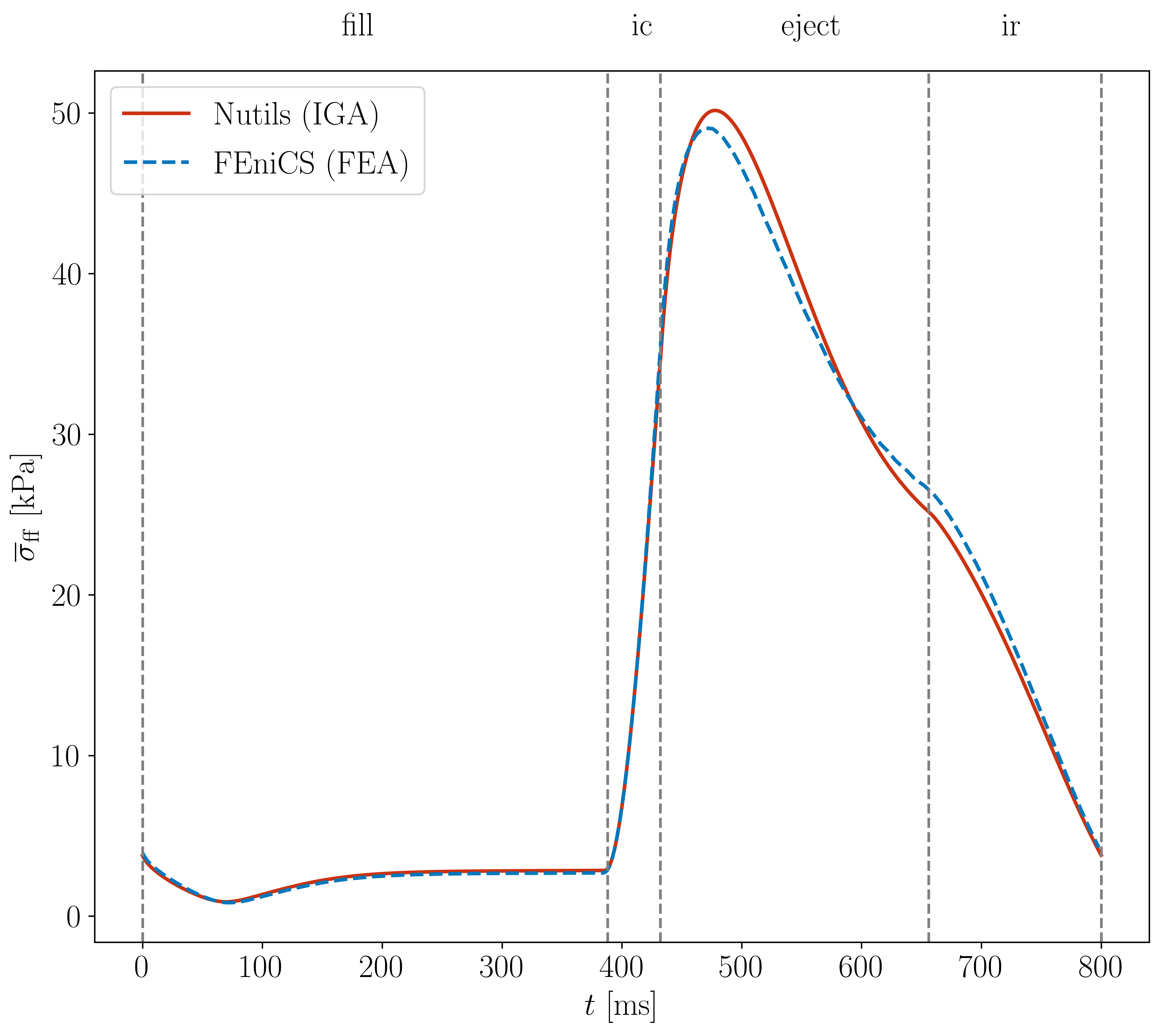}
                  \caption{}
                  \label{fig:StressValidationAverage}
     \end{subfigure}
        \caption{Comparison between the FEniCS finite element analysis (FEA) and the Nutils isogeometric analysis (IGA) result. (a) Contour plots of the total Cauchy fiber stress component $\sigma_{\mathrm{ff}}$ visualized on the left ventricle half in the reference configuration, $\Omega_0$. The equator plane is denoted by $\Gamma^{\mathrm{eq}}_0$ and located at $z=0$. The rows indicate the spatial discretization for the displacement field and the columns the time-instances. (b) The total Cauchy fiber stress component is spatially averaged over the equatorial plane $\overline{\sigma}_{\mathrm{ff}}$ and monitored in time. The vertical dashed lines indicate the different cardiac phases.}
        \label{fig:StressValidation}
\end{figure*}

\begin{figure*}[!t]
     \centering
     \begin{subfigure}[b]{0.5\textwidth}
         \centering
             \includegraphics[width=\textwidth]{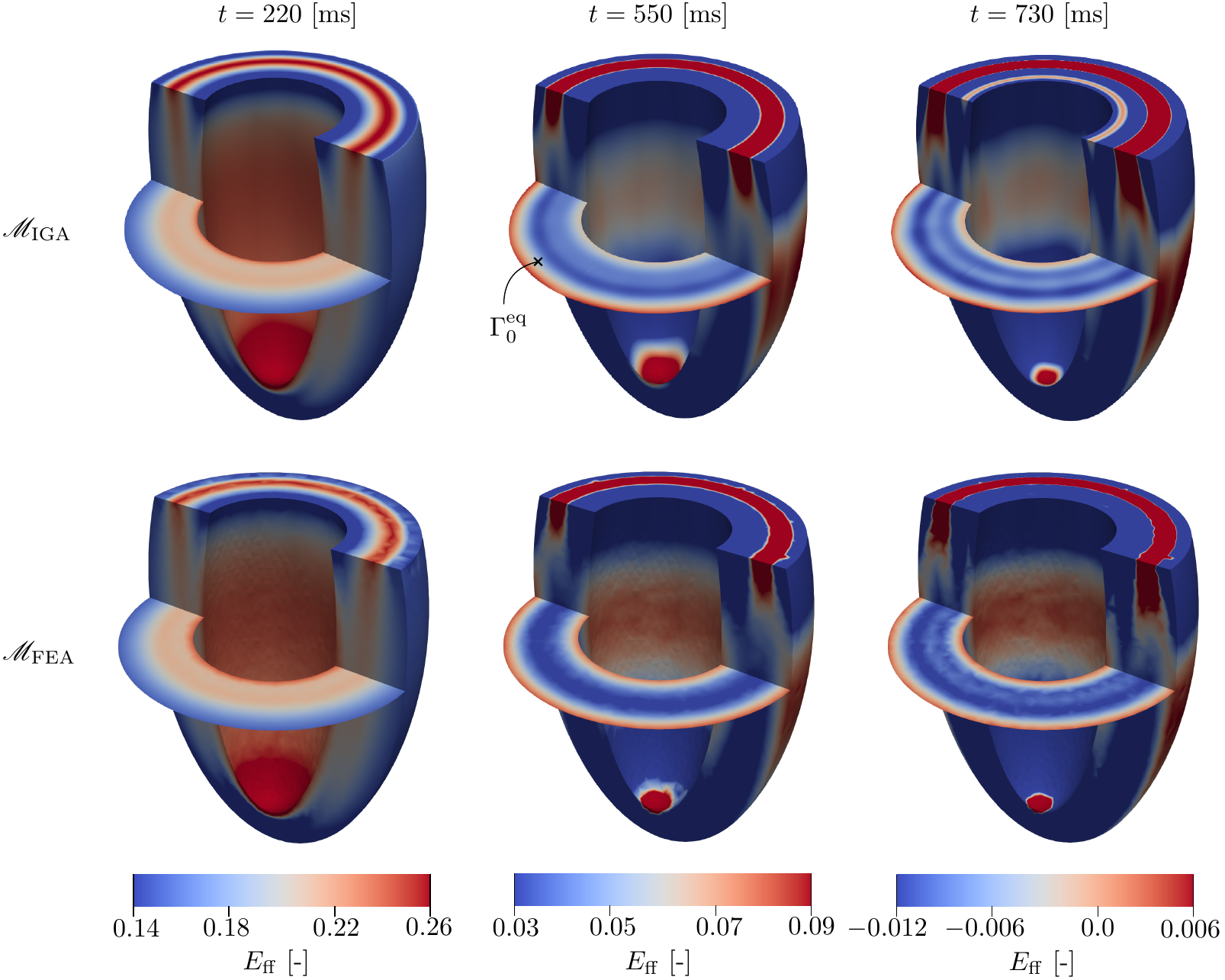}
         \caption{}
         \label{fig:ContourStrainValidationLV} 
     \end{subfigure}
     \hfill
     \begin{subfigure}[b]{0.45\textwidth}
         \centering
    \includegraphics[width=\textwidth]{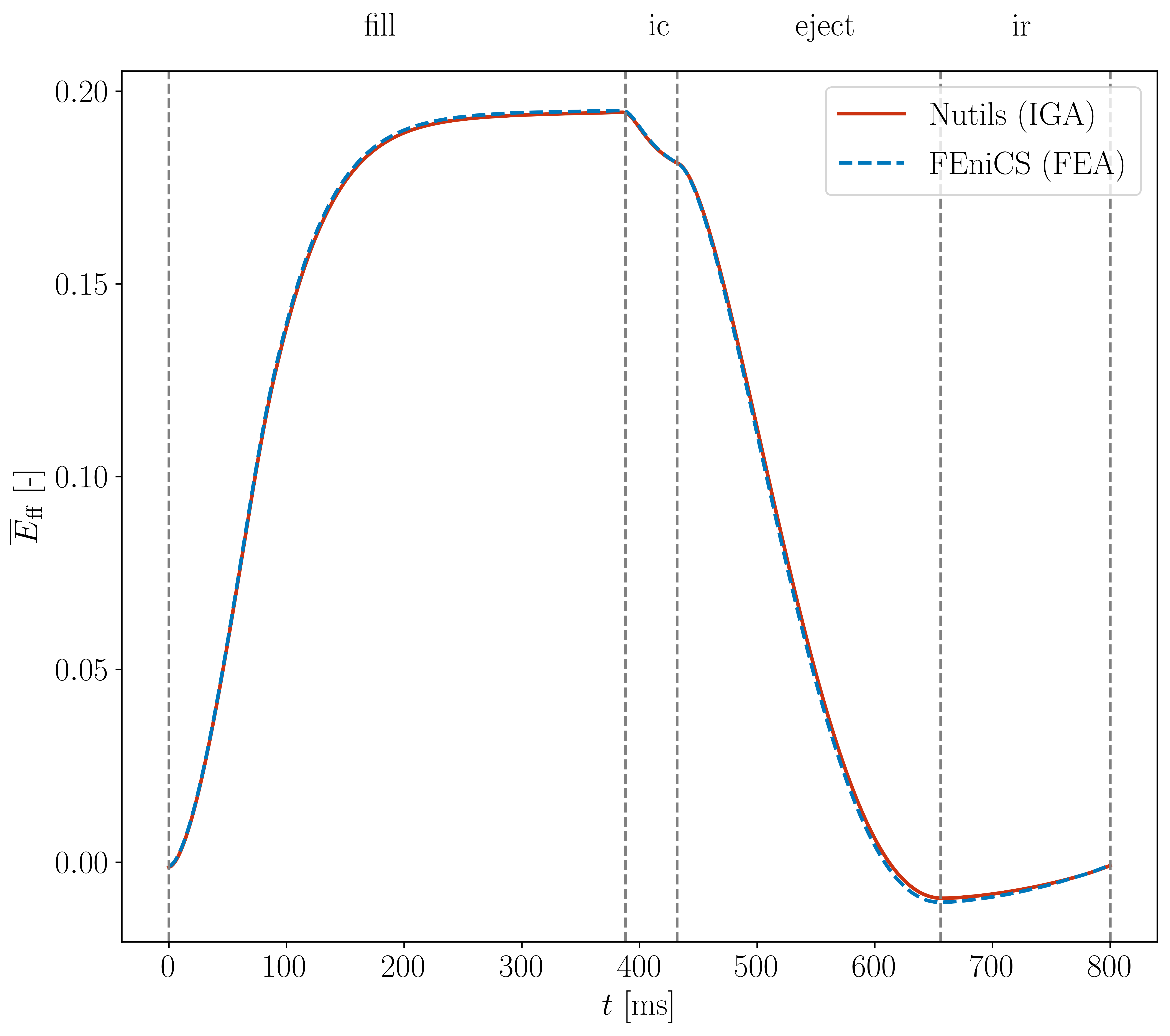}
                  \caption{}
                  \label{fig:StrainValidationAverage}
     \end{subfigure}
        \caption{Comparison between the FEniCS finite element analysis (FEA) and the Nutils isogeometric analysis (IGA) result. (a) Contour plots of the Green-Lagrange strain fiber component $E_{\mathrm{ff}}$ visualized on the left ventricle half in the reference configuration, $\Omega_0$. The equator plane is denoted by $\Gamma^{\mathrm{eq}}_0$ and located at $z=0$. The Green-Lagrange strain tensor is defined with respect to the reference configuration. The rows indicate the spatial discretization for the displacement field and the columns the time-instances. (b) The Green-Lagrange strain fiber component is spatially averaged over the equatorial plane $\overline{E}_{\mathrm{ff}}$ and monitored in time. The vertical dashed lines indicate the different cardiac phases.}
        \label{fig:StrainValidation}
\end{figure*}

\begin{figure*}[!t]
     \centering
    \begin{subfigure}[b]{0.5\textwidth}
         \centering
    \includegraphics[width=\textwidth]{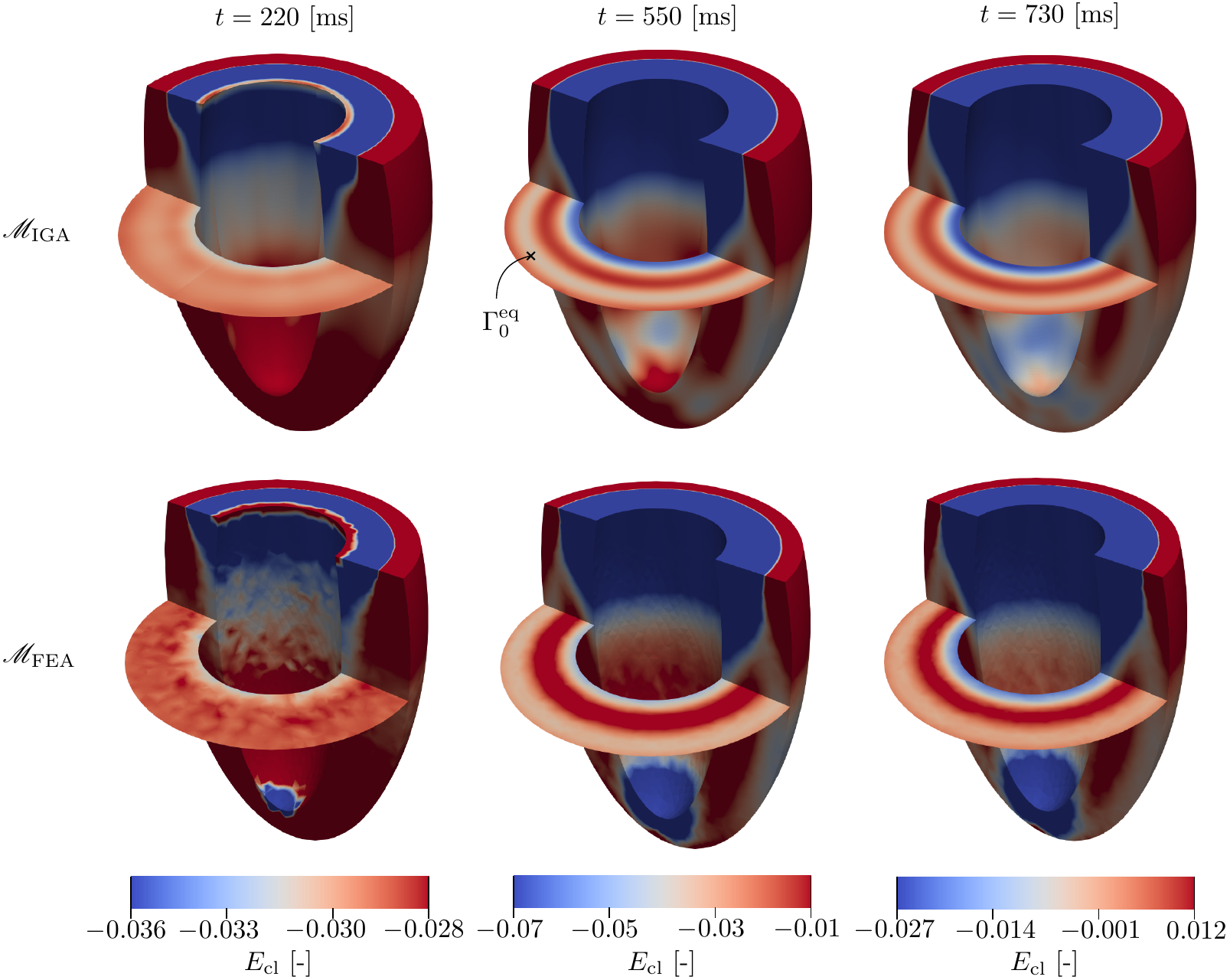}
                  \caption{}
                  \label{fig:ContourShearValidationLV}
     \end{subfigure}
     \hfill
    \begin{subfigure}[b]{0.45\textwidth}
         \centering
    \includegraphics[width=\textwidth]{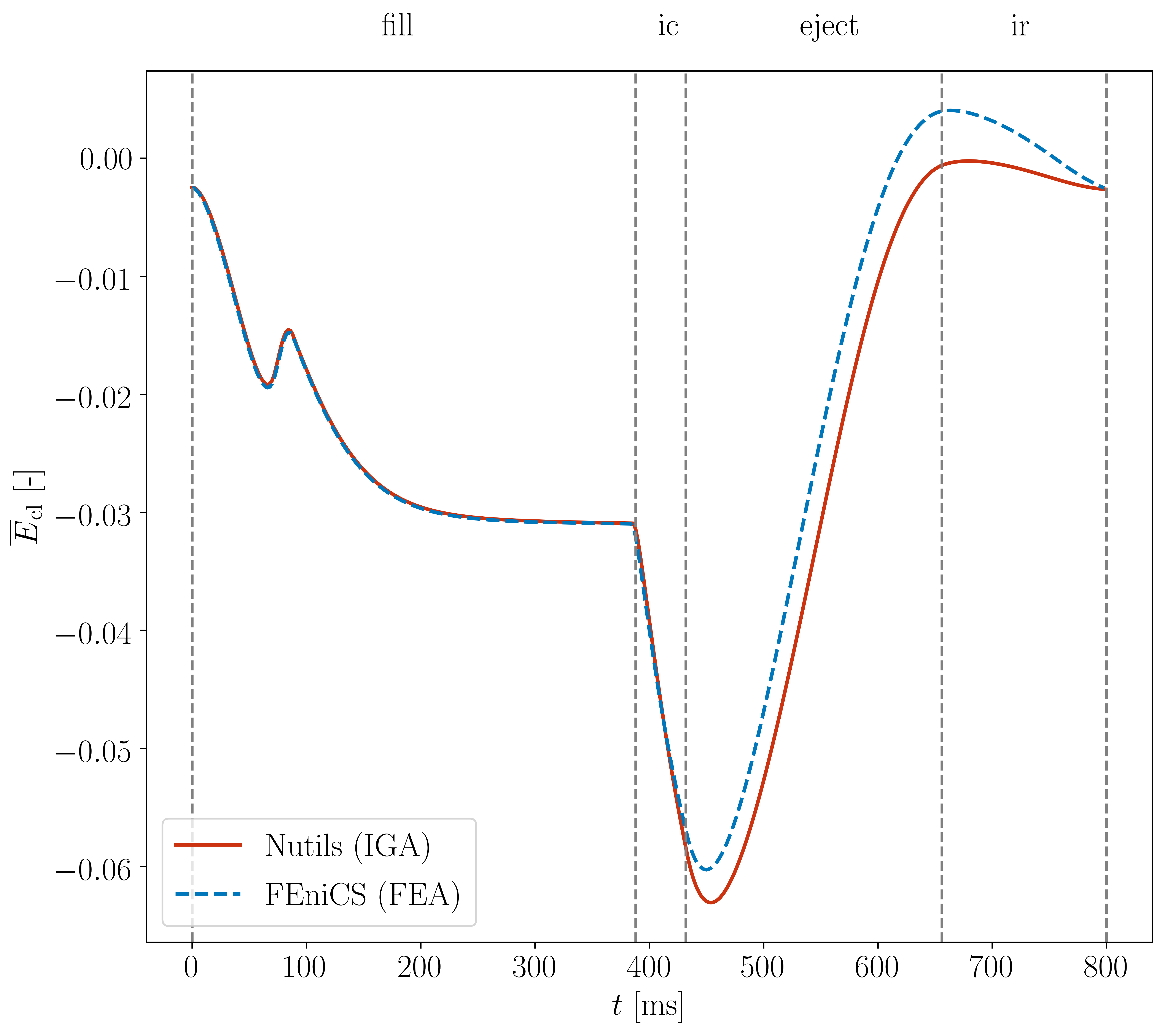}
                  \caption{}
                  \label{fig:ShearValidationAverage}
     \end{subfigure}\\
          \begin{subfigure}[b]{0.5\textwidth}
         \centering
             \includegraphics[width=\textwidth]{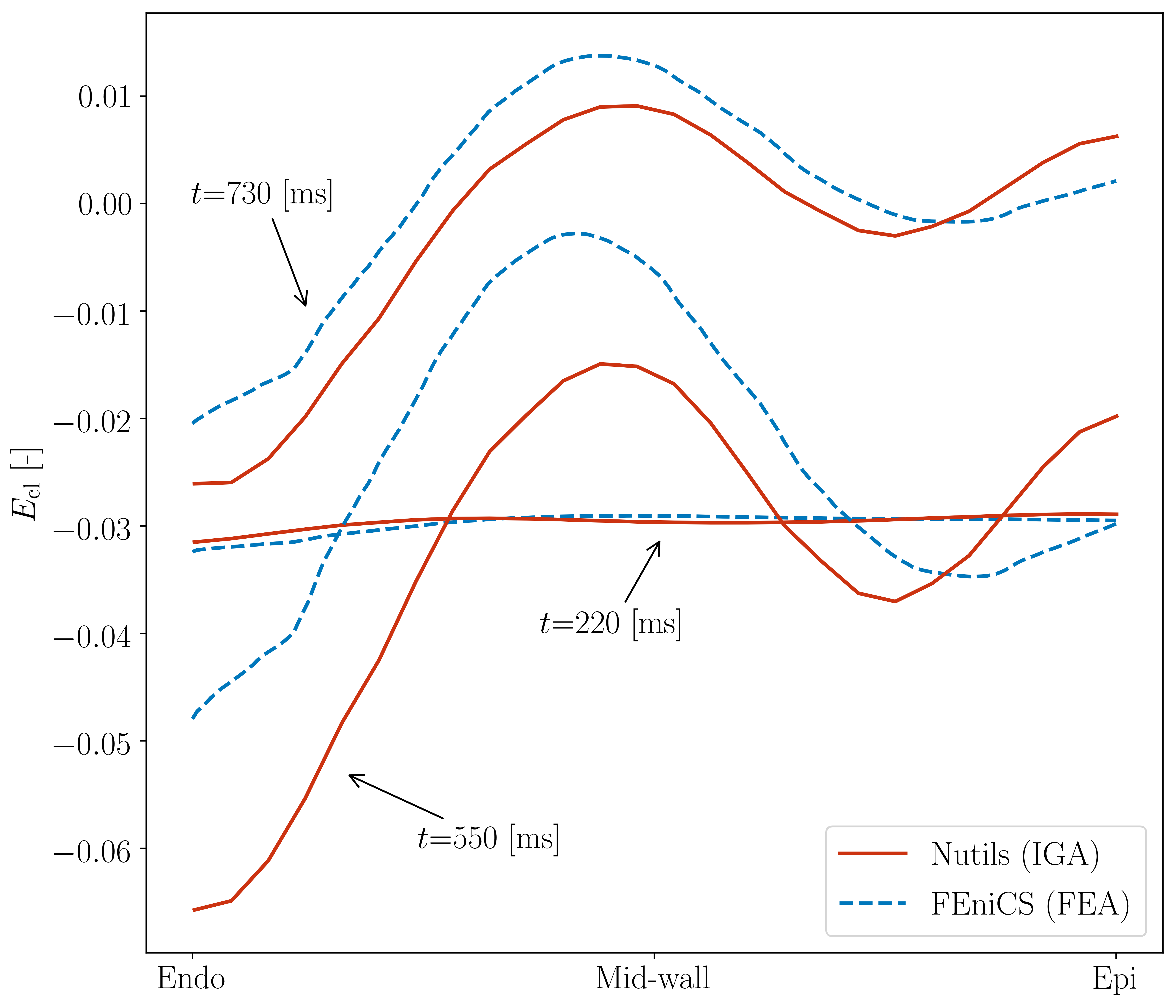}
         \caption{}
         \label{fig:ShearValidationRadial} 
     \end{subfigure}
        \caption{Comparison between the FEniCS finite element analysis (FEA) and the Nutils isogeometric analysis (IGA) result. (a) Contour plots of the Green-Lagrange strain longitudinal-circumferential shear component $E_{\mathrm{cl}}$. The shear component is visualized on the left ventricle half in the reference configuration, $\Omega_0$, where the equator plane is denoted by $\Gamma^{\mathrm{eq}}_0$ and located at $z=0$. The Green-Lagrange strain shear component is defined with respect to the reference configuration. The rows indicate the spatial discretization for the displacement field and the columns the time-instances. (b) The Green-Lagrange strain shear component is spatially averaged over the equatorial plane $\overline{E}_{\mathrm{cl}}$ and monitored in time. The vertical dashed lines indicate the different cardiac phases. Note that the jagged pattern in the lower-left figure of (a) is pronounced by the selected narrow range of the color bar.}
        \label{fig:ShearValidation}
\end{figure*}

The hemodynamics of the left ventricle are visualized in Figure~\ref{fig:HemodynamicsValidation}, which shows that the coarse IGA model is capable of matching the FEniCS model almost exactly. This is similar to the convergence comparison, Figure~\ref{fig:ComparisonLVHemo}, which shows that the hemodynamics are primarily dependent on the cavity volume, which is a geometrical quantity and thus remains exact for the IGA case. The FEniCS result, on the other hand, is influenced by mesh refinement, due to the linear tetrahedrons that approximate the geometry, also shown in \ref{app:testcase} for the thick-walled sphere.

Besides the hemodynamic results, we analyze the mechanical results again for the fiber Cauchy stress and Green-Lagrange strain, both visualized in Figure~\ref{fig:StressValidation} and \ref{fig:StrainValidation}. The results for both the Cauchy stress and Green-Lagrange strain show a good agreement when averaged over the equatorial plane and monitored in time. The minor deviations observed are expected when comparing identical models, both discretized and implemented in different software. The contour plots yield similar results, albeit that subtle differences are noticed. The FEA results tend to create a sharp and often irregular behavior, most noticeable near the boundaries, for the stress and strain quantities. This is inherent to the inter-element $C^0$-continuity in combination with the unstructured mesh~\cite{Barbarotta2021}. Furthermore, the IGA result shows multiple 'rings' or value oscillations on the base boundary at $t$=550 and $t$=730~[ms], while the FEA only shows one. The same is observed on the equatorial plane, albeit less pronounced. With the limitations of the considered software frameworks, it remains inconclusive whether either the IGA or FEA results correctly capture these oscillations.

The shear component is discussed separately since it does show a noticeable difference, similar to the convergence behavior in Figure~\ref{fig:ShearConvergence}. Upon investigating the radial distribution in Figure~\ref{fig:ShearValidationRadial}, both the Nutils and FEniCS results exhibit a similar oscillating behavior. Nonetheless, the magnitude deviates, which is most likely attributed to the not fully converged results and the difference in element type, \emph{i.e.}, hexahedron (Nutils) and tetrahedron (FEniCS). As shown in Figure~\ref{fig:ShearConvergence} of the previous section, the IGA model is unable to achieve fully converged shear components without considering an impractical number of elements. This slow shear convergence is also observed for the FEA model, suggesting that significant through-the-thickness variations occur for this particular quantity of interest. Furthermore, tetrahedron elements typically underperform with respect to hexahedron elements in terms of convergence rate, as indicated in \ref{app:testcase}, and shear loads \cite{TetvsHex2022,TetvsHex2011}. Whether the observed discrepancy in Figure~\ref{fig:ShearValidation} is attributed to both the unconverged result and element type remains inconclusive. However, regarding the purpose of the proposed IGA model, we are primarily interested in the general mechanical behavior. In view of this, the observed deviations in specific mechanical quantities are considered not to be restrictive.\label{sec:LVanalysis}
\subsection{Bi-ventricle model}
We now consider the application of the benchmarked IGA model to a bi-ventricle geometry, utilizing the workflow of Figure~\ref{fig:workflow}. More specifically, this section first leverages the fast geometry creation procedure for various anatomical (geometrical) variations, without the need for meshing. Three variations are considered with changing left ventricle (LV) and right ventricle (RV) shapes while maintaining constant cavity volumes. This section is concluded by a numerical analysis of the anatomical variations in terms of the hemodynamics and mechanical stress-strain behavior. The hemodynamic parameters, as illustrated in Figure~\ref{fig:lumpedscheme}, were calibrated and fixed for all simulations such that the model produced a hemodynamically physiological result. It should, however, be emphasized that the focus of the presented simulations is on demonstrating the efficient IGA workflow as depicted in Figure~\ref{fig:workflow}, rather than on the physiological interpretation of the simulation results.

\subsubsection*{Geometrical variations}
\begin{figure*}[!tp]
\centering
\includegraphics[width=.85\textwidth]{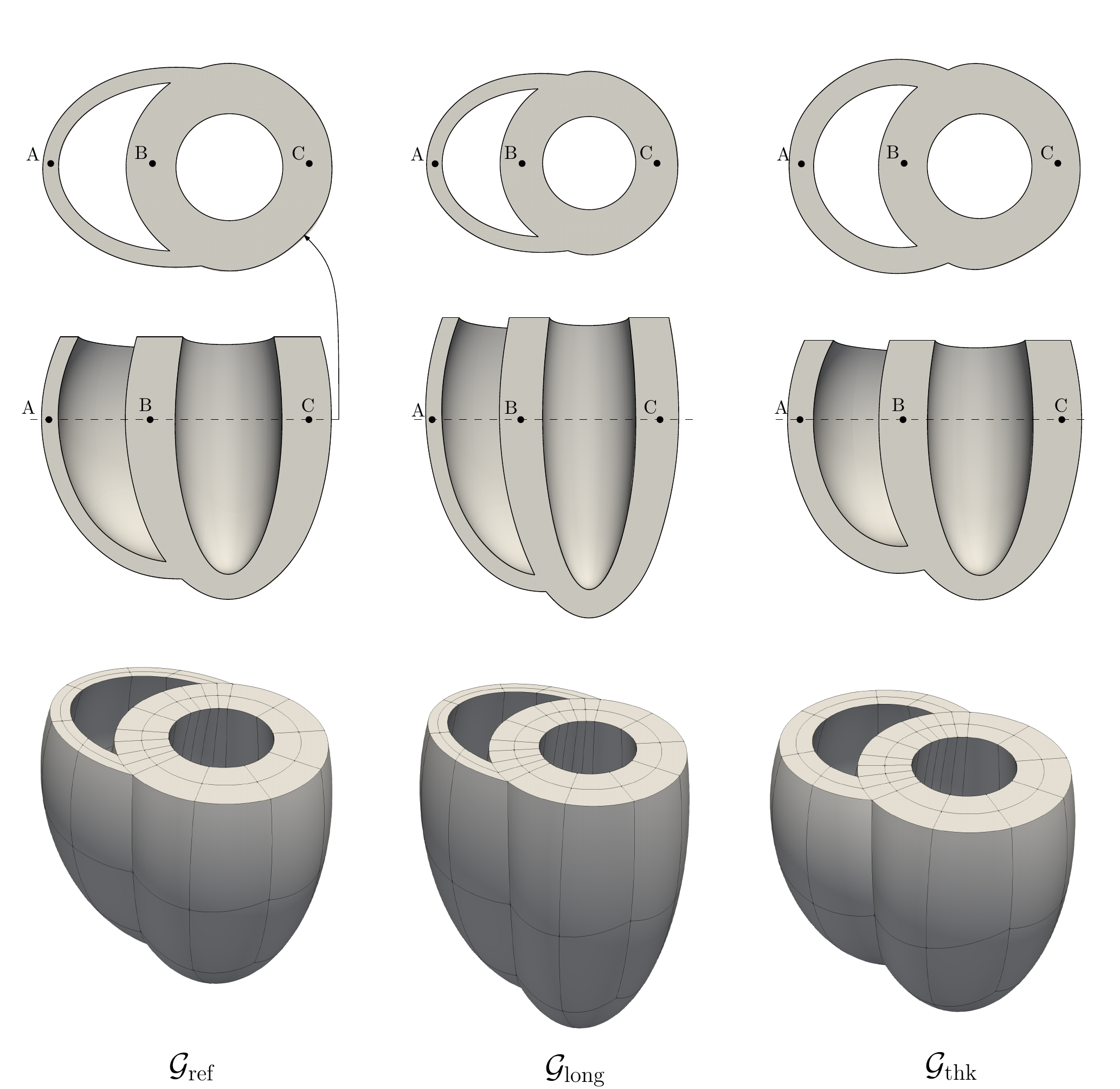}
  \caption{\label{fig:geomvariations} Visualization of the reference bi-ventricle geometry, $\mathcal{G}_{\mathrm{ref}}$, and two variations of it, $\mathcal{G}_{\mathrm{long}}$ and $\mathcal{G}_{\mathrm{thk}}$. The variations consist of an elongated or vertically stretched reference variant, $\mathcal{G}_{\mathrm{long}}$, and a thickened right ventricle wall variant, $\mathcal{G}_{\mathrm{thk}}$. Three mid-wall points are defined on the equatorial plane ($z$=0) in the right ventricle, A, the septum, B, and the left ventricle mid-wall C. These points are used to evaluate mechanical quantities of interest.} 
\end{figure*}

The reference bi-ventricle geometry, denoted by $\mathcal{G}_{\mathrm{ref}}$, is an extension of the previously discussed left ventricle geometry, as it employs the same dimensions with the addition of a right ventricle. The dimensions of the stress-free right ventricle, \emph{i.e.}, wall thickness, attachment location, etc., are based on Ref.~\cite{Pluijmert2017}. The resulting reference bi-ventricle geometry is depicted in Figure~\ref{fig:geomvariations}.

We consider two geometrical variations of the reference bi-ventricle geometry, $\mathcal{G}_{\mathrm{ref}}$, as shown in Figure~\ref{fig:geomvariations}. The first variation, $\mathcal{G}_{\mathrm{long}}$, is a vertical elongation of the ventricles, which is a typical bi-ventricle shape mode observed in the population \cite{Mauger2019}. The second variation, $\mathcal{G}_{\mathrm{thk}}$, has a thickened right ventricle wall with slightly adjusted radii, which could for example represent right ventricle hypertrophy \cite{Lorell2000}. All three geometries have an identical left and right cavity volume of $44$~[ml], which is similar to the left ventricle in Section~\ref{sec:LVanalysis}. The wall volume of the reference and elongated variation, $\mathcal{G}_{\mathrm{ref}}$ and $\mathcal{G}_{\mathrm{long}}$ is similar, while the variation, $\mathcal{G}_{\mathrm{thk}}$, has a slightly larger wall volume due to the thickened right ventricle wall.

The spatial discretization is set identical for all geometries and consists of 120 elements, which is chosen coarse on purpose as it results in fast computations with sufficient accuracy on account of the considered spline discretization. The displacement field employs cubic B-spline basis functions and quadratic B-spline basis functions for the contractile length. This results in a nonlinear system containing 4,237 degrees of freedom, which is iteratively solved using Newton-Raphson iterations and Crank-Nicolson time-integration ($\theta$=0.5 in Equation~\eqref{eq:thetamethod}) with a fixed time step of $2$~[ms]. The cardiac cycle duration is set to $75$ beats-per-minute (or $800$~[ms]), which requires a minimum of $7$ cardiac cycles before a cyclic steady-state behavior is obtained (stroke volume of the left and right ventricle deviate $\leq$0.2\%). The results provided in the next section pertain to the $8$th cycle.

\subsubsection*{Analysis}
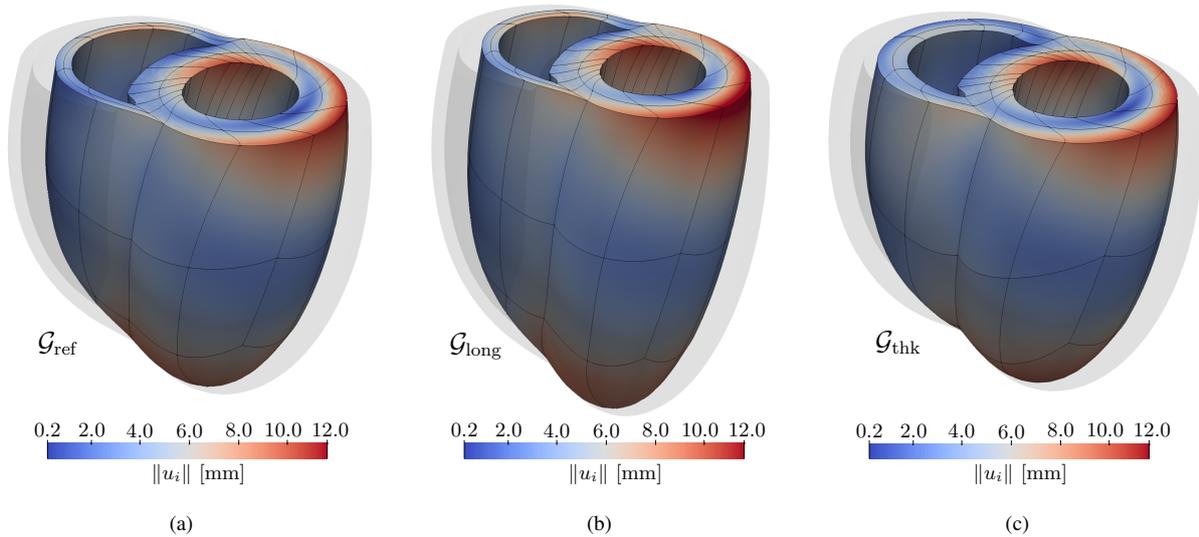
\begin{figure*}[ht]
\begin{tikzpicture}
    \node[anchor=south west,inner sep=0] at (0,0) {\includegraphics[width=\textwidth]{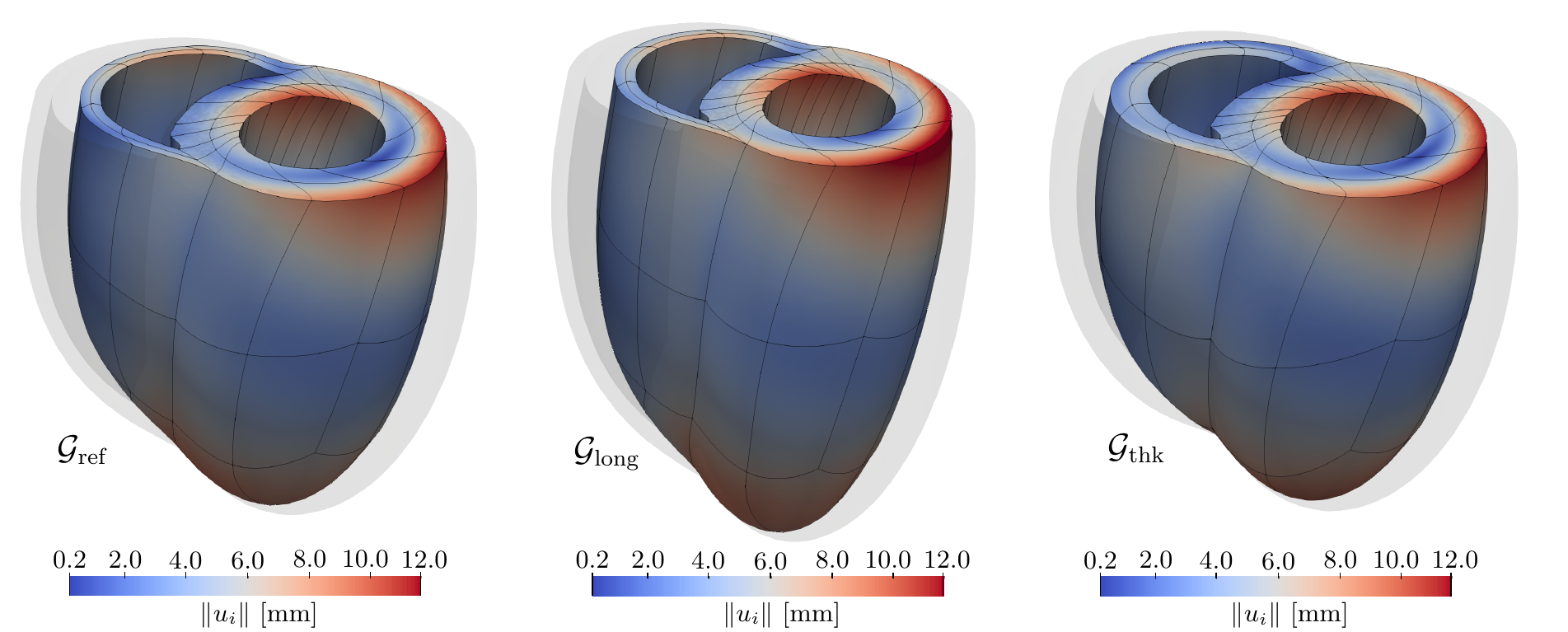}};
    \node at (2.5, -0.5) {\footnotesize  (a)};
    \node at (8 , -0.5) {\footnotesize  (b)};
    \node at (13.5 , -0.5) {\footnotesize  (c)};
\end{tikzpicture}
  \caption{\label{fig:BVdisplacementMAG} Displacement solutions of the three anatomical variations with the spatial discretization (mesh) visualized. The colored part represents the deformed bi-ventricle at end-systole, while the opaque grey represents the bi-ventricle at end-diastole.}    
\end{figure*}

The numerical solutions of the anatomical variations are analyzed in terms of the hemodynamics and mechanical stress and strain behavior. Before we discuss these quantities, we commence with a study of the spatial bi-ventricle displacement. The displacement field is visualized in Figure~\ref{fig:BVdisplacementMAG} for the end-systolic phase (after contraction), while the end-diastolic phase (before contraction) is shown in opaque grey. The results show the complex interaction of torsion in the LV and its influence on the RV. As the LV twists, it starts pulling (tension) at the RV at the front attachment area (from the observer perspective used in Figure~\ref{fig:BVdisplacementMAG}), while it starts pushing (compression) at the other side. This mechanism has an effect on the shear component inside the septum (the wall that divides the two ventricles) relative to the free-wall (the wall opposite to the RV) and would not be captured by the single left ventricle simulation.

\begin{figure*}[!tp]
\centering
\includegraphics[width=.5\textwidth]{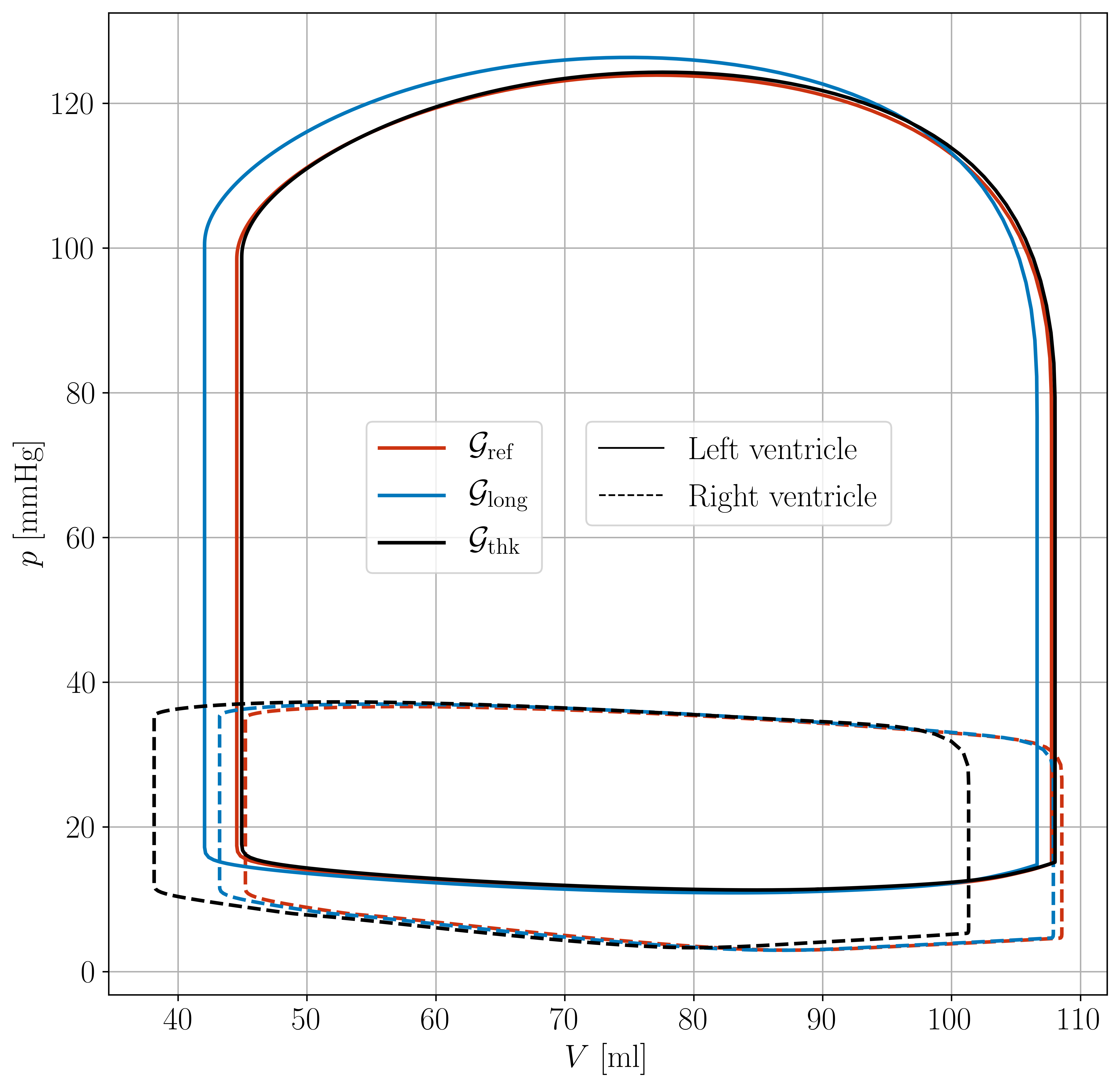}
  \caption{\label{fig:BVhemodynamics} The hemodynamic results (cavity pressure-volume) of the different anatomical variations $\mathcal{G}$.} 
\end{figure*}

The hemodynamic pressure-volume results are visualized in Figure~\ref{fig:BVhemodynamics} and a corresponding set of cardiac function indicators are listed in Table~\ref{tab:cardiacoutput}. Based on both the figure and table, it is observed that both the LVs of $\mathcal{G}_{\mathrm{ref}}$ and $\mathcal{G}_{\mathrm{thk}}$ exhibit similar behavior, which is expected since both configurations share identical LV dimensions. The RV cardiac function is, however, different for both cases, which is a result of the thickened RV wall of $\mathcal{G}_{\mathrm{thk}}$. The thickened wall results in a less compliant (or stiffer) ventricle, which prevents the right cavity to obtain an end-diastolic volume similar to $\mathcal{G}_{\mathrm{ref}}$. It is also observed that the elongated case, $\mathcal{G}_{\mathrm{long}}$, shows an improved cardiac function in terms of ejection fraction (EF) and total LV work, which increased by $4.5$\% (see Table~\ref{tab:cardiacoutput}). This observation is remarkable since the elongated case shares a similar wall volume as the reference case, $\mathcal{G}_{\mathrm{ref}}$. This could, however, be attributed to a more favorable geometrical shape for the considered fiber distribution in the elongated case. Nonetheless, we consider the further interpretation of the observed physiological differences beyond the scope of this manuscript, restricting our presentation to illustrating how the proposed IGA simulation workflow enables the efficient study of anatomical variations.

\begin{table}[ht]
\caption{Overview of several cardiac function indicators for the different anatomical variations $\mathcal{G}$: The total work $W$, the ejection fraction EF, and the stroke volume $V_{\mathrm{stroke}}$.}\label{tab:cardiacoutput}
\centering
\begin{tabular}{lccccc}
\cline{2-6}
                              & \multicolumn{2}{c}{$W$ {[}J{]}} & \multicolumn{2}{c}{EF {[}\%{]}} & $V_{\mathrm{stroke}}$ {[}ml{]} \\ \cline{2-6} 
                              & left           & right          & left           & right          &                                \\ \hline
$\mathcal{G}_{\mathrm{ref}}$  & 0.88           & 0.25           & 58.65          & 58.34          & 63.27                          \\
$\mathcal{G}_{\mathrm{long}}$ & 0.92           & 0.26           & 60.56          & 59.94          & 64.63                          \\
$\mathcal{G}_{\mathrm{thk}}$  & 0.88           & 0.25           & 58.40          & 62.35          & 63.13                          \\ \hline
\end{tabular}
\end{table}

\begin{figure*}[!t]
     \centering
     \begin{subfigure}[b]{0.3\textwidth}
         \centering
             \includegraphics[width=\textwidth]{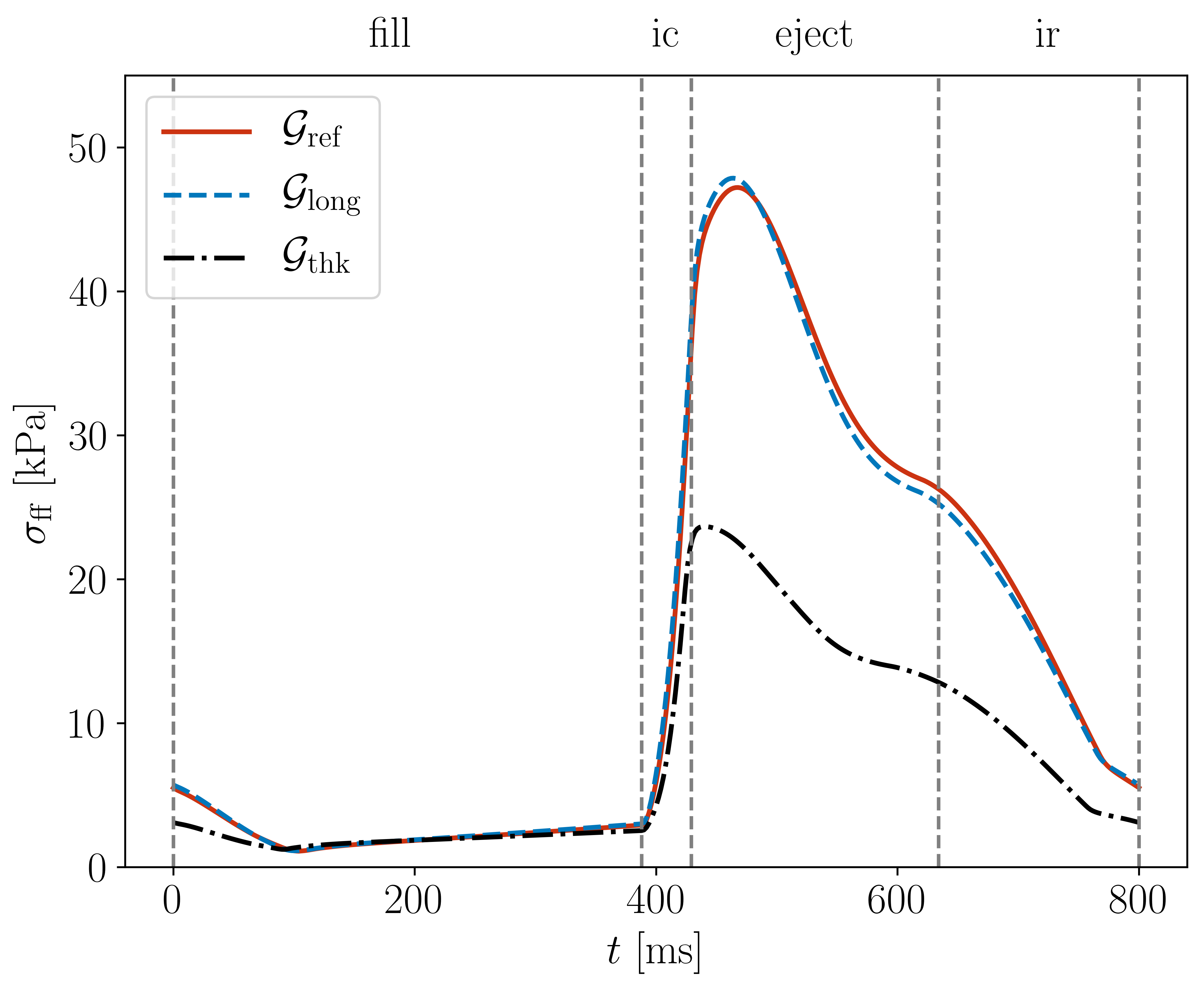}
         \caption{}
         \label{fig:Stressff_right} 
     \end{subfigure}
     \hfill
     \begin{subfigure}[b]{0.3\textwidth}
         \centering
             \includegraphics[width=\textwidth]{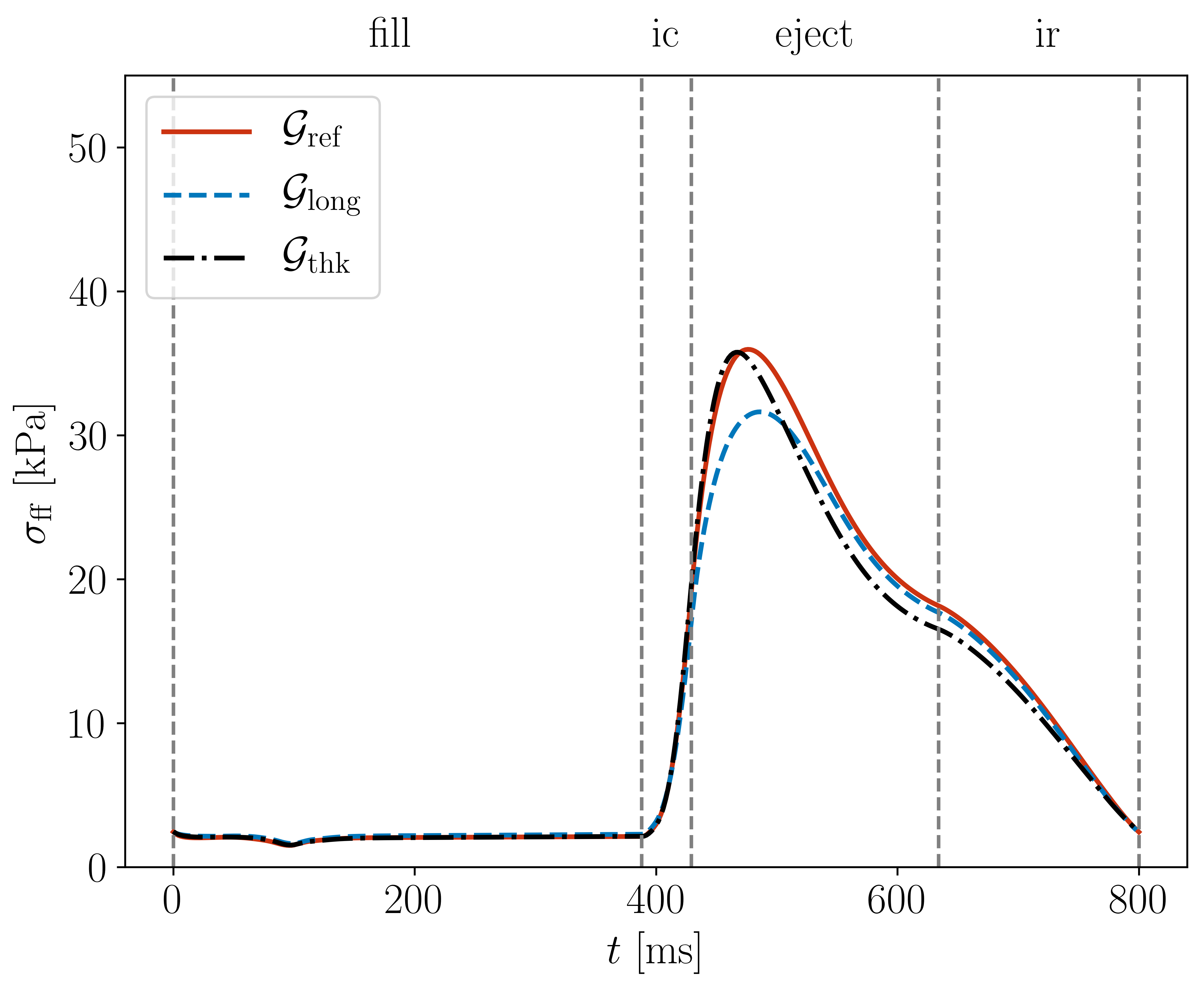}
         \caption{}
         \label{fig:Stressff_sept} 
     \end{subfigure}
     \hfill     
     \begin{subfigure}[b]{0.3\textwidth}
         \centering
             \includegraphics[width=\textwidth]{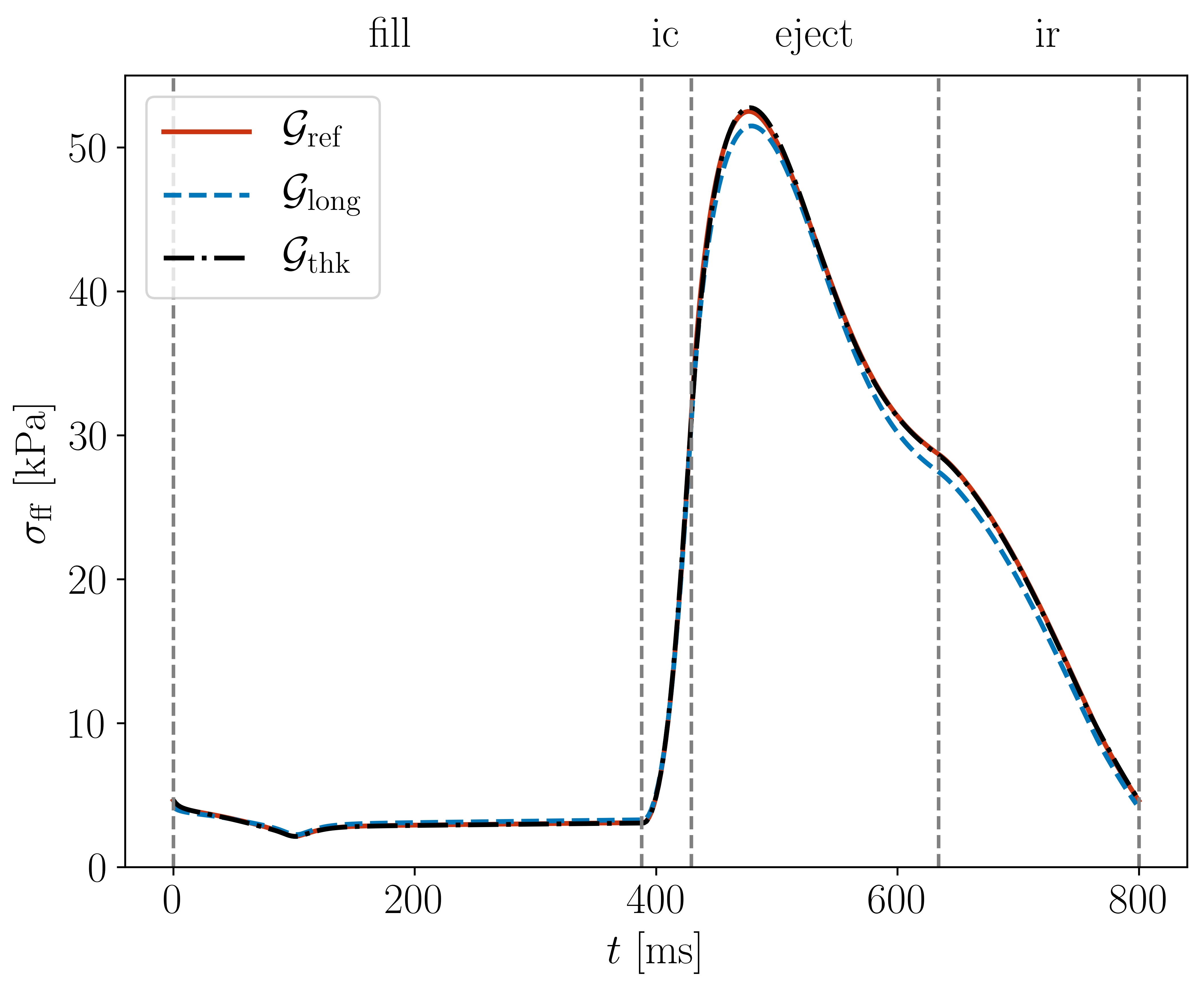}
         \caption{}
         \label{fig:Stressff_left} 
     \end{subfigure}
        \caption{The Cauchy fiber stress results of the anatomical variations $\mathcal{G}$ at three spatial mid-wall locations: (a) the right ventricle (point A in Figure~\ref{fig:geomvariations}), (b) the septum (point B in Figure~\ref{fig:geomvariations}), (c) the left ventricle free-wall (point C in Figure~\ref{fig:geomvariations}).}
        \label{fig:BVstressff}
\end{figure*}

\begin{figure*}[!t]
     \centering
     \begin{subfigure}[b]{0.3\textwidth}
         \centering
             \includegraphics[width=\textwidth]{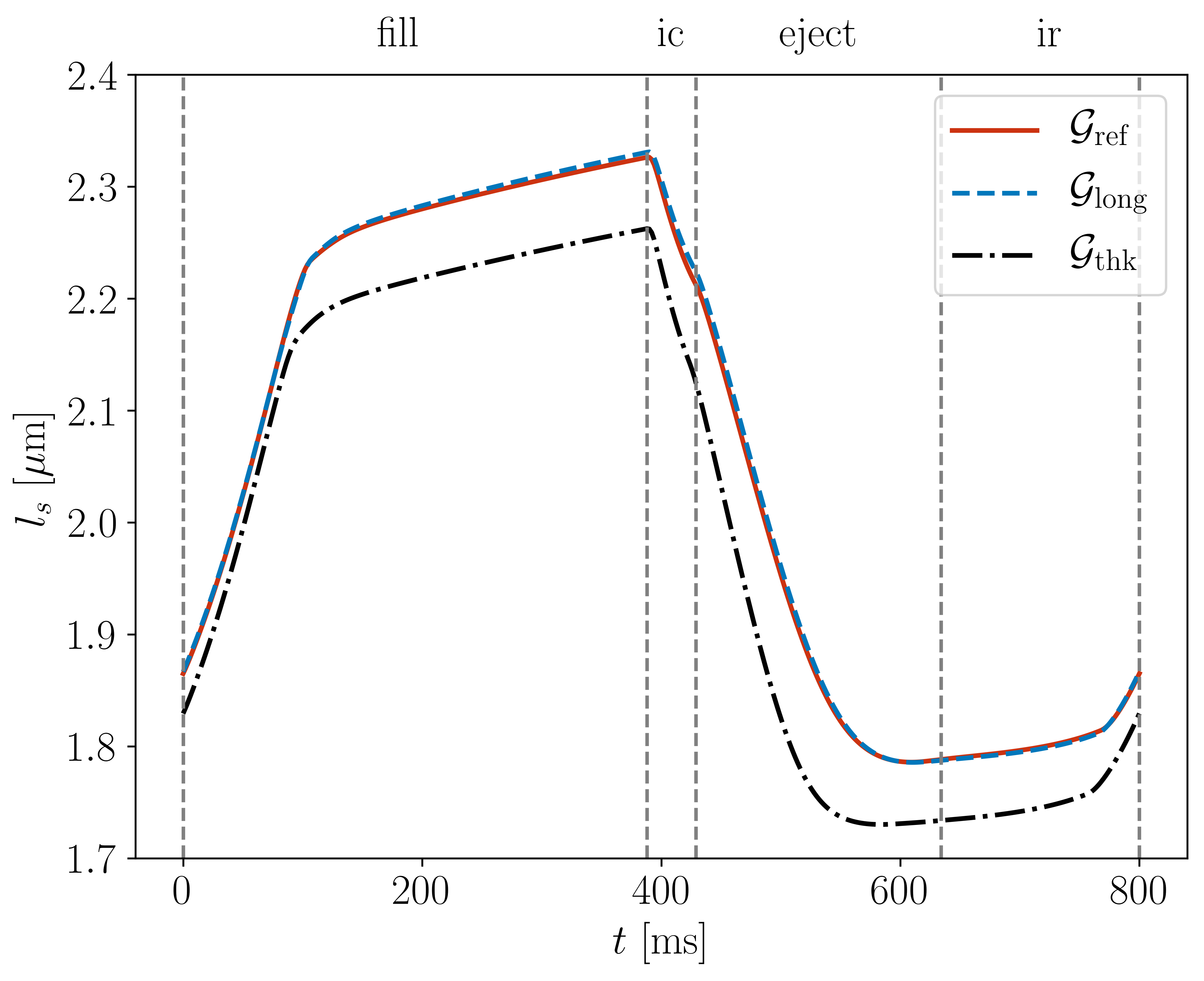}
         \caption{}
         \label{fig:sarcomereL_right} 
     \end{subfigure}
     \hfill
     \begin{subfigure}[b]{0.3\textwidth}
         \centering
             \includegraphics[width=\textwidth]{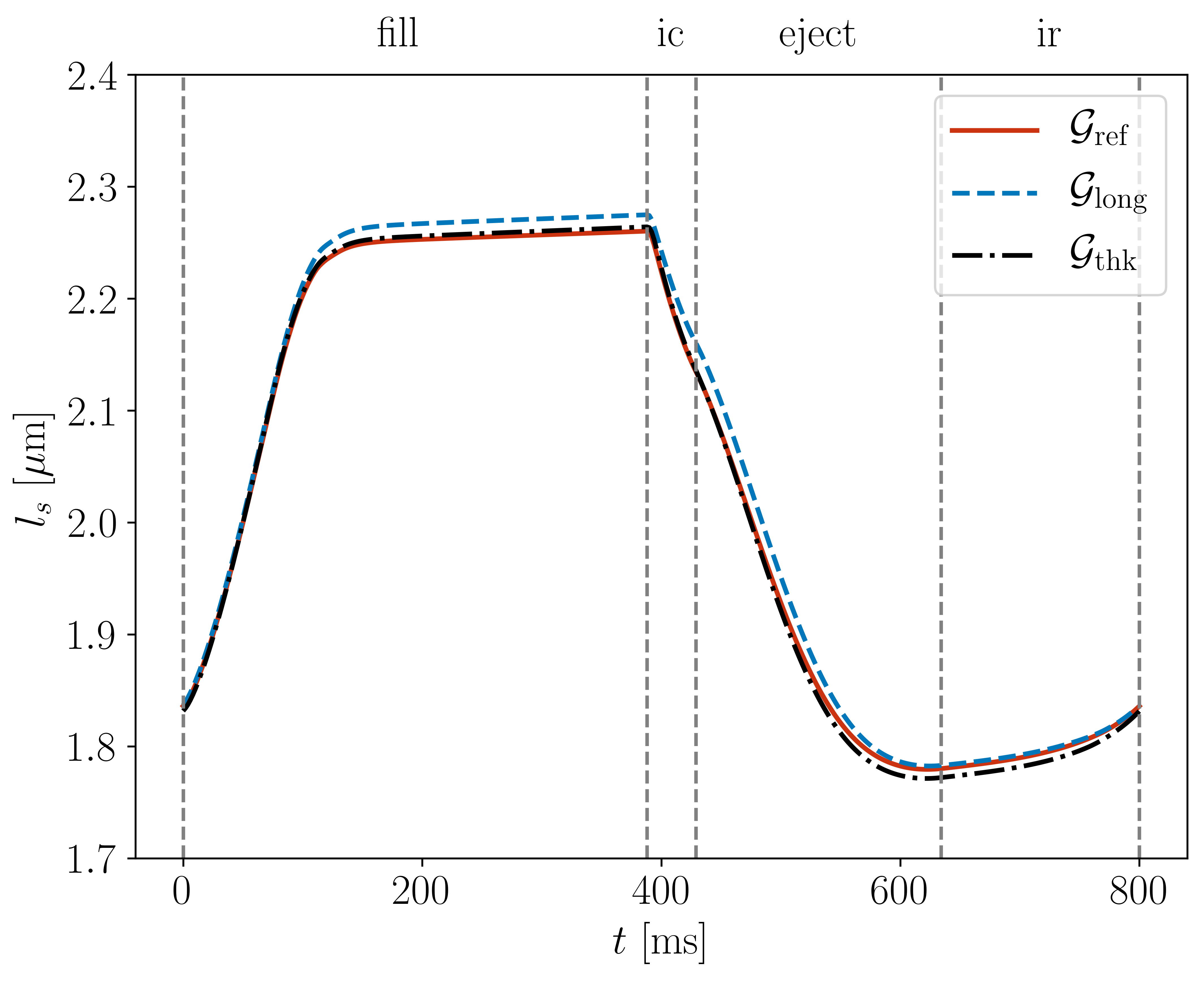}
         \caption{}
         \label{fig:sarcomereL_sept} 
     \end{subfigure}
     \hfill     
     \begin{subfigure}[b]{0.3\textwidth}
         \centering
             \includegraphics[width=\textwidth]{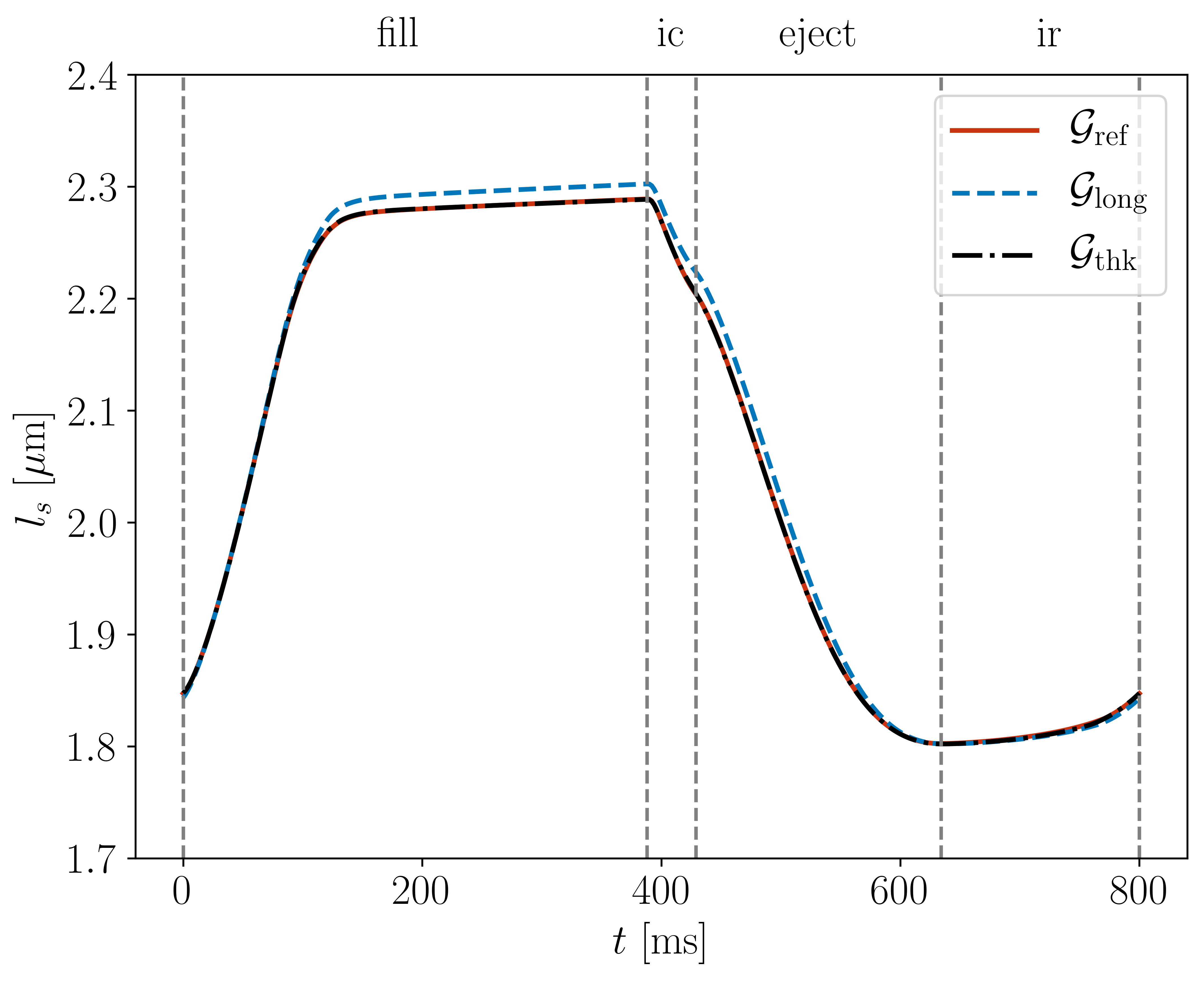}
         \caption{}
         \label{fig:sarcomereL_left} 
     \end{subfigure}
        \caption{The sarcomere length results of the anatomical variations $\mathcal{G}$ at three spatial mid-wall locations: (a) the right ventricle (point A in Figure~\ref{fig:geomvariations}), (b) the septum (point B in Figure~\ref{fig:geomvariations}), (c) the left ventricle free-wall (point C in Figure~\ref{fig:geomvariations}).}
        \label{fig:BVsarcomereL}
\end{figure*}

The mechanical fiber stress and sarcomere length results are monitored in time at $3$ mid-wall positions: the RV wall A, the septum B, and the LV free-wall C, indicated in Figure~\ref{fig:geomvariations}. The fiber Cauchy stress is visualized in Figure~\ref{fig:BVstressff}, in which near identical results are observed between the different cases in the LV free-wall (C) and septum (B). However, a distinct difference is seen in the right ventricle mid-wall (A), Figure~\ref{fig:Stressff_right}, as the thickened RV wall experiences lower fiber stresses when compared to the other cases. This decrease may be affected by the Frank-Starling effect, which is an intrinsic dependence of myocyte stress development on the sarcomere length, which is incorporated into our model. To illustrate this effect, the sarcomere length is provided in Figure~\ref{fig:BVsarcomereL}, which clearly shows the reduction in sarcomere (fiber) elongation in the RV wall for the $\mathcal{G}_{\mathrm{thk}}$ case. As a result, the produced active stress is lower in comparison to the other cases. Nonetheless, the resulting peak cavity pressure remains similar to the other cases due to the thicker wall, Figure\ref{fig:BVhemodynamics}.

\begin{figure*}[!t]
     \centering
     \begin{subfigure}[b]{0.3\textwidth}
         \centering
             \includegraphics[width=\textwidth]{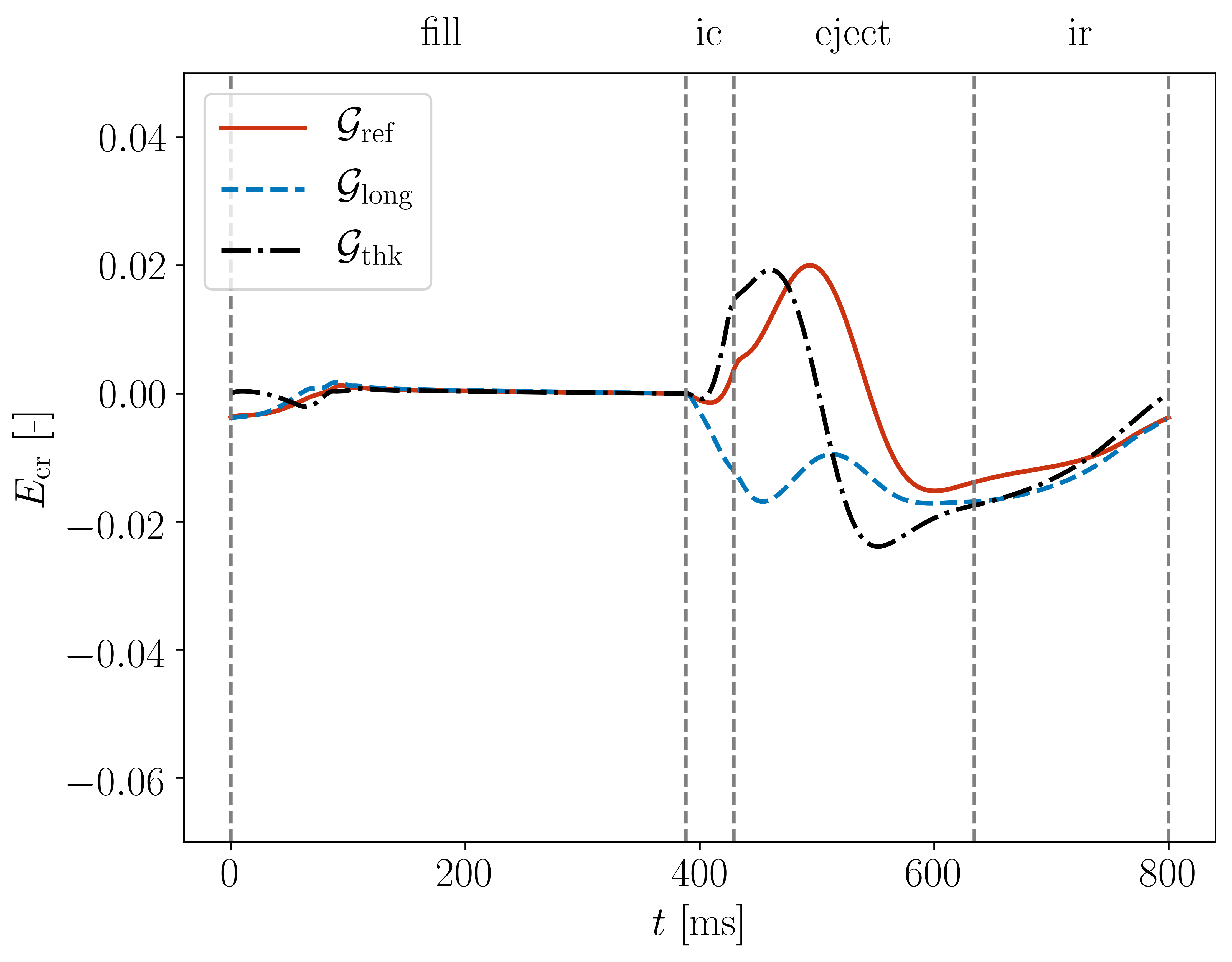}
         \caption{}
         \label{fig:Shearcr_right} 
     \end{subfigure}
     \hfill
     \begin{subfigure}[b]{0.3\textwidth}
         \centering
             \includegraphics[width=\textwidth]{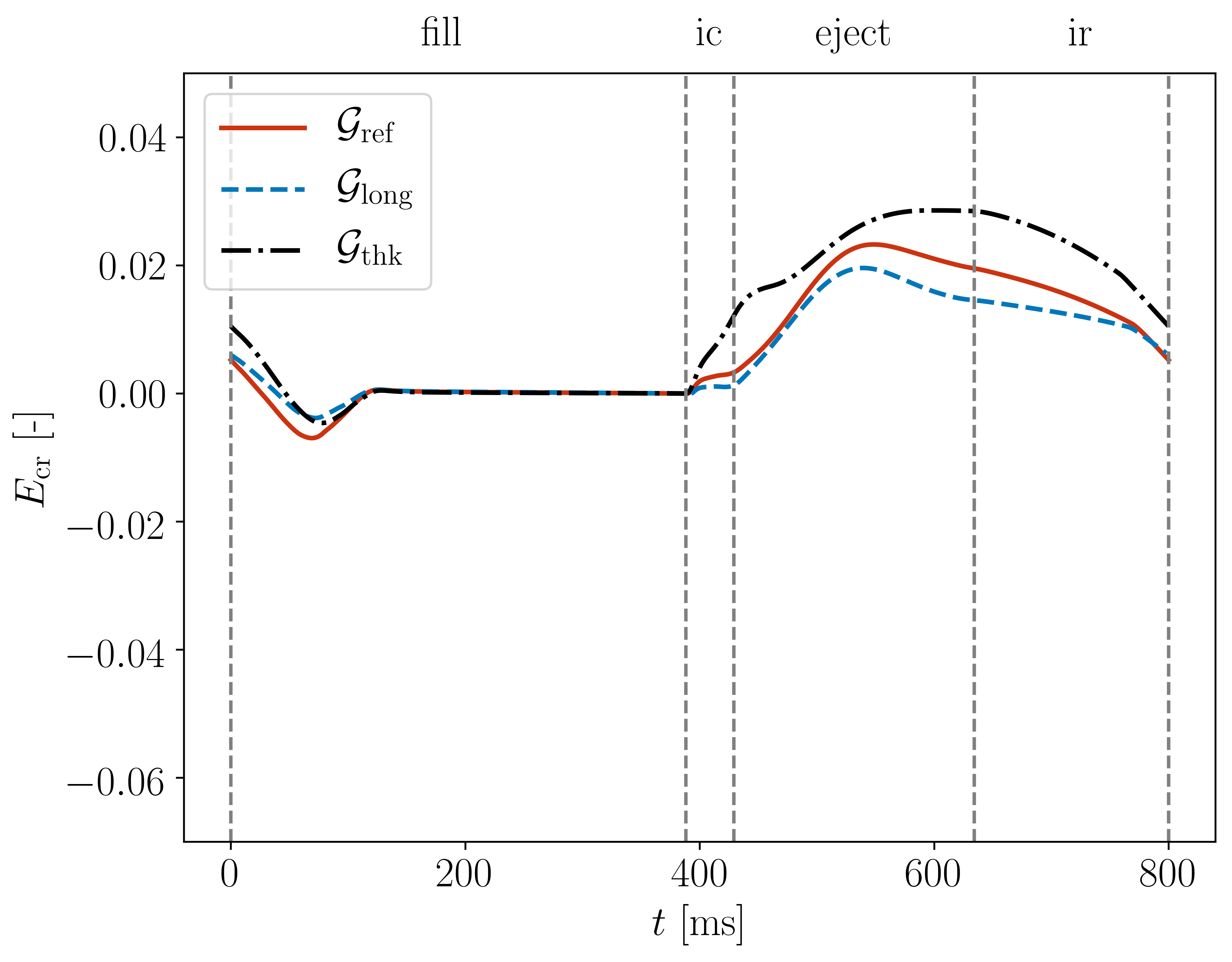}
         \caption{}
         \label{fig:Shearcr_sept} 
     \end{subfigure}
     \hfill     
     \begin{subfigure}[b]{0.3\textwidth}
         \centering
             \includegraphics[width=\textwidth]{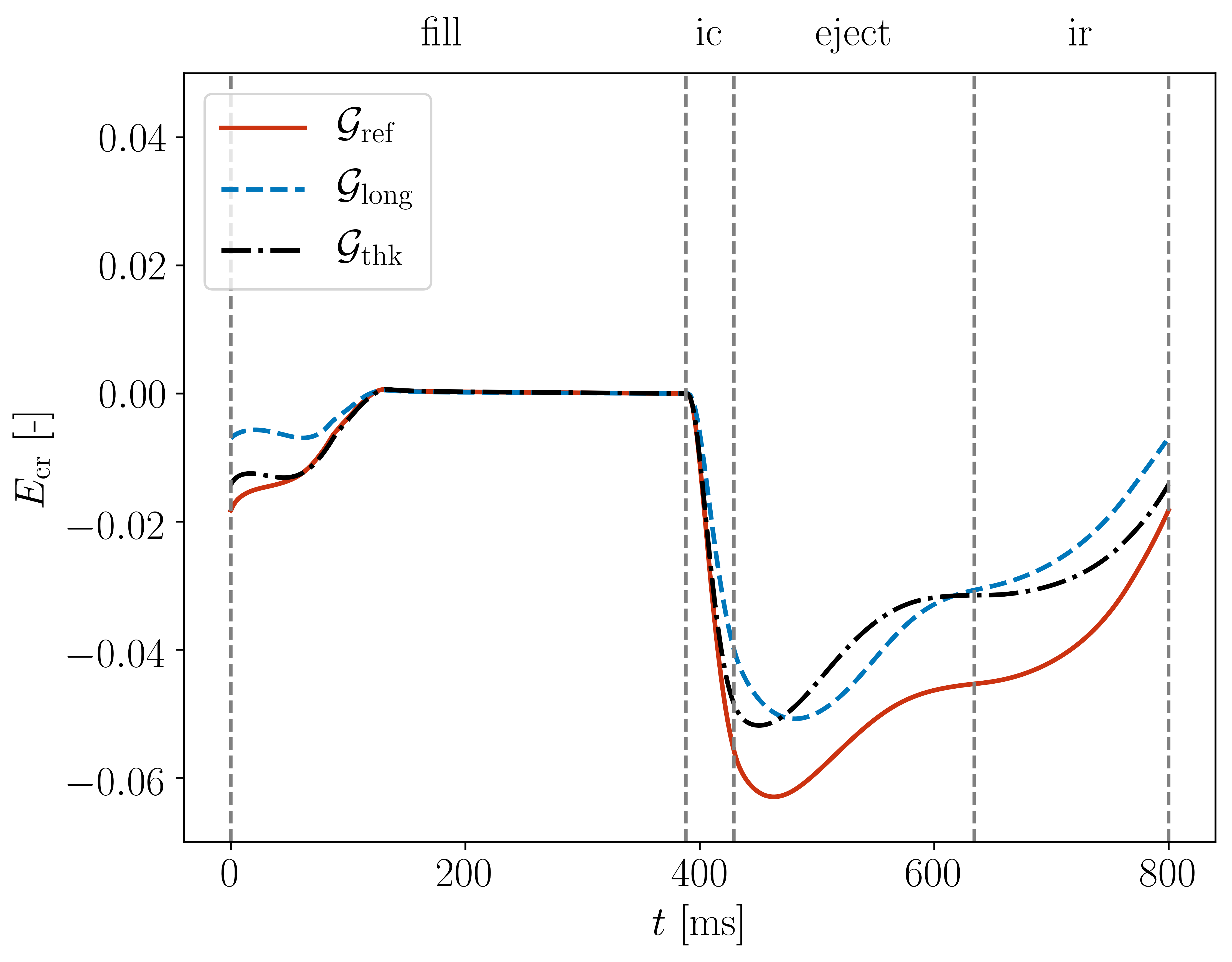}
         \caption{}
         \label{fig:Shearcr_left} 
     \end{subfigure}
        \caption{The Green-Lagrange shear strain results of the anatomical variations $\mathcal{G}$ at three spatial mid-wall locations: (a) the right ventricle (point A in Figure~\ref{fig:geomvariations}), (b) the septum (point B in Figure~\ref{fig:geomvariations}), (c) the left ventricle free-wall (point C in Figure~\ref{fig:geomvariations}).}
        \label{fig:BVshearcr}
\end{figure*}

Although the fiber stress and sarcomere length exhibit a similar pattern in the three considered locations, albeit, with different magnitudes, a noticeable difference is observed for the shear behavior. More specifically, the circumferential-radial Green-Lagrange strain component, $E_{\mathrm{cr}}$, visualized in Figure~\ref{fig:BVshearcr}, shows a different behavior when comparing the septum and the LV free-wall, Figure~\ref{fig:Shearcr_sept} and \ref{fig:Shearcr_left}, respectively. Such a difference between the septum and the LV free-wall would not be observed for any mechanical quantity if the single idealized LV was analyzed due to the axisymmetric nature of the single LV geometry. This difference clearly indicates the effect of the RV on the LV wall shear. However, whether the addition of the RV is of added value is dependent on the research question the model should answer.\label{sec:BVanalysis}

\section{Conclusions}\label{sec:ConcRec}
We have proposed a simulation workflow based on the isogeometric analysis paradigm, suitable for the cardiac analysis of single- and bi-ventricle heart models under the consideration of sparse anatomical data. In the isogeometric approach, the mathematical-physical model -- in this case the cardiac model -- is discretized directly using the spline-based parametrization of the geometry. This avoids the introduction of a large number of elements to capture the geometry, as can be the case in traditional finite element analyses. This efficient capturing of the geometry makes the approach very suitable for studying anatomical variations, thereby working toward a patient-specific simulation workflow. In the successful application of the isogeometric approach, it is essential to attain a robust analysis-suitable parametrization of the geometry, making the proposed geometry-construction procedure a key contribution.

To ensure the suitability of the spline geometry for isogeometric analysis it is important to limit the distortion of the spline patches. For the considered anatomies, especially the bi-ventricle model, it is a necessity to consider a multi-patch spline parametrization. To construct the geometry model using a limited number of anatomical parameters, a construction procedure is developed that systematically traverses through the dimensions of the problem (\emph{i.e.}, first patch vertices are defined, then patch curves, then patch surfaces, and finally patch volumes). In this procedure, use is made of a combination of standard CAD construction techniques, in particular surface lofting, and tailored geometry construction algorithms. The key contribution of this work in relation to the geometry construction is the usage of constrained optimization problems to define patch edges. This geometry construction component is essential in ensuring well-defined connectivity between the patches, which is in particular far from trivial in the bi-ventricle scenario.

We have implemented a state-of-the-art biomechanical cardiac model, encompassing a rule-based fiber orientation model, an anisotropic hyperelastic material model, a model taking into account the active stress contributions, and a 0-dimensional Windkessel model for the circulatory system. The Galerkin framework on which isogeometric analysis is built allows for a direct translation of the finite element discretization of the cardiac model to the proposed isogeometric framework. It is noted that the considered model lacks a detailed description of cardiac electrophysiology but assumes simultaneous mechanical activation of the myocytes. The suitability of isogeometric analysis for incorporating more detailed electrophysiology models has recently been demonstrated \cite{IGAquarteroni,PEGOLOTTI201952,PATELLI2017248}. Our results also suggest that other model enhancements can be beneficial for the stability of the simulations. Most notably this pertains to the diode model used for the valves, where additional (temporal) smoothness is expected to improve robustness. These model improvements are of a generic nature, and not specific to the employed isogeometric approach.

The presented simulation results convey that the isogeometric framework reproduces the finite element benchmark with a high degree of accuracy at a significantly reduced number of degrees of freedom. When considering the hemodynamics, very coarse isogeometric discretizations provide solutions that are virtually indistinguishable from the benchmark. An observation that is particularly noteworthy in this regard, is that the preservation of the geometry parametrization under mesh refinement in the case of isogeometric analysis removes the geometry-related inaccuracies in the hemodynamical response observed using standard finite element meshes. When considering mechanical quantities of interest, in particular stresses and strains, it is observed that the isogeometric discretizations overall match well with the benchmark. For the shear stresses, a substantial mismatch between the isogeometric model and the finite element benchmark is observed, which is attributed to the significant variation of these stresses through the thickness of the wall. Although the quantification of this particular inaccuracy remains inconclusive, it conveys that coarse isogeometric discretization of relatively low degrees (quadratic or cubic) inherently lack the resolution to represent high-frequency solution variations. Adaptive (local) mesh refinement strategies developed for IGA (\emph{e.g.}, \cite{DIVI2022}) may have the potential to enhance the proposed modeling framework with capabilities of representing such variations, without dramatically increasing the computation time (as would be the case when considering uniform mesh refinements).

The presented anatomy-variation study serves the purpose of demonstrating the robustness of the developed isogeometric framework concerning geometry changes. The 15 anatomical parameters underlying the spline parametrization of the bi-ventricle geometry already enable the tailoring of the model to patient-specific data to a large extent. This parametrization of the anatomy can be extended, for example by taking into account the radius of curvature and thickness of the septum. In future work, we aim at studying the capabilities of the model with respect to patient-specific data, where we will focus particularly on the data-scan scenario in which the spline template interpolates the anatomy in regions where data is absent.

\subsection*{Acknowledgements}
This publication is part of the COMBAT-VT project (project no. 17983) of the research program High Tech Systems and Materials which is partly financed by the Dutch Research Council (NWO). Additionally, this work was performed within the IMPULS framework under the Picasso project (reference no. TKI HTSM/20.0022) of the Eindhoven MedTech Innovation Center (e/MTIC, incorporating Eindhoven University of Technology, Philips Research, and Catharina Hospital), including a PPS-supplement from the Dutch Ministry of Economic Affairs and Climate Policy. The research of the second author is sponsored by the European Union's Horizon 2020 research and innovation program under grant agreement 874827 (BRAVE).

\newpage
\appendix
\section{Fiber field model}\label{app:fiberfield}
The fiber field is obtained by the method of Rossi \emph{et al.}~\cite{ThermoQuarteroni,IntegratedHeartQuart}, which consists of several steps, visualized in Figure~\ref{fig:RRBM}. Step~(I) in Figure~\ref{fig:RRBM} assigns the boundary names, which are required for solving the two Laplace boundary-value (LBV) problems in Step~(II). Note that in this section, we make use of the index notation, which corresponds to the Cartesian coordinate system, $i \in \{x,y,z\}$. The LBV problems then state:
\begin{equation}\label{eq:laplace1}
    \begin{cases}\text { Find } \phi \text { such that: } & \\ 
    \Delta \phi=0 & \text { in } {\Omega}_0, \\ 
    \phi=0 & \text { at } {\Gamma}^{\rm{lv,endo}}_{0},  {\Gamma}^{\rm{rv,endo}}_{0} \\ 
    \phi=1 & \text { at } {\Gamma}^{\rm{lv,epi}}_{0},{\Gamma}^{\rm{rv,epi}}_{0}, {\Gamma}^{\rm{rv,endo-sep}}_{0},  \\ 
    n_{0_i} \left( \nabla_{0_i} \phi \right)=0 & \text { at } {\Gamma}^{\rm{base}}_0,\end{cases}
\end{equation}
and
\begin{equation}\label{eq:laplace2}
    \begin{cases}\text { Find } \gamma \text { such that: } & \\ 
    \Delta \gamma=0 & \text { in } {\Omega}_0, \\ 
    \gamma=1 & \text { at }  {\Gamma}_{0}^{\text{lv,endo}} \\ 
    \gamma=-1 & \text { at } {\Gamma}_{0}^{\mathrm{rv,endo}}, \\ 
    n_{0_i} \left( \nabla_{0_i} \gamma \right)=0 & \text { at } {\Gamma}^{\mathrm{base}}_0, {\Gamma}^{\mathrm{epi}}_0.\end{cases}
\end{equation}
In computing the numerical solution to the LBVs, the strong imposition of the Dirichlet boundary condition in Equation~\eqref{eq:laplace1} should be handled with care. More specifically, the Dirichlet conditions conflict at the interface $\Gamma^{\rm{rv,endo}}_0 \cap {\Gamma}^{\rm{rv,endo-sep}}_{0}$. Because of this, a strong imposition of the Dirichlet condition causes oscillatory behavior in the solution. It is, therefore, recommended to impose them weakly, for example using Nitsche's method \cite{Nitsche1971,BAZILEVSNitsche2007}.

The solution of Equation~\eqref{eq:laplace1} is used to derive the local basis $\{\mathbf{e}^{\rm{c0}},\mathbf{e}^{\rm{t0}},\mathbf{e}^{l0}\}$ as follows,
\begin{equation}
 \left\{\begin{aligned} 
       \mathrm{e}^{\rm{t0}}_i&=\frac{\nabla_i \phi}{\left\| \nabla_j \phi \right\|}, \\ 
       \mathrm{e}^{l\rm{0}}_i&=\frac{z_{0_i} - \left( z_{0_j} \ \mathrm{e}^{\rm{t0}}_j \right) \mathrm{e}^{\rm{t0}}_i }{\left\| z_{0_i} - \left( z_{0_j} \ \mathrm{e}^{\rm{t0}}_j \right) \mathrm{e}^{\rm{t0}}_i \right\|},\\
       \mathrm{e}^{\rm{c0}}_i&=\mathrm{e}^{l\rm{0}}_i \times \mathrm{e}^{\rm{t0}}_i,\\
 \end{aligned}\right.
\end{equation}
in which $z_{0_i}$ is the unit reference vector in the longitudinal direction, chosen to be equal to the normal direction at the base, $\Gamma^{\mathrm{base}}_0$, but not restricted to the base boundary, visualized in Figure~\ref{fig:RRBM}. The fiber direction is then obtained by rotating the circumferential vector, $\mathbf{e}^{\rm{c0}}$, along the transmural vector, $\mathbf{e}^{\rm{t0}}$, given a rotation matrix $\mathcal{R}(\cdot)$:
\begin{equation}\label{eq:fiberRot}
    \mathrm{e}^{\mathrm{f0}}_i = \mathcal{R}^{\mathrm{e}^{\mathrm{t0}}}_{ij}(\alpha)  \mathrm{e}^{\mathrm{c0}}_j.
\end{equation}
The rotation matrix is dependent on the helix angle $\alpha$, and is defined according to Rodriguez formula~\cite{IntegratedHeartQuart},
\begin{equation}
    \mathcal{R}^{\mathrm{e}^{\mathrm{t0}}}_{ij}(\alpha)=\delta_{ij}+\sin (\alpha) \left[\mathrm{e}^{\mathrm{t0}}_{\times}\right]_{ij}+2 \sin ^{2}(\theta / 2)\left[\mathrm{e}^{\mathrm{t0}}_i \mathrm{e}^{\mathrm{t0}}_j -\delta_{ij}\right]
\end{equation}
where $\left[\mathrm{e}^{\mathrm{t0}}_{\times}\right]_{ij}$ is the cross-product matrix defined as
\begin{equation}
    \left[\mathrm{e}^{\mathrm{t0}}_{\times}\right]_{ij}=\left(\begin{array}{ccc}
0 & -\mathrm{e}^{\mathrm{t0}}_3 & \mathrm{e}^{\mathrm{t0}}_2 \\
\mathrm{e}^{\mathrm{t0}}_3 & 0 & -\mathrm{e}^{\mathrm{t0}}_1 \\
-\mathrm{e}^{\mathrm{t0}}_2 & \mathrm{e}^{\mathrm{t0}}_1 & 0
\end{array}\right).
\end{equation}
Note that the transmural vector $\mathrm{e}^{\mathrm{t0}}_i$ has three components or indices, defined in the Cartesian coordinate system $i = \{1,2,3\} = \{x,y,z\}$. The helix angle $\alpha$ is obtained by interpolating known histological angles at the endo- and epicardium of the left and right ventricles, using the solution of Equation~\eqref{eq:laplace1} and~\eqref{eq:laplace2} such that
\begin{equation}
    \alpha = \frac{1}{2} \left( \alpha^{\rm{lv}}\left( \phi \right) \left( \gamma + 1 \right) - \alpha^{\rm{rv}}\left( \phi \right) \left( \gamma - 1 \right)\right),
\end{equation}
where
\begin{equation}
    \alpha^{i}\left( \phi \right) = \alpha^{i}_{\rm{endo}} \left( 1 - \phi \right) + \alpha^{i}_{\rm{epi}} \phi \qquad i = \rm{lv}, \ \rm{rv}. 
\end{equation}
Note that the sheet $\mathbf{e}^{\rm{s0}}$ and sheet-normal $\mathbf{e}^{\rm{n0}}$ directions can be obtained in a similar way as Equation~\eqref{eq:fiberRot}, but are not required for the considered cardiac model.
\section{Weak formulation and temporal discretization}\label{app:discretization}
In this appendix we derive the weak formulation of the coupled 3D-0D problem, Equation~\eqref{eq:setofeqs}, using Galerkin's method for the equilibrium equation and the sarcomere dynamics model. We conclude with the temporal discretization of the sarcomere dynamics and circulation models. Once again, we make use of the index notation, which corresponds to the Cartesian coordinate system, $i \in \{x,y,z\}$.

\subsection{Equilibrium equation}
Given the strong formulation of the coupled 3D-0D problem, Equation~\eqref{eq:setofeqs}, we first derive the weak formulation of the equilibrium equation, which solves for the displacement field $u_i$. An appropriate setting for the displacement field is the Sobolev space, $\boldsymbol{H}^1\left(  \Omega \right)$, in which the displacement field satisfies the Dirichlet boundary condition at the basal plane, $\Gamma^{\rm{base}}$. The corresponding trial and test spaces are given by, $\boldsymbol{\mathcal{U}}  = \{ u_i \ | \ u_i \in \boldsymbol{H}^1\left( \Omega \right) , \ u_i n_i = 0 \ \text{on} \ \Gamma^{\rm{base}} \}$ and $\boldsymbol{\mathcal{W}}  = \{ w_i \ | \ w_i \in \boldsymbol{H}^1\left( \Omega \right) , \ w_i n_i = 0 \ \text{on} \ \Gamma^{\rm{base}} \}$, respectively. Multiplying the equilibrium equation with a test function, $w_i \subset \boldsymbol{\mathcal{W}}$, and integration by parts over the current domain, $\Omega$, yields the following weak formulation
\begin{equation}
      \int \frac{\partial w_j}{\partial x_i}  \sigma^{\phantom{}}_{ij} \ \text{d}{\Omega} = \int w_j n_i \sigma^{\phantom{}}_{ij} \ \text{d}{\Gamma} \qquad \forall w_j \in \boldsymbol{\mathcal{W}}.  
\end{equation}
Application of the Neumann and Robin boundary conditions to the derived weak formulation gives,
\begin{equation}
\begin{aligned}
\int \frac{\partial w_j}{\partial x_i}  \sigma^{\phantom{}}_{ij} \ \text{d}_{\Omega}
= &- \int w_j p^{\rm{lv}}\left(t\right) n_j \ \text{d}{\Gamma^{\rm{lv,endo}}} \\
&- \int w_j p^{\rm{rv}}\left(t\right) n_j \ \text{d}{\Gamma^{\rm{rv,endo}}} \\
&- \int w_j k^{\rm{peri}}u_j \ \text{d}{\Gamma^{\rm{epi}}}  \qquad \forall w_j \in \boldsymbol{\mathcal{W}}.
\end{aligned}
\end{equation}
By mapping the derivatives, integrals, and normals to the reference domain, and rewriting the equation as a residual, we obtain
\begin{equation}
\begin{aligned}
    \mathscr{R}^{u} &=  \int \frac{\partial w_j}{\partial x_{0_k}} F^{-1}_{ki} \sigma^{\phantom{}}_{ij} J \ \text{d}{\Omega_0} \\
    &- \int w_j p^{\rm{lv}}\left(t\right) n_{0_k} F^{-1}_{kj} J \ \text{d}{\Gamma^{\rm{lv,endo}}_0} \\ & - \int w_j p^{\rm{rv}}\left(t\right) n_{0_k} F^{-1}_{kj} J \ \text{d}{\Gamma^{\rm{rv,endo}}_0}  \\ &- \int w_j k^{\rm{peri}} u_j J \ \text{d}{\Gamma^{\rm{epi}}_0} = 0 \qquad \forall w_j \in \boldsymbol{\mathcal{W}},
\end{aligned}    
\end{equation}
where the subscript $'0'$, denotes that a quantity is defined in the reference configuration and $J$ is the determinant of the gradient of deformation tensor $F_{ij}$.

The Cauchy stress tensor is still defined in the current domain. By applying the transformation based on the Kirchhoff stress, as defined in Equation~\eqref{eq:secondPstress}, we obtain the following weak formulation for the equilibrium equation
\begin{equation}
\begin{aligned}
   \mathscr{R}^{u} &=  \int \frac{\partial w_j}{\partial x_{0_k}} \left[ S^{\rm{pas}}_{ki} + S^{\rm{act}}_{kj} \right] F^{\phantom{}}_{ki} \ \text{d}{\Omega_0} \\ &- \int w_j p^{\rm{lv}}\left(t\right) n_{0_k} F^{-1}_{kj} J \ \text{d}{\Gamma^{\rm{lv,endo}}_0} \\ & - \int w_j p^{\rm{rv}}\left(t\right) n_{0_k} F^{-1}_{kj} J \ \text{d}{\Gamma^{\rm{rv,endo}}_0}  \\ &- \int w_j k^{\rm{peri}} u_j J \ \text{d}{\Gamma^{\rm{epi}}_0} = 0 \qquad \forall w_j \in \boldsymbol{\mathcal{W}}.
\end{aligned}    
\end{equation}
The passive contribution of the second Piola-Kirchhoff stress, defined according to Equation~\eqref{eq:EnergyDensity}, is written as
\begin{equation*}
    S^{\rm{pas}}_{ij} = \frac{\partial \psi^{S}}{\partial E_{ij}} + \frac{\partial \psi^{V}}{\partial E_{ij}},
\end{equation*}
where
\begin{equation*}
   \frac{\partial \psi^{S}}{\partial E_{ij}} = C \left( 2 a_1 I_1 \frac{\partial I_1}{\partial E_{ij}}- a_2 \frac{\partial I_2}{\partial E_{ij}} + 2 a_3 I_4 \frac{\partial I_4}{\partial E_{ij}} \right) \ \text{exp}\left(Q \right),
\end{equation*}
with
\begin{subequations}
    \begin{align}
       \frac{\partial I_1}{\partial E_{ij}} &= \delta_{ij}, \\
       \frac{\partial I_2}{\partial E_{ij}} & = I_1 \delta_{ij} - E_{ji},\\
       \frac{\partial I_4}{\partial E_{ij}} &= \mathrm{e}^{\mathrm{f0}}_{i} \mathrm{e}^{\mathrm{f0}}_{j}.
\end{align}
\end{subequations}
The volumetric contribution is given by,
\begin{equation}
   \frac{\partial \psi^{V}}{\partial E_{ij}} = 4 \kappa J^2 C^{-1}_{ji} \left( J^2 - 1 \right)  \quad \text{with} \quad C_{ij} = F_{ki} F_{kj}.
\end{equation}
The active second Piola-Kirchhoff stress tensor, $S^{\rm{act}}_{ij}$, is already defined in Equation~\eqref{eq:Sact}.

\subsection{Sarcomere dynamics}
Given the evolution equation of the contractile sarcomere length, $l^{\rm{c}}$, in Equation~\eqref{eq:lengths}, we require this field to be square integrable, \emph{i.e.}, $l^{\rm{c}} \in {L}^2\left( \Omega \right)$. However, we note that the additional smoothness obtained from higher-order functions is desirable when interpolating this field. The corresponding trial and test functions are then defined as, ${\mathcal{L}}  = \{ l^{\rm{c}} \ | \ l^{\rm{c}} \in {L}^2\left( \Omega \right)\}$ and $\mathcal{Q}  = \{ q_i \ | \ q_i \in L^2\left( \Omega \right) \}$, respectively. The weak formulation defined on the current domain, $\Omega$, is then given by,
\begin{equation}
\mathscr{R}^{l^{\rm{c}}} = \int q,  \frac{d l^{\mathrm{c}}}{dt} \ \text{d}{\Omega_0} - \int q, \left[ E^{\mathrm{a}} \left( l^{\mathrm{s}} - l^{\mathrm{c}} \right) - 1\right] v^{\mathrm{0}} \ \text{d}{\Omega_0} = 0 \qquad \forall q \in {\mathcal{Q}}, 
\end{equation}
and is identical when defined on the reference configuration. Use is made of the $\theta$-method, Equation \eqref{eq:thetamethod}, for the temporal discretization, such that
\begin{equation}\label{eq:resLcapp}
\begin{aligned}
   \mathscr{R}^{l^{\rm{c}}} &= \int q \left( l^{\mathrm{c}}_{n+1} - l^{\mathrm{c}}_n \right) \ \text{d}{\Omega_0} \\&- \left( 1 - \theta^{l^{\mathrm{c}}} \right) \ \Delta t \int q \ f_{n} \left( u_{i}, l^{\mathrm{c}} \right) \ \text{d}{\Omega_0}\\
    & - \theta^{l^{\mathrm{c}}} \ \Delta t \int q \ f_{n+1} \left( u_{i}, l^{\mathrm{c}} \right) \ \text{d}{\Omega_0}  \qquad \forall q \in {\mathcal{Q}},
\end{aligned}
\end{equation}
with,
\begin{equation}
    f_n\left( u_{i}, l^{\mathrm{c}} \right) = \left[ E^{\mathrm{a}} \left( l^{\mathrm{s}}(u_{i_n}) - l^{\mathrm{c}}_n \right) - 1\right] v^{\mathrm{0}},
\end{equation}
where $n$ is the time-increment and $\theta$ is the weight that controls the integration method. For convenience, we decompose the residual according to Equation~\eqref{eq:resdecomp}, which for the contractile length equals
\begin{equation}
    \mathscr{R}^{l^{\rm{c}}}= \mathscr{P}^{l^{\rm{c}}}_{n} + \mathscr{C}^{l^{\rm{c}}}_{n+1}.
\end{equation}
When applied to Equation~\eqref{eq:resLcapp}, we obtain,
\begin{equation}
    \begin{aligned}
    \mathscr{P}^{l^{\rm{c}}}_{n} &= - \int q \left( l^{\rm{c}}_n + \Delta t \ \theta^{l^{\rm{c}}}  f( u_{i_n}, l_n^{\rm{c}})  \right) \ \text{d}{\Omega_0}, \\
    \mathscr{C}^{l^{\rm{c}}}_{n+1} &= \int q \left( l^{\rm{c}}_{n+1} - \Delta t \ \left( 1 - \theta^{l^{\rm{c}}} \right)  f(u_{i_{n+1}}, l_{n+1}^{\rm{c}}) \right) \ \text{d}{\Omega_0}.
    \end{aligned}
\end{equation}

\subsection{Circulatory model}
The circulatory model does not need to be discretized in space, since the pressures, $\tilde{p} = \{p^{\mathrm{lv}}, p^{\mathrm{rv}}, p^{\mathrm{art,P}} , p^{\mathrm{art,S}}, p^{\mathrm{ven,S}}\}$, are defined in 0D, $\tilde{p} \in \mathbb{R}^5$. The temporal discretization of Equation~\eqref{eq:circulat} has already been given in Equation~\eqref{eq:resP} in compact form. Elaboration yields: 

\begin{equation}\label{eq:circdis}
    \begin{aligned}
    \mathscr{R}^{\tilde{p}} = \mathscr{C}_{n+1}^{\tilde{p}}+\mathscr{P}_{n}^{\tilde{p}}= \left\{\begin{aligned} 
          V^{\mathrm{lv}}_{n+1}\left( u_i \right) -  \theta^{\tilde{p}} \Delta t \left(q^{\rm{ven,P}}_{n+1} - q^{\rm{art,S}}_{n+1} \right) \quad &  -V^{\mathrm{lv}}_{n}\left( u_i \right) - \left( 1 - \theta^{\tilde{p}} \right) \Delta t \left(q^{\rm{ven,P}}_{n} - q^{\rm{art,S}}_{n} \right),  \\
          V^{\mathrm{rv}}_{n+1}\left( u_i \right)  -  \theta^{\tilde{p}} \Delta t \left(q^{\rm{ven,S}}_{n+1} - q^{\rm{art,P}}_{n+1} \right)  \quad  &   -V^{\mathrm{rv}}_{n}\left( u_i \right)- \left( 1 - \theta^{\tilde{p}} \right) \Delta t \left(q^{\rm{ven,S}}_{n} - q^{\rm{art,P}}_{n} \right), \\
          C^{\mathrm{art,P}} p^{\mathrm{art,P}}_{n+1}- \theta^{\tilde{p}} \Delta t \left(q^{\rm{art,P}}_{n+1} - q^{\rm{per,P}}_{n+1} \right) \quad  & - C^{\mathrm{art,P}} p^{\mathrm{art,P}}_{n} -  \left( 1 - \theta^{\tilde{p}} \right)  \Delta t  \left(q^{\rm{art,P}}_{n} - q^{\rm{per,P}}_{n} \right),  \\
          C^{\mathrm{art,S}} p^{\mathrm{art,S}}_{n+1} -  \theta^{\tilde{p}} \Delta t \left(q^{\rm{art,S}}_{n+1} - q^{\rm{per,S}}_{n+1} \right) \quad  & - C^{\mathrm{art,S}} p^{\mathrm{art,S}}_{n} -  \left( 1 - \theta^{\tilde{p}} \right)  \Delta t   \left(q^{\rm{art,S}}_{n} - q^{\rm{per,S}}_{n} \right),  \\
          \underbrace{C^{\mathrm{ven,S}} p^{\mathrm{ven,S}}_{n+1} - \theta^{\tilde{p}} \Delta t  \left(q^{\rm{per,S}}_{n+1} - q^{\rm{art,S}}_{n+1} \right) }_{\mathscr{C}^{\tilde{p}}_{n+1}} \quad  & \underbrace{- C^{\mathrm{ven,S}} p^{\mathrm{ven,S}}_{n} -  \left( 1 - \theta^{\tilde{p}} \right)  \Delta t  \left(q^{\rm{per,S}}_{n} - q^{\rm{art,S}}_{n} \right)}_{\mathscr{P}^{\tilde{p}}_{n}},  
         \end{aligned}\right.
\end{aligned}
\end{equation}
The volume flows, $q^i$, are defined in Equation~\eqref{eq:volumeflow} and the residual defined in Equation~\eqref{eq:circdis} has also been decomposed in a current, $\mathscr{C}_{n+1}^{\tilde{p}}$, and previous, $\mathscr{P}_{n}^{\tilde{p}}$, time-increment residual, in accordance with Equation~\eqref{eq:resdecomp}. This decomposition is used in Section~\ref{sec:spatialdisc}. 
\section{Test case: thick-walled sphere}\label{app:testcase}
The complexity of the cardiac model, as proposed in Section~\ref{sec:CardiacModel}, does not allow for verification using an analytic solution that incorporates all model components. As a result, the individual model components are verified separately using a simplified test case. In this section, the geometry construction procedure and the passive material response of the cardiac model are verified using a \textit{thick-walled sphere} test case for which an analytic solution exists. First, the geometry is constructed using the procedure as outlined in Section~\ref{sec:multipatch}, from which potential limitations are discussed. Second, the passive nonlinear material response, as explained in Section~\ref{sec:CardiacModel}, is verified using an analytical solution after inflating the sphere. The section is concluded by a comparison between the isogeometric analysis (IGA) and finite element analysis (FEA) results, both in terms of approximation errors and convergence rates.

\begin{figure*}[!t]
     \centering
     \begin{subfigure}[b]{0.45\textwidth}
         \centering
         \includegraphics[width=\textwidth]{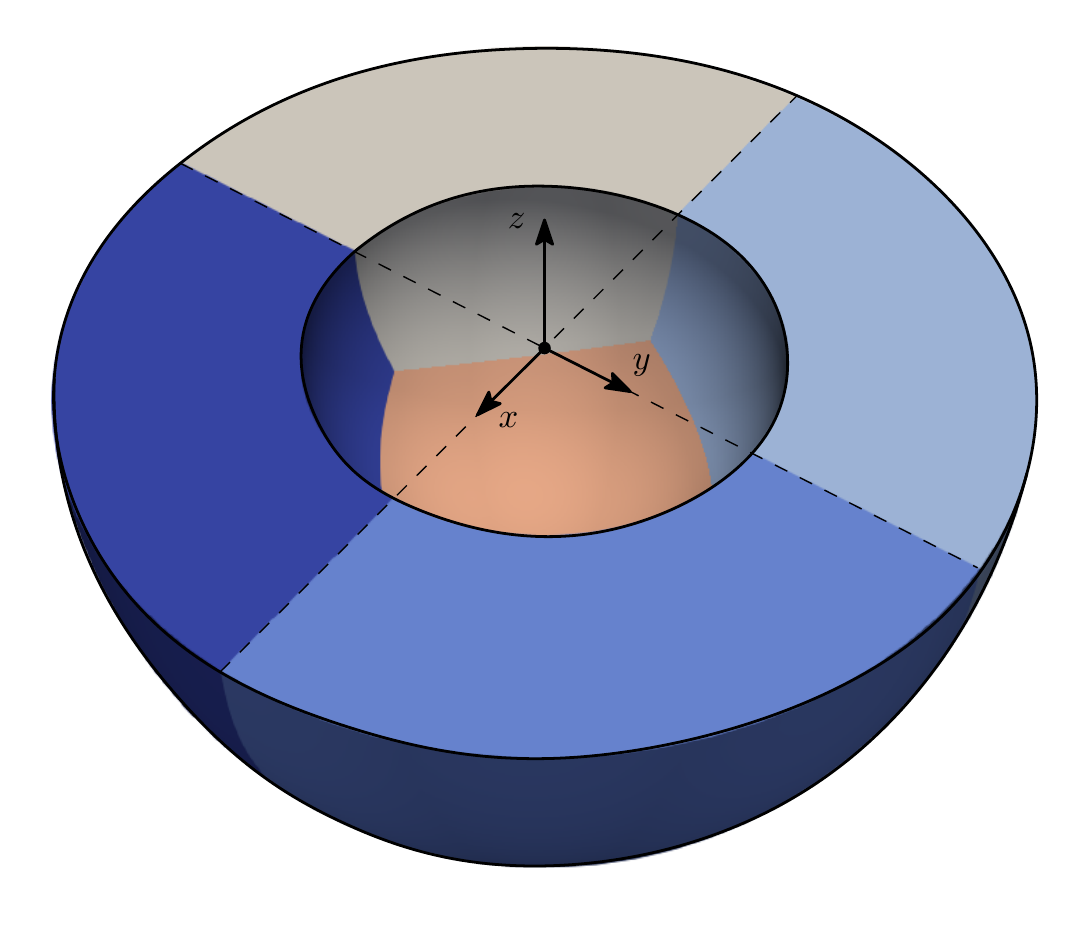}
                  \caption{}
                  \label{fig:spherePatches}
     \end{subfigure}
     \hfill
     \begin{subfigure}[b]{0.4\textwidth}
         \centering
         \includegraphics[width=\textwidth]{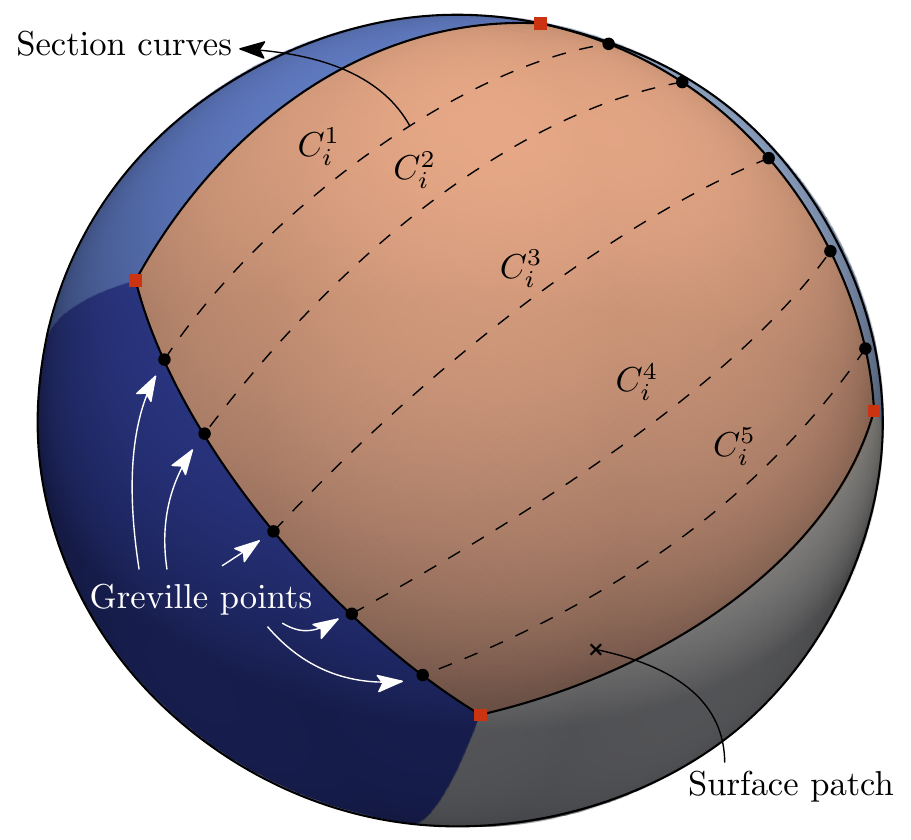}
                  \caption{}
                  \label{fig:sphereSeccurves}
     \end{subfigure}
        \caption{(a) The multi-patch thick-walled (half) sphere, where each color represents a different patch. The geometry is obtained according to the procedure outlined in Section~\ref{sec:multipatch}. (b) A set of 5 section curves are visualized for the middle patch and used to perform the lofting procedure. The boundaries of the section curves are located at the Greville points of the patch boundary. }
        \label{fig:Spheremultipatch}
\end{figure*}

\subsection{Geometry verification}
When constructing an analysis-suitable geometry, it is often simplified before conducting any numerical analyses, \emph{i.e.}, (axi)symmetry, periodicity, \emph{etc}. Such simplifications are governed by the balance equations to be solved and the corresponding boundary conditions applied to them. When considering a sphere, this can be represented by a 2D axisymmetric geometry, a 3D one-eighth periodic geometry, or other variations, as long as the balance equations and boundary conditions allow for it. In this analysis, we want to demonstrate the geometry construction procedure which makes use of multi-patch NURBS to construct the bi-ventricle template, Figure~\ref{fig:Templates}. As a result, only half of the sphere is modeled, which is similar to the left ventricle template (Figure~\ref{fig:LVTemplate}), resulting in a thick-walled multi-patch geometry as shown in Figure~\ref{fig:spherePatches}. It is noted that the proposed construction procedure computes an approximation of the actual geometry. As a result, we conduct this test case to quantify any introduced approximation errors, albeit very small.

\begin{figure*}[!t]
     \centering
     \begin{subfigure}[b]{0.3\textwidth}
         \centering
         \includegraphics[width=\textwidth]{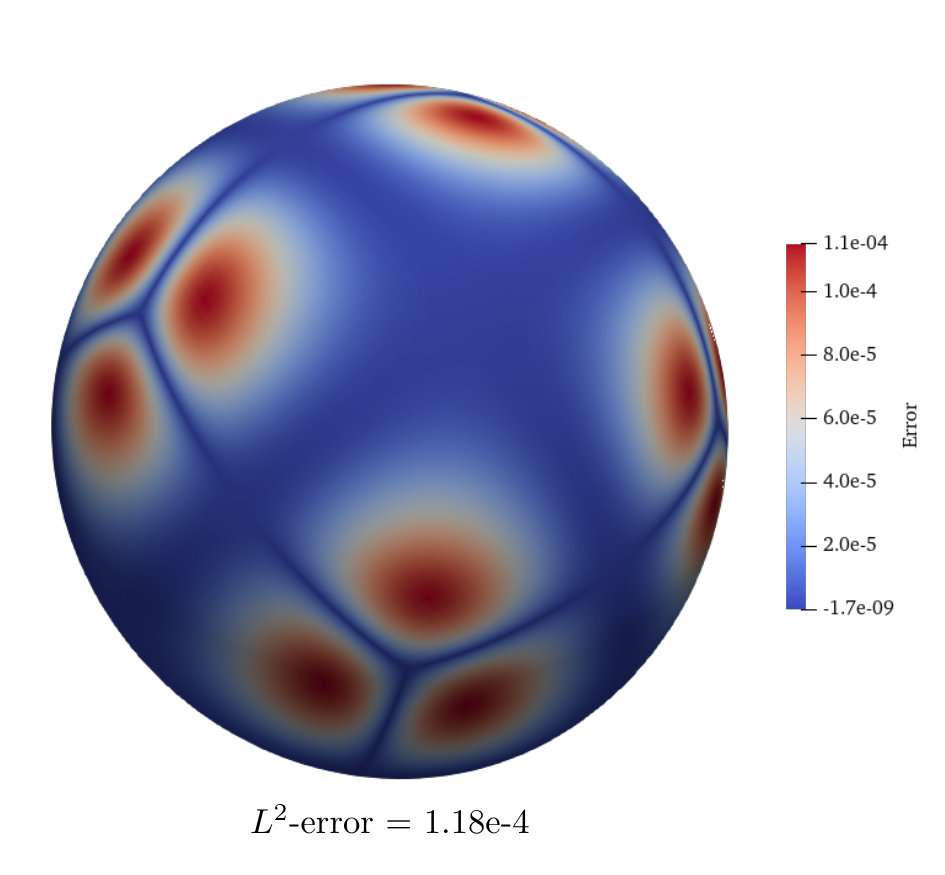}
                  \caption{}
                  \label{fig:sphereL1}
     \end{subfigure}
     \hfill
     \begin{subfigure}[b]{0.3\textwidth}
         \centering
         \includegraphics[width=\textwidth]{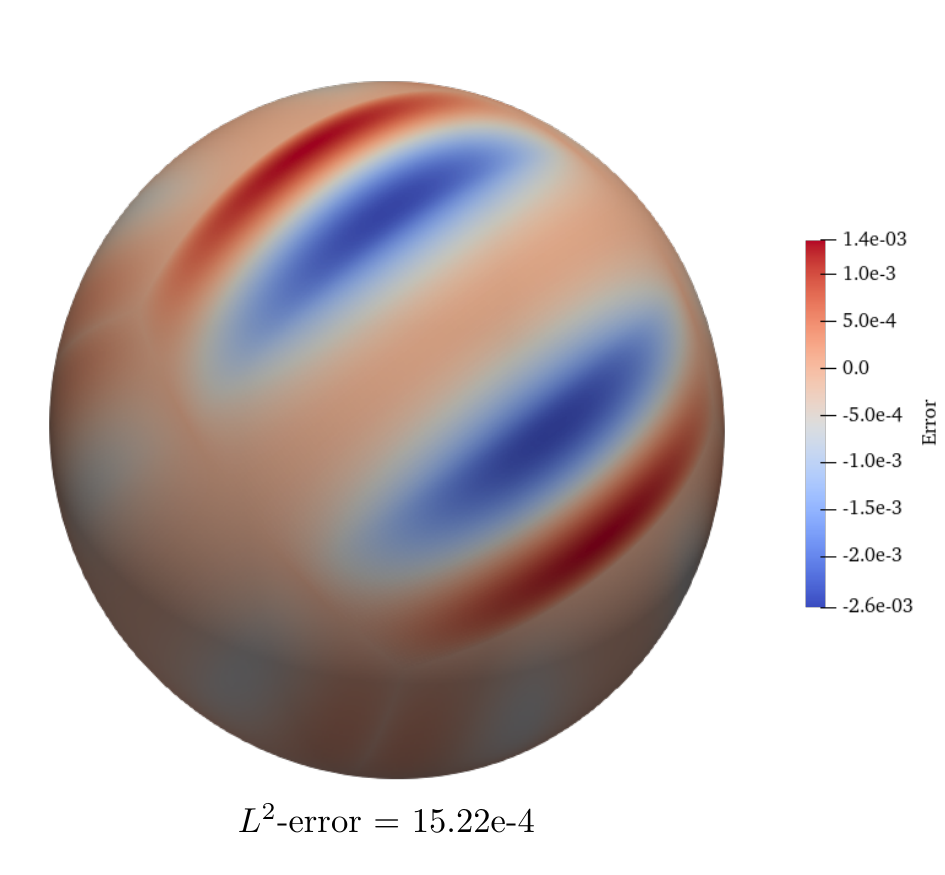}
                  \caption{}
                  \label{fig:sphereL3}
     \end{subfigure}
     \hfill
    \begin{subfigure}[b]{0.3\textwidth}
         \centering
         \includegraphics[width=\textwidth]{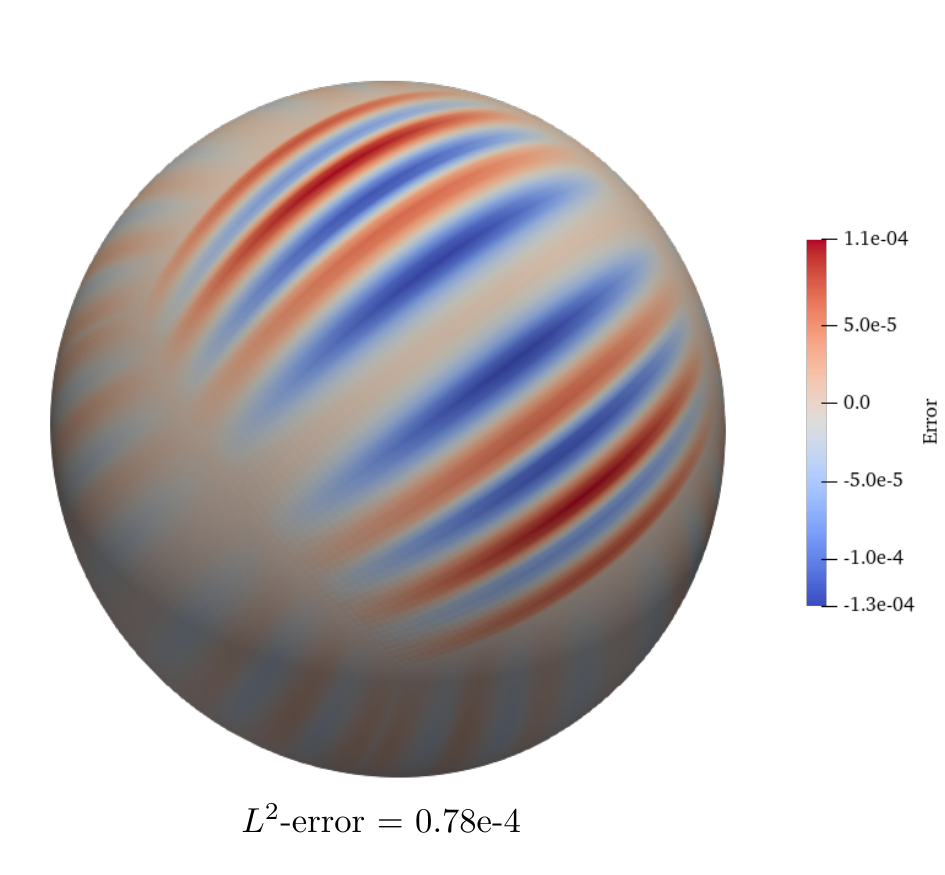}
                  \caption{}
                  \label{fig:sphereL7}
     \end{subfigure}
        \caption{Effect of increasing the number of section curves on the spatial error distribution, Error$=R^{\mathrm{outer}} - \|\mathbf{x}^{\mathrm{sphere}}\|$, visualized for the outer sphere surface. Spatial error distribution for (a) only 1 section curve, (b) 3 section curves, and (c) 7 section curves in each patch, including the $L^2$-error norm. Note that the range of the color scale is different for each of the plots.}
        \label{fig:SphereLofting}
\end{figure*}

The thick-walled sphere is constructed by specifying the different radii, constraining the sphere at its center to the Cartesian origin, and specifying the patch vertex locations, \emph{step (i)}. For the current analysis, the radii are set to 0.5 and 1 for the inner and outer radii, respectively. The curves that describe the patch surfaces are obtained by solving Equation~\ref{eq:argmin}, in which the mathematical description of the sphere is the only constraint required, \emph{step (ii)}. Combining the resulting curves and converting them to surfaces is achieved by lofting, \emph{step (iii)}, which gives the user the freedom to specify the number of section curves as visualized in Figure~\ref{fig:SphereLofting} (the reader is referred to \emph{step~(iii)} in Section~\ref{sec:multipatch} for an explanation of lofting). The influence of the number of section curves is visualized in Figure~\ref{fig:SphereLofting}, which shows the spatial error on the outer surface of the thick-walled sphere. 

Figure~\ref{fig:SphereLofting} clearly shows the effect of a change in the number of section curves. First, it is observed that for all cases, the \emph{worst} approximation of the sphere deviates by only a maximum of $0.26$\% with respect to the radius. Such deviation is well within the acceptable range, given the purpose of the proposed model. Second, the results clearly show that each section curve is located exactly on the sphere, \emph{i.e.}, the error is nearly equal to zero, which is expected from the construction procedure. It does however result in an oscillating behavior observed on the surface after lofting, albeit with a relatively small magnitude. Based on the comparison, it is clear that the coarsest mesh, \emph{i.e.}, the lowest number of section curves, yields a sufficiently accurate approximation. In the remainder of this section, we will use the geometry that is constructed with only 1 section curve, Figure~\ref{fig:SphereLofting}~(a).

\subsection{Hyperelastic material model verification}
The thick-walled sphere is numerically analyzed by applying a uniform normal traction to the inner (cavity) boundary and a zero traction boundary condition at the outer boundary. The resulting deformation is then verified with an analytic solution, from which the convergence rates are computed based on different element sizes. The analytic solution assumes a fully incompressible hyperelastic isotropic material and is available in the literature~\cite{Ogden2001}. A distinction is made between the radial coordinate defined in the reference configuration, $R$, and the current configuration, $r$. The radial coordinates in both configurations are then bounded by
\begin{subequations}
    \begin{align}
       R_{\mathrm{in}} \leq & \ R \leq R_{\mathrm{out}},  \\
       r_{\mathrm{in}} \leq & \ r \leq r_{\mathrm{out}}, 
    \end{align}
\end{subequations}
where the subscripts refer to the inner and outer boundary of the sphere. For a fully incompressible hyperelastic isotropic material described by an energy density function, $\psi$, the analytic solution then reads,
\begin{equation}\label{eq:analyticP}
    p^{\mathrm{cav}} = - \int^{\lambda_{\mathrm{out}}}_{\lambda_{\mathrm{in}}} \frac{1}{\lambda^3- 1} \frac{\partial \psi^{\mathrm{pas}}}{\partial \lambda}  \ \text{d} \lambda, 
\end{equation}
with
\begin{equation}
    \lambda= \frac{r}{R}, \quad \lambda_{i} = \frac{r_{i}}{R_{i}}, \quad \text{for} \quad {i}=\mathrm{in},\mathrm{out};
\end{equation}
The analytic solution in Equation~\ref{eq:analyticP} relates the cavity pressure, $p^{\mathrm{cav}}$, to the deformation or stretch of the wall, $\lambda$, corresponding the specified energy density function $\psi^{\mathrm{pas}}$. The computed cavity pressure for a specified deformation yields the following displacement vector in spherical coordinates $[r,\theta,\phi]$,
\begin{equation}\label{eq:analyticU}
    u^{\mathrm{an}}_i\left(R\right) = \left[ \left( R^3 + r_{\mathrm{in}}^3 - R_{\mathrm{in}}^3 \right)^{\frac{1}{3}} - R, \ 0, \ 0 \right].
\end{equation}

\begin{figure*}[!t]
\centering
    \includegraphics[width = 0.5\linewidth]{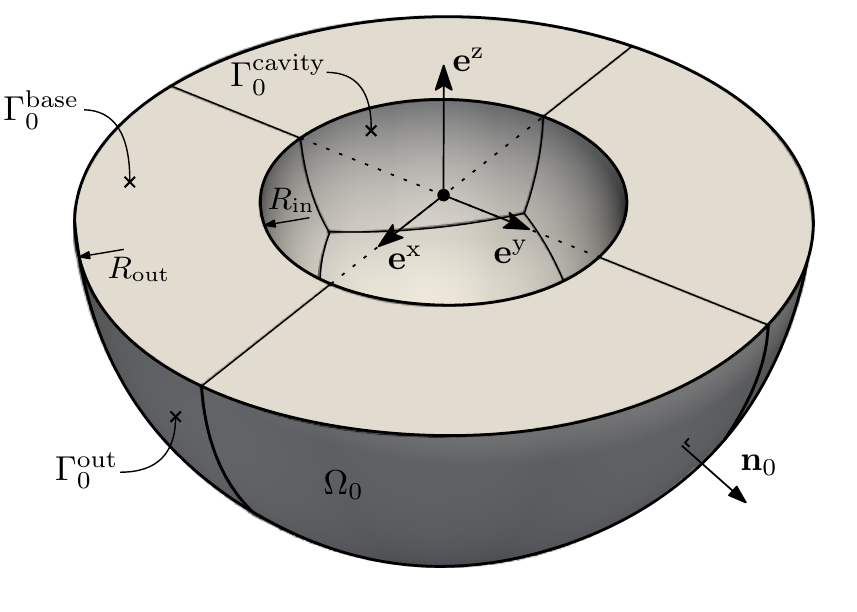}
    \caption{Numerical domain in the reference configuration, $\Omega_0$, of the thick-walled sphere consisting of three boundaries, ${{\Gamma}_0 = {\Gamma}^{\mathrm{base}}_0 \cup \ {\Gamma}^{\mathrm{cavity}}_0\cup \ {\Gamma}_{\mathrm{out}}}_0$. The problem is defined in the Cartesian coordinate system, $\{ \mathbf{e}^{\mathrm{x}}, \mathbf{e}^{\mathrm{y}}, \mathbf{e}^{\mathrm{z}} \}$.}
    \label{fig:sphereNumdomain}
\end{figure*}

\begin{figure*}[!t]
     \centering
     \begin{subfigure}[b]{0.45\textwidth}
         \centering
         \includegraphics[width=\textwidth]{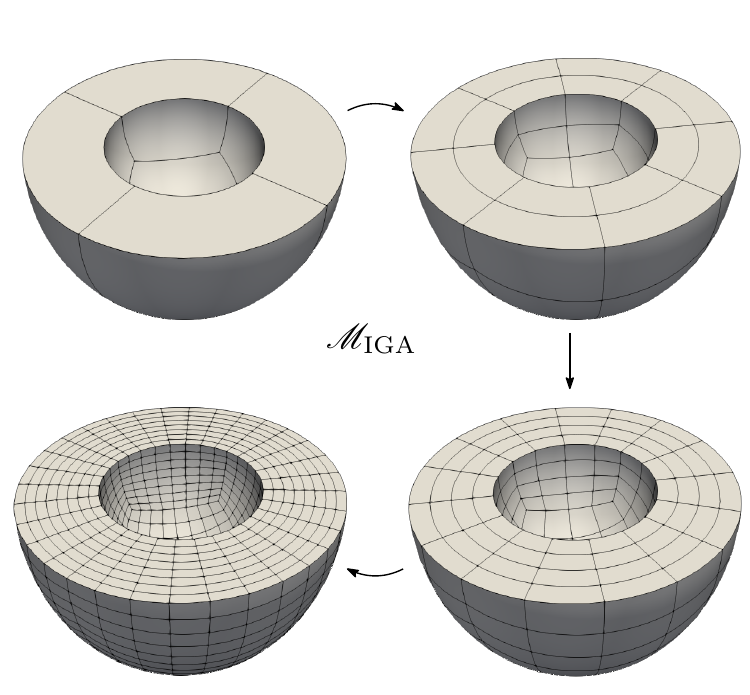}
                  \caption{}
                  \label{fig:sphereMeshIGA}
     \end{subfigure}
     \hfill
     \begin{subfigure}[b]{0.45\textwidth}
         \centering
         \includegraphics[width=\textwidth]{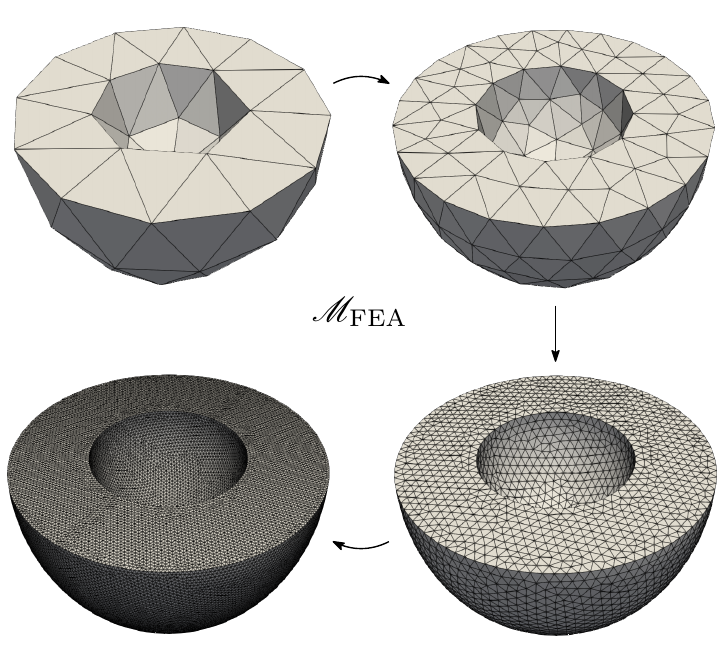}
         \caption{}
         \label{fig:sphereMeshFEA}
     \end{subfigure}
        \caption{Different mesh resolutions for the isogeometric analysis (IGA) and the finite element analysis (FEA) discretization. Refinements for both methods are uniform.}
        \label{fig:sphereMesh}
\end{figure*}

The same problem is then numerically analyzed for which the domain is visualized in Figure~\ref{fig:sphereNumdomain}, where we compute the solution of the displacement vector $u_i$. For completeness, we solve for both the single-field formulation, as proposed in Equation~\eqref{eq:passivePK}, and the mixed-field formulation, which omits the volumetric contribution to the energy density function, $\psi^{\mathrm{V}}$. The mixed formulation enforces the incompressibility constraint, \emph{i.e.}, $\text{det}(\mathbf{F})=J=1$, using the hydrostatic pressure which acts as a Lagrange multiplier. Both models then state:
\begin{equation}\label{eq:sphereSingle}
    \left( \mathrm{Single} \right)\begin{cases}\text { Find } u_i \text { such that: } & \\ 
    \frac{\partial}{\partial x_i} \sigma^{\mathrm{Single}}_{ij} = 0_j & \text { in } \Omega, \\ 
    \sigma_{ij} n_j = -p^{\mathrm{cav}} n_i & \text { at } \Gamma^{\mathrm{cavity}},  \\ 
    \sigma_{ij} n_j = 0_i & \text { at } \Gamma^{\mathrm{out}},  \\ 
    u_i n_i = 0 & \text { at } \Gamma^{\mathrm{base}},  \\ 
    u_i \mathrm{e}^{\mathrm{x}}_i = 0 & \text { at } \{ \Gamma^{\mathrm{base}} \cap \Gamma^{\mathrm{cavity}} \cap y =0 \}, \\ 
    u_i \mathrm{e}^{\mathrm{y}}_i = 0 & \text { at } \{ \Gamma^{\mathrm{base}} \cap \Gamma^{\mathrm{cavity}} \cap x =0 \},\end{cases}
\end{equation}
and
\begin{equation}\label{eq:sphereMixed}
    \left( \mathrm{Mixed} \right)\begin{cases}\text { Find } (u_i,p) \text { such that: } & \\ 
    \frac{\partial}{\partial x_i} \left( -p \delta_{ij}^{\phantom{}} + \sigma^{\mathrm{Mixed}}_{ij} \right) = 0_j & \text { in } \Omega, \\ 
    J = 1 & \text { in } \Omega, \\ 
    \sigma_{ij} n_j =- p^{\mathrm{cav}} n_i & \text { at } \Gamma^{\mathrm{cavity}},  \\ 
    \sigma_{ij} n_j = 0_i & \text { at } \Gamma^{\mathrm{out}},  \\ 
    u_i n_i = 0 & \text { at } \Gamma^{\mathrm{base}},  \\ 
    u_i \mathrm{e}^{\mathrm{x}}_i = 0 & \text { at } \{ \Gamma^{\mathrm{base}} \cap \Gamma^{\mathrm{cavity}}  \cap y =0 \}, \\ 
    u_i \mathrm{e}^{\mathrm{y}}_i = 0 & \text { at } \{ \Gamma^{\mathrm{base}} \cap \Gamma^{\mathrm{cavity}} \cap x =0 \},\end{cases}
\end{equation}
where for both models, the normal displacement at the base plane is constrained, including four nodes in the $x$- and $y$-direction, corresponding to the Cartesian coordinate system as visualized in Figure~\ref{fig:sphereNumdomain}. These Dirichlet boundary conditions satisfy the periodicity of the sphere and also prevent any rigid body motion. The Cauchy stress for this problem is related to the second Piola-Kirchhoff stress such that,
\begin{equation}
    \sigma^{t} = \frac{1}{J} F^{\phantom{}}_{ik} \frac{\partial \psi^t}{\partial E_{kn}} F^{\phantom{}}_{jn}, \quad \text{for} \quad t=\mathrm{Single}, \mathrm{Mixed};
\end{equation}
in which the energy density function, $\psi^t$, is defined according to Equation~\eqref{eq:EnergyDensity}, but slightly different for both the single and mixed field formulations,
\begin{subequations}
    \begin{align}
       \psi^{\mathrm{Single}} &= \psi^{\mathrm{S}} + \psi^{\mathrm{V}},  \\
       \psi^{\mathrm{Mixed}} &= \psi^{\mathrm{S}}.
    \end{align}
\end{subequations}
The parameters associated with the energy density functions, $\psi^{\mathrm{S}}$ and $\psi^{\mathrm{V}}$, are identical to Table~\ref{tab:params}, apart from $a_3=0$, which is related to the fiber direction.

The numerical domain is discretized in two fundamentally different ways: The NURBS multi-patch geometry (Figure~\ref{fig:Spheremultipatch}) is discretized according to the IGA paradigm by B-spline basis functions for the displacement vector, while the mesh is obtained by uniform grid refinement. In accordance with the FEA paradigm, we employ Lagrangian basis functions for the displacement vector, while the mesh consists of unstructured linear tetrahedrons. The FEA geometry and mesh are created using GMSH \cite{gmsh}, a separate meshing tool. This meshing tool guarantees that the exact geometry is obtained under mesh refinement, while the IGA geometry has a fixed geometrical error, visualized in Figure~\ref{fig:SphereLofting}. However, one should use an impractical amount of linear tetrahedrons for the FEA mesh, in order to obtain a lower geometry error than introduced by the IGA geometry. Furthermore, an unstructured linear tetrahedron mesh is chosen, which is similar to the FEniCS and Nutils cardiac model comparison in Section~\ref{sec:FEniCSvsNutils}.

Given the spatial volumetric mesh of the multi-patch geometry, a suitable set of B-spline basis functions is employed to discretize the displacement vector. The employed univariate B-spline basis functions of degree p, $B^{\mathrm{p}}_{i,d}$, are then associated with the three knot-vectors $\Xi_d, d=1,2,3$, which are related to the three spatial dimensions. The tensor-product of the B-spline basis functions on the 3D volumetric mesh is then defined as,
\begin{equation}
    B^{\mathrm{p}}_{ijk} := B^{\mathrm{p}}_{i1} \otimes B^{\mathrm{p}}_{j2} \otimes B^{\mathrm{p}}_{k3},     
\end{equation}
with $i=1,...,n_1, \ j=1,...,n_2, \  k=1,...,n_3$, where $n_i$ are the number of functions in the $i$-th spatial direction. We define the tensor-product B-spline space as the span of these basis functions,
\begin{equation}
    S^{\mathrm{p}}_{\alpha} := \text{span} \left\{  B^{\mathrm{p}}_{ijk}  \right\}^{n_1,n_2,n_3}_{i=1,j=1,k=1} ,
\end{equation}
where it is assumed that all directions share the same degree p, and regularity $\alpha$. The regularity is a measure of inter-element smoothness bounded by, $0\leq \alpha \leq p-1$, and is related to the multiplicities of the knot vector $\Xi$. It is often chosen to be maximal when smoothness is desired for the solution. However, stability criteria exist when multiple spline spaces are used~\cite{Buffa2011}, \emph{i.e.}, the mixed formulation, which requires a compatible spline space to be chosen. For the numerical analysis, we consider three different spaces: 
\begin{subequations}
    \begin{align}\label{eq:spaces}
       \mathcal{B}^{\mathrm{p}}_{\mathrm{M}} &= \left\{ S^{\mathrm{p}}_{\mathrm{p}-2} \times S^{\mathrm{p}}_{\mathrm{p}-2} \times S^{\mathrm{p}}_{\mathrm{p}-2} , \ S^{\mathrm{p-1}}_{\mathrm{p}-2} \right\} ,  \\
       \mathcal{B}^{\mathrm{p}}_{\mathrm{S}}& = S^{\mathrm{p}}_{\mathrm{p}-1} \times S^{\mathrm{p}}_{\mathrm{p}-1} \times S^{\mathrm{p}}_{\mathrm{p}-1} ,  \\
       \mathcal{L}^{\mathrm{p}}_{\mathrm{S}}& = L^{\mathrm{p}} \times L^{\mathrm{p}} \times L^{\mathrm{p}},
    \end{align}
\end{subequations}
where use is made of the Taylor-Hood element family for the mixed field (B-spline) space, $\mathcal{B}^{\mathrm{p}}_{\mathrm{M}}$, which employs a displacement-pressure space of different degree but with identical regularity, $\alpha=p-2$ \cite{Buffa2011}. The single field (B-spline) space, $\mathcal{B}^{\mathrm{p}}_{\mathrm{M}}$, only comprises a displacement (vector) space, where the regularity is chosen to be maximal. The single field space associated to the FEA discretization, $\mathcal{L}^{\mathrm{p}}_{\mathrm{S}}$, employs Lagrangian basis functions of degree p, $L^{\mathrm{p}}$, for the displacement vector. These three spaces are then associated with the meshes visualized in Figure~\ref{fig:sphereMesh}, to obtain the displacement results of the problem stated in Equations~\eqref{eq:sphereSingle} and \eqref{eq:sphereMixed}. 

\begin{figure}[H]
    \centering
    \includegraphics[width = 0.5\linewidth]{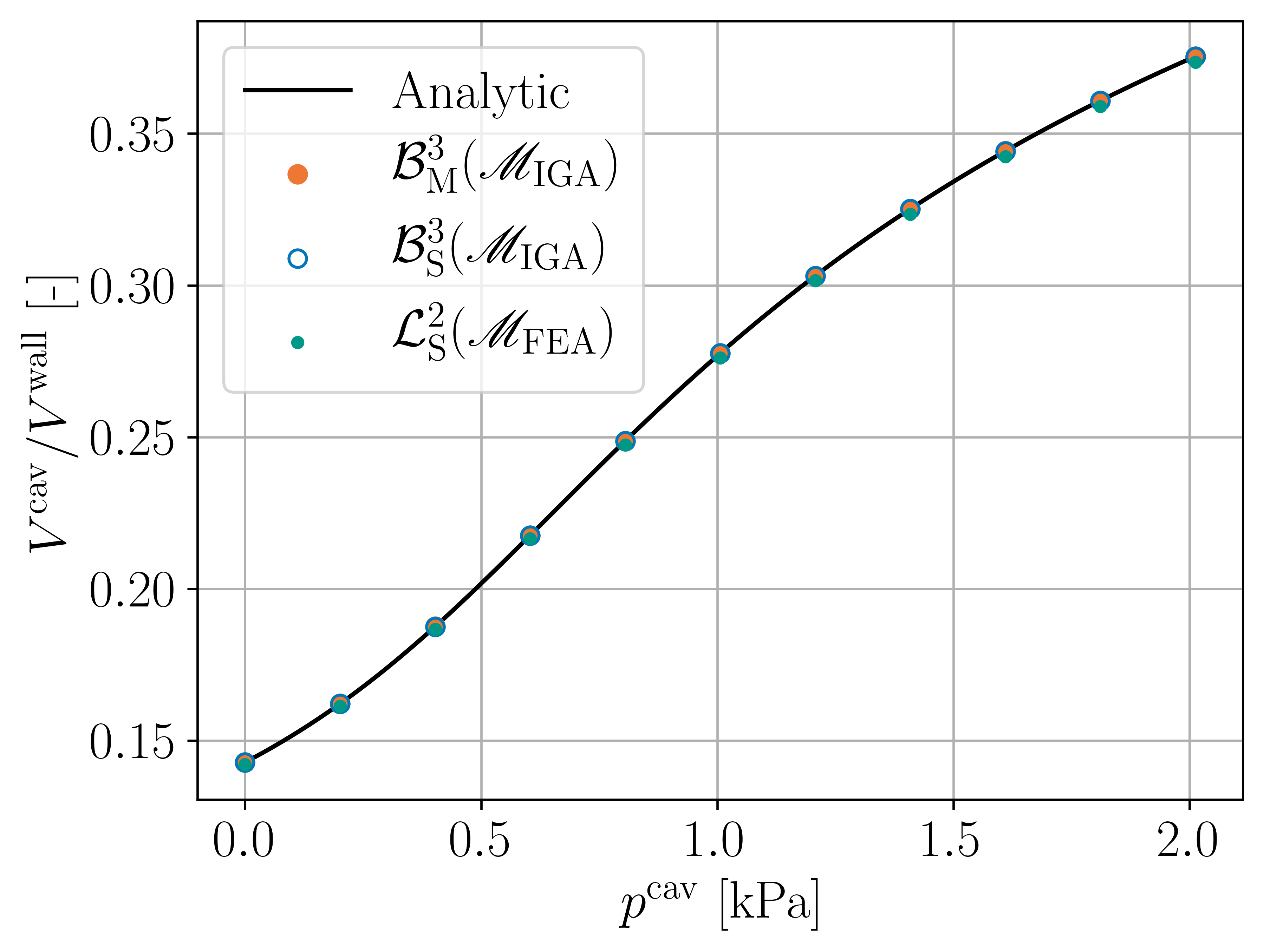}
    \caption{Comparison between the pressure-volume relation of the analytic solution and numerical results, where the cavity volume is normalized by the wall volume. The numerical results are solved on the second mesh refinement (third mesh) in Figure~\ref{fig:sphereMesh}, for both the IGA and FEA case. This results 12,720 degrees-of-freedom (DOFs) for $\mathcal{B}^{3}_{\mathrm{M}}$, 4,053 DOFs for $\mathcal{B}^{3}_{\mathrm{S}}$, and 170,040 DOFs for $\mathcal{L}^{2}_{\mathrm{S}}$.}
    \label{fig:sphereTypical}
\end{figure}

The analytic solution, Equation~\eqref{eq:analyticP}, for a varying cavity pressure is visualized in Figure~\ref{fig:sphereTypical}. The analytic solution is compared to the three numerical results, which correspond to the three spaces in Equation~\eqref{eq:spaces}, and are solved for a fixed mesh size. The results clearly show the nonlinear material behavior for increasing cavity pressure, but also a good agreement between the numerical and analytic solutions. A slight deviation is observed for the FEA result, which is related to the geometry approximation introduced by the linear tetrahedrons during the meshing procedure.

Next, we quantify the convergence at a fixed cavity pressure by the spatial displacement error, defined as,
\begin{equation}
    \varepsilon_i = u^{\mathrm{an}}_i - u^{\phantom{}}_i
\end{equation}
where $u^{\mathrm{an}}_i$ is the analytic displacement vector, Equation~\eqref{eq:analyticU}, and $u^{\phantom{}}_i$ the numerical result, both in spherical coordinates. The spatial error distribution is then quantified by the $L^2$- and $H^1$-error norms such that,
\begin{equation}
    L^2\text{-error} := \left( \int \varepsilon_k \varepsilon_k  \ \text{d} \Omega_0 \right)^{0.5},
\end{equation}
and
\begin{equation}
    H^1\text{-error} := \left( \int \varepsilon_k \varepsilon_k + \frac{ \partial \varepsilon_i }{\partial x_j} \frac{ \partial \varepsilon_i }{\partial x_j}   \ \text{d} \Omega_0 \right)^{0.5}.
\end{equation}

\begin{figure*}[!t]
     \centering
     \begin{subfigure}[b]{0.45\textwidth}
         \centering
         \includegraphics[width=\textwidth]{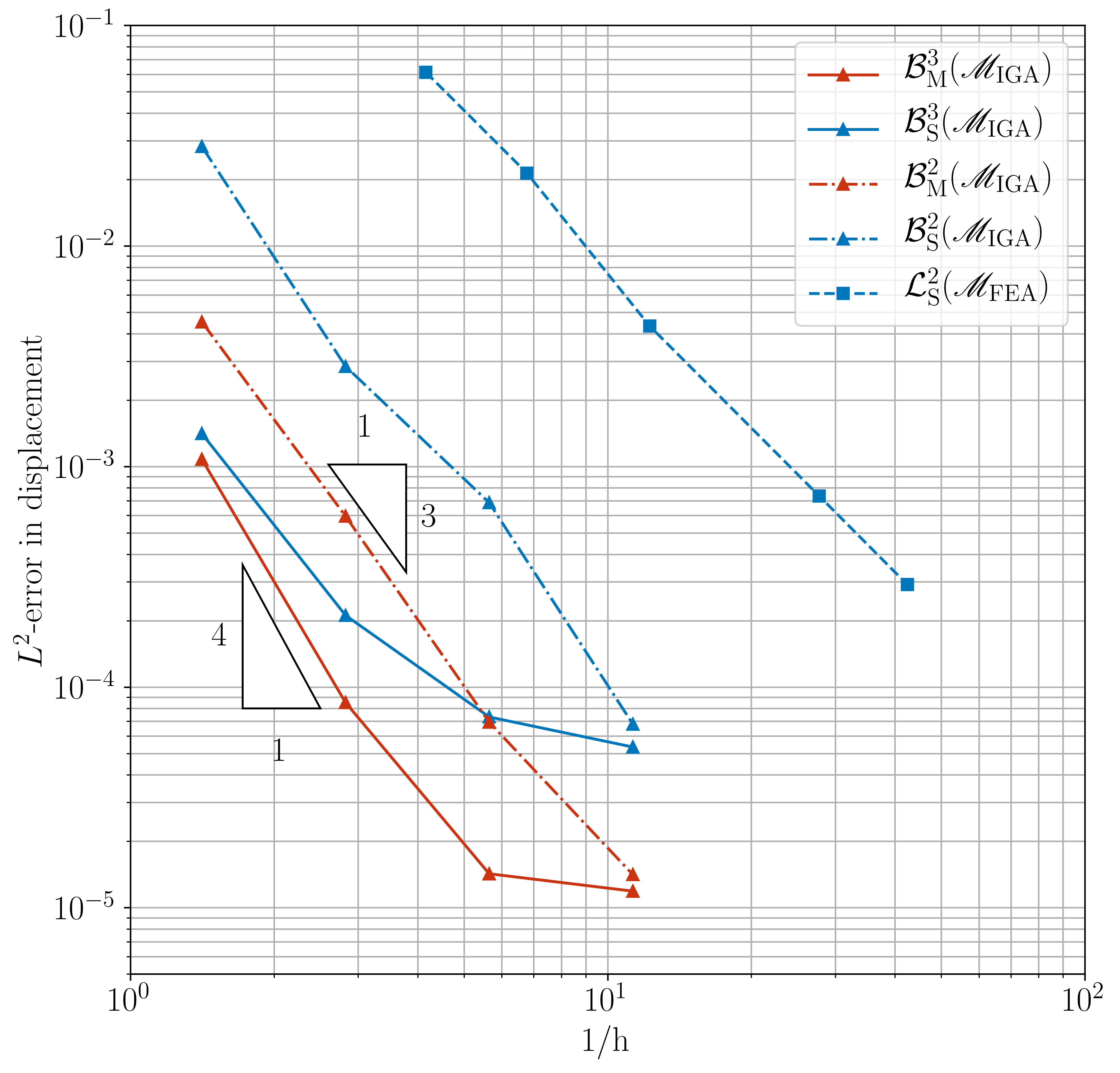}
                  \caption{}
                  \label{fig:L2sphere}
     \end{subfigure}
     \hfill
     \begin{subfigure}[b]{0.45\textwidth}
         \centering
         \includegraphics[width=\textwidth]{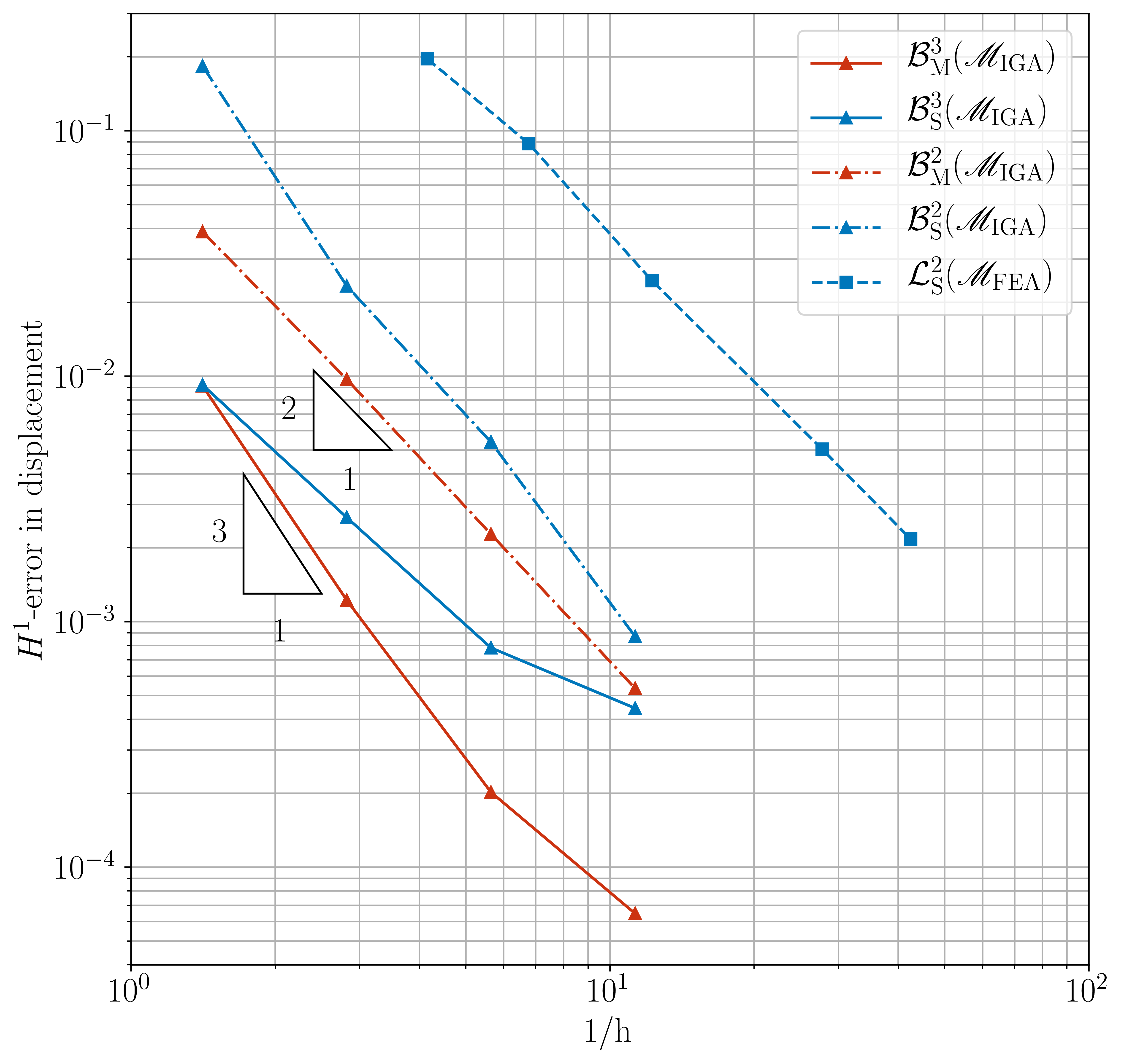}
         \caption{}
         \label{fig:H1sphere}
     \end{subfigure}
        \caption{$L^2$- and $H^1$-error norms of the displacement vs. the mesh size $h$ for the inflated thick-walled sphere at a cavity pressure of {$p^{\mathrm{cav}}=2$~[kPa]}. Two methods are compared, the Isogeometric analysis (IGA) and finite element analysis (FEA), where optimal convergence rates are obtained for the IGA mixed-formulation, while the single field formulations behave slightly sub-optimal. The FEA results show a good convergence rate but require a sufficiently small element size.}
        \label{fig:Sphereconvergence}
\end{figure*}

The convergence results are provided in Figure~\ref{fig:Sphereconvergence}, and plotted against the inverse of the average mesh {size~$h$}. The results show that the IGA model is capable of approximating the exact solution with far fewer elements when compared to the FEA result. This even holds when adopting identical basis function degrees for both the B-splines and Lagrangian functions, \emph{i.e.}, quadratic. This difference is expected to be related to the geometry approximation of the linear tetrahedrons. The IGA result does, however, show a plateau behavior at smaller element sizes, which is caused by two factors: First, the analytic solution assumes a fully incompressible material, while the single field formulation solves for a nearly incompressible material. The difference in the equations that are solved prevents the single-field formulation to approximate the analytic solution more exactly. Second, the mixed formulation, which does solve for a fully incompressible material, also shows a plateau, but at a smaller error norm value. This is caused by the fixed geometrical error introduced by the lofting procedure, visualized in Figure~\ref{fig:SphereLofting}, which prevents any improvement of the approximate displacement solution. Nonetheless, the geometrical error introduced by lofting is substantially smaller when compared to the geometrical errors introduced by the linear tetrahedrons, \emph{i.e.}, the FEA result. The results also show the benefit of using higher degree (cubic) B-spline basis functions over quadratic, especially when observing the $H^1$-norm. In future analyses, it is desired to employ cubic B-splines with a refinement similar to the second refined mesh as visualized in Figure~\ref{fig:sphereMeshIGA} (third mesh).

\section{Left ventricle model}\label{app:lvmodel}
This appendix explains the boundary conditions when applying the bi-ventricular IGA cardiac model of Section~\ref{sec:CardiacModel} to the left ventricle. The left ventricle IGA model is then benchmarked against an established FEA model, as discussed in Section~\ref{subsec:leftventricle}. Additional results of the IGA and FEA comparison are provided at the end of this Appendix. 

\subsection{Geometry and fiber field}\label{appS:fiberfield}
The left ventricle is modeled as a truncated ellipsoid, which is parameterized in prolate coordinates $\{ \xi, \theta, \phi \}$ such that,
\begin{equation}\label{eq:prolate}
    \begin{aligned}x^{\mathrm{ell}}_i=
    \left\{\begin{aligned} 
          x & = C \mathrm{sinh}(\xi)\mathrm{sin}(\theta) \mathrm{cos}(\phi) \\ 
          y & = C \mathrm{sinh}(\xi)\mathrm{sin}(\theta) \mathrm{sin}(\phi) \\ 
          z & = C \mathrm{cosh}(\xi)\mathrm{cos}(\theta), 
         \end{aligned}\right.
\end{aligned}
\end{equation}
where $C$ is the focal length set to be $43.0$ [mm]. The considered ellipsoid employs a value of $\xi^{\mathrm{end}}$=0.371 for the endocardium and $\xi^{\mathrm{epi}}$=0.678 for the epicardium such that $\xi\in[\xi^{\mathrm{end}},\xi^{\mathrm{epi}}]$. The ellipses are then truncated at the truncation height $H$=$24.0$ [mm] such that $\theta \in [\mathrm{arccos}(H/\mathrm{cosh}(\xi)), \pi]$ and $\phi\in[0,2\pi]$. The fiber field can then be specified analytically deriving the local basis $\{\mathbf{e}^{\mathrm{c0}}, \mathbf{e}^{l\mathrm{0}}, \mathbf{e}^{\mathrm{t0}}\}$, such that
\begin{equation}
    \mathrm{e}^{\mathrm{c0}}_i = \frac{\partial x^{\mathrm{ell}}_i}{\partial \phi} \left\| \frac{\partial x^{\mathrm{ell}}_j}{\partial \phi}  \right\|^{-1} = \mathrm{cos}(\phi) e^{\mathrm{y}}_i - \mathrm{sin}(\phi) e^{\mathrm{x}}_i,
\end{equation}
\begin{equation}
\begin{aligned}
    \mathrm{e}^{l\mathrm{0}}_i = \frac{\partial x^{\mathrm{ell}}_i}{\partial \theta} \left\| \frac{\partial x^{\mathrm{ell}}_j}{\partial \theta}  \right\|^{-1} &= \left( \mathrm{sinh}(\xi) \mathrm{cos}(\theta) \mathrm{cos}(\phi) e^{\mathrm{x}}_i \right. \\ & +\mathrm{sinh}(\xi) \mathrm{cos}(\theta) \mathrm{sin}(\phi) e^{\mathrm{y}}_i \\
    & \left. -\mathrm{cosh}(\xi) \mathrm{sin}(\theta) e^{\mathrm{z}}_i \right) \left( \mathrm{sinh}^2(\xi) + \mathrm{sin}^2(\theta)\right)^{-\frac{1}{2}},
    \end{aligned}
\end{equation}
\begin{equation}
\begin{aligned}
    \mathrm{e}^{\mathrm{t0}}_i = \frac{\partial x^{\mathrm{ell}}_i}{\partial \xi} \left\| \frac{\partial x^{\mathrm{ell}}_j}{\partial \xi}  \right\|^{-1} & = \left( \mathrm{cosh}(\xi) \mathrm{sin}(\theta) \mathrm{cos}(\phi) e^{\mathrm{x}}_i \right. \\ & +\mathrm{cosh}(\xi) \mathrm{sin}(\theta) \mathrm{sin}(\phi) e^{\mathrm{y}}_i \\
    & \left. +\mathrm{sinh}(\xi) \mathrm{cos}(\theta) e^{\mathrm{z}}_i \right) \left( \mathrm{sinh}^2(\xi) + \mathrm{sin}^2(\theta)\right)^{-\frac{1}{2}},
\end{aligned}
\end{equation}
where $\{\mathbf{e}^{\mathrm{x}}, \mathbf{e}^{\mathrm{y}}, \mathbf{e}^{\mathrm{z}} \}$ are the Cartesian unit vectors, $i \in \{x,y,z\}$. The local basis is rotated according to the fiber field defined in \cite{bovendeerdlvstrain}.

\subsection{Model description}
\begin{figure}[H]
    \centering
    \includegraphics[width = 0.6\linewidth]{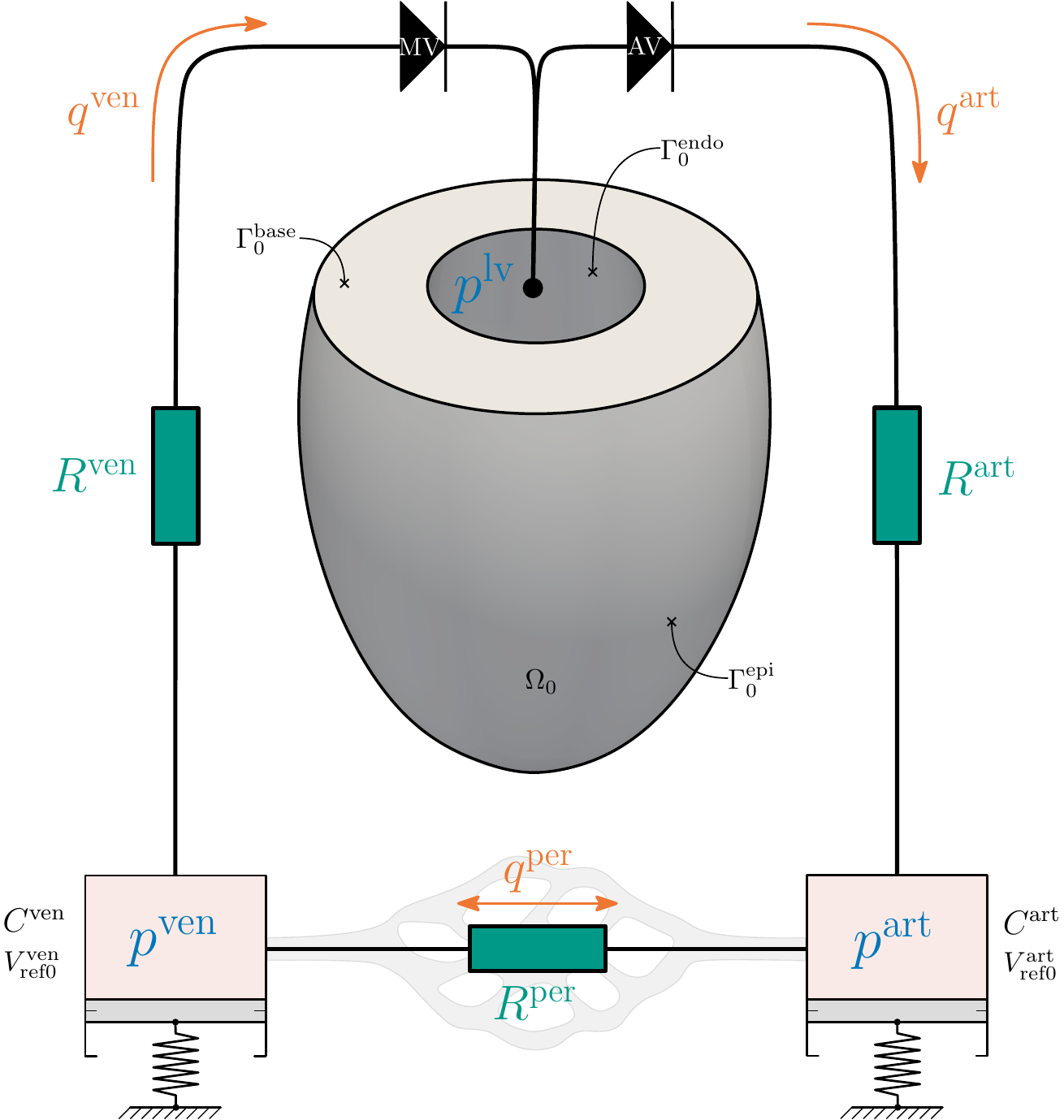}
    \caption{Schematic overview of the 0D circulatory system coupled to the 3D left ventricle domain. The circulatory system is modeled as two Windkessel compartments in series, characterized by a compliance, $C^i$, and a volume at zero reference pressure, $V^i_{\rm{ref0}}$. The compartments and ventricles are linked via resistances, $R^i$ (green), through which arterial (art), venous (ven), and peripheral (per) volume flow occurs, $q^i$ (orange), caused by the compartment pressure, $p^i$ (blue). The two heart valves: the mitral valve (MV) and the aortic valve (AV), are idealized such that they open and close instantaneously and prevent back-flow. The ventricle and compartment pressures, illustrated in blue, are the quantities that are solved in the circulatory model.}
    \label{fig:circulatorySchemeLV}
\end{figure}

There are two differences in the applied boundary conditions when compared to the bi-ventricle model, Section~\ref{sec:CardiacModel}: The pericardium and circulatory system boundary conditions. The pericardium boundary condition ate the epicardium is removed and replaced by a set of nodal Dirichlet conditions. These constrained nodes or points are chosen such that, $\Gamma^{\mathrm{base}}_0 \cap \Gamma^{\mathrm{endo}}_0 \cap { x_{0_k} = 0 }$ for $k=\{1,2\}=\{x,y\}$, \emph{i.e.}, patch-vertices located on the x- and y-axes. The circumferential displacement is then constrained which in the axisymmetric case prevents any rigid body motion. Furthermore, due to the absence of a right ventricle, the circulatory model is simplified to only two compartments in which the right ventricle is lumped. A visualization of the circulatory model is given in Figure~\ref{fig:circulatorySchemeLV}.

The cardiac model adjusted for the left ventricle then states

\begin{equation}\label{eq:LVcardiacmodel}
    \left( \text{Strong} \right) \begin{cases}\text { Find } \left( u_k, l^{\rm{c}},\tilde{p} \right) \text { for } t\in \left( 0,T\right], & \text{such that: } \\
    \frac{\partial}{\partial x_i} \left[ {\sigma}^{\rm{pas}}_{ij} ( u_k ) + \sigma_{ij}^{\rm{act}}( l^{\rm{c}}, u_k, t )\right] = 0_j  & \text { in } \Omega \times(0, T], \\ 
    \mathcal{A} ( l^{\rm{c}}, u_k)  = 0 & \text { in } \Omega\times(0, T], \\ 
    \mathcal{M} (\tilde{p}^{\mathrm{0D}}, u_k )  = \left\{0\right\} & \text { in } (0, T],\\
    &\\
\text{Subject to:} & \\
    n_i \cdot \sigma^{\phantom{}}_{ij} = -p^{\mathrm{lv}}(u_k) \ n_j & \text { on } \Gamma^{\mathrm{endo}} \times(0, T], \\
    u_0 = 0  & \text { on } \Gamma^{\mathrm{base}} \cap \Gamma^{\mathrm{endo}} \cap { x_{0} = 0 } ,\\
    u_1 = 0  & \text { on } \Gamma^{\mathrm{base}} \cap \Gamma^{\mathrm{endo}} \cap { x_{1} = 0 } ,\\
    n_i u_i  = 0 & \text { on } \Gamma^{\mathrm{base}},\\
    \tilde{p} = \tilde{p}_{\mathrm{0}}, l^{\rm{c}}=l^{\rm{s0}}, \ u_i = 0_i & \text { at } t=0.
\end{cases}
\end{equation}

The model components are identical to Equation~\eqref{eq:setofeqs} with the exception of the circulatory model $\mathcal{M}(\cdot)$, which is simplified to only 3 compartments: the venous compartment and the arterial compartment, this yields the following relation for the total blood volume,
\begin{equation}\label{eq:totalvolumeLV}
    V^{\mathrm{tot}} = V^{\mathrm{lv}} + V^{\mathrm{ven}} V^{\mathrm{art}}.
\end{equation}

Combining the pressure-volume relations for all compartments yields the following circulatory model:
Find $\tilde{p} = \{p^{\mathrm{lv}}, p^{\mathrm{art}} \} \in \left(0, T\right]$, such that
\begin{equation}\label{eq:circulatlv}
    \mathcal{M} (\tilde{p}, u_i ) =
    \left\{\begin{aligned} 
          \frac{d V^{\mathrm{lv}}\left( u_i \right)}{dt} - q^{\mathrm{ven}} + q^{\mathrm{art}} & = 0, \\
          C^{\mathrm{art}}\frac{d p^{\mathrm{art}}}{dt} - q^{\mathrm{art}} + q^{\mathrm{per}} & = 0, \\
          C^{\mathrm{ven}} p^{\mathrm{ven}}  - V^{\mathrm{tot}} + V^{\mathrm{lv}} + V^{\mathrm{art}} +  V^{\mathrm{ven}}_{\rm{ref0}}   & = 0,
         \end{aligned}\right.
\end{equation}
The venous pressure need not be solved for since this can be computed based on the total blood volume, Equation~\eqref{eq:totalvolumeLV}. The volume flows are defined as
\begin{equation}\label{eq:volumeflowLV}
    \left\{\begin{aligned} 
          q^{\mathrm{ven}} & = \langle \frac{p^{\mathrm{ven}} - p^{\mathrm{lv}}}{R^{\mathrm{ven}}} \rangle , \\
          q^{\mathrm{art}} & = \langle \frac{p^{\mathrm{lv}} - p^{\mathrm{art}}}{R^{\mathrm{art}}} \rangle , \\
          q^{\mathrm{per}} & =  \frac{p^{\mathrm{art}} - p^{\mathrm{ven}}}{R^{\mathrm{per}}} ,
         \end{aligned}\right.
\end{equation}
The heart valves, \emph{i.e.,} the mitral and aortic valve, are modeled as ideal diodes: They open and close instantaneously and without any back-flow. Parameter values for the circulatory model are listed in Table~\ref{tab:paramslumpedLV} and are based on \cite{bovendeerdlvstrain}.

\begin{table}[]
\caption{\label{tab:paramslumpedLV} Circulatory system model parameters based on \cite{bovendeerdlvstrain}.}
\centering
\begin{tabular}{lllll}
\hline
                           & Parameter                          &   Value              &  Unit            \\ \hline
Arterial resistance        & $R^{\mathrm{art}}$                 & $1.00\times 10^{-2}$ & kPa s ml$^{-1}$ \\
Peripheral resistance      & $R^{\mathrm{per}}$                 & $1.20\times 10^{-1}$ & kPa s ml$^{-1}$ \\
Venous resistance          & $R^{\mathrm{ven}}$                 & $2.00\times 10^{-3}$ & kPa s ml$^{-1}$ \\
Arterial compliance        & $C^{\mathrm{art}}$                 & $2.50\times 10^{1}$  & ml kPa$^{-1}$   \\
Venous compliance          & $C^{\mathrm{ven}}$                 & $6.00\times 10^{2}$  & ml kPa$^{-1}$   \\
Arterial stressfree volume & $V_{\mathrm{ref0}}^{\mathrm{art}}$ & $5.00\times 10^{2}$  & ml              \\
Venous stressfree volume   & $V_{\mathrm{ref0}}^{\mathrm{ven}}$ & $3.00\times 10^{3}$  & ml              \\ 
Total blood volume         & $V^{\mathrm{tot}}$                 & $5.00\times 10^{3}$  & ml             
\end{tabular}
\end{table}

The remaining model parameters, as listed in Table~\ref{tab:params}, are identical with the exception of the pericardium spring stiffness, $k^{\rm{peri}}$, which is set to zero, and the reference active stress level which is set to $T^0=140$~[kPa]. The latter is a consequence of the unstiffening of the epicardium due to the removal of the pericardium stiffness. 

The temporal settings are fixed for all left ventricle simulations: Explicit time-integration for the circulatory system and sarcomere dynamics, a time step of $2$~[ms], and a cardiac cycle of $800$~[ms], \emph{i.e.}, activation of the entire heart every $800$th [ms].

\subsection{Mechanical strains}
\begin{figure*}[!tp]
\centering
\includegraphics[width=1.\textwidth]{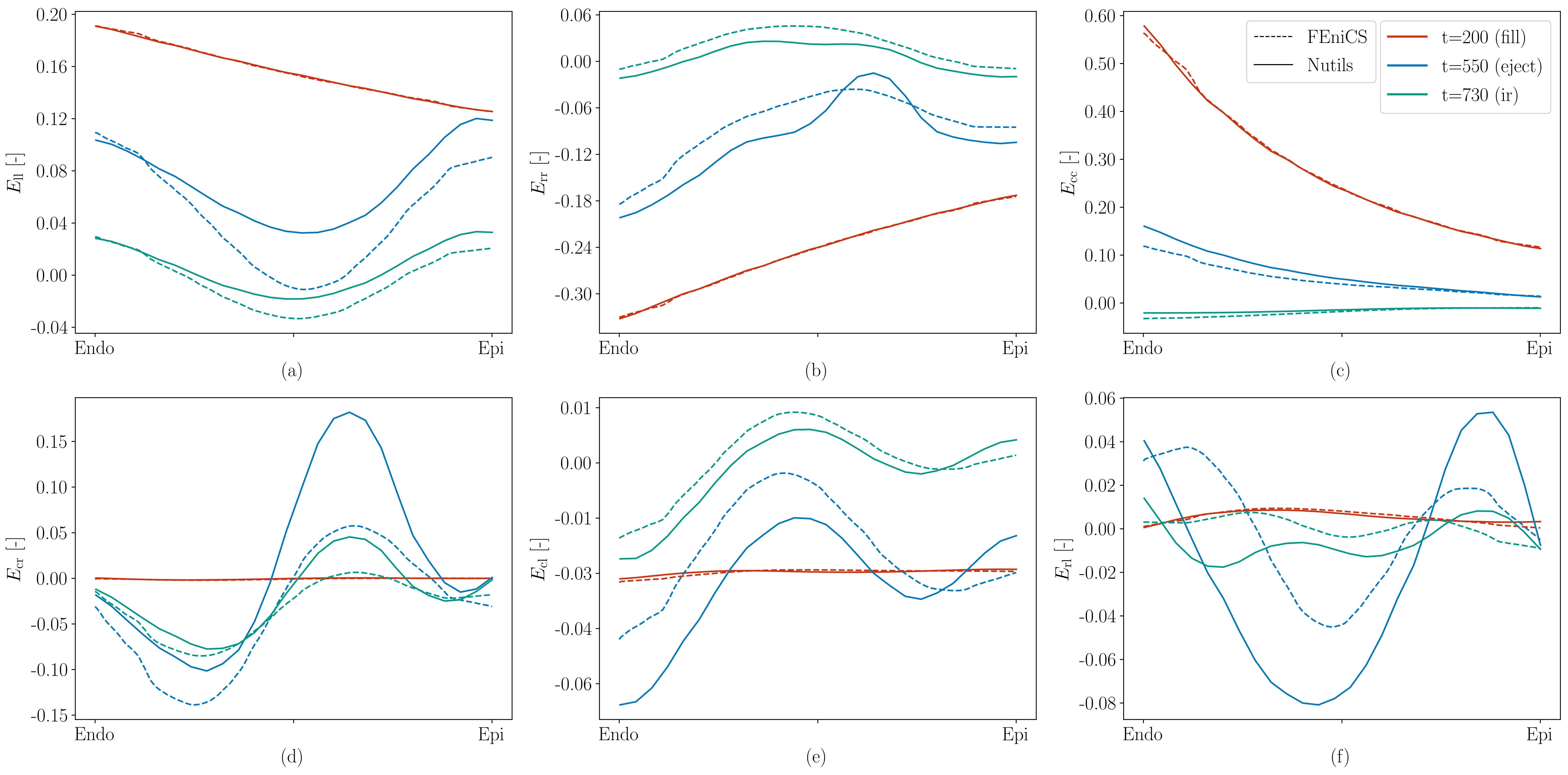}
  \caption{\label{fig:Localradialstrains} Green-Lagrange strain components for a cylindrical coordinate system with respect to the reference configuration. The results of FEniCs (FEA) and Nutils (IGA) are compared at three different time instances as a function of the radius. } 
\end{figure*}

The Green-Lagrange strain components of the IGA left ventricle model are compared to the FEniCs results at the equatorial plane. The strain tensor components are given for a cylindrical coordinate system $\{\mathbf{e}^{\rm{r}}, \mathbf{e}^{\rm{c}}, \mathbf{e}^{l}\}$, which is identical to the local basis $\{\mathbf{e}^{\rm{t0}}, \mathbf{e}^{\rm{c0}}, \mathbf{e}^{l0}\}$ of Figure~\ref{fig:RRBM} at the equatorial plane ($z$=0). The results are given in Figure~\ref{fig:Localradialstrains}.

\bibliographystyle{elsarticle-num} 
\bibliography{References}

\end{document}